\documentclass[twocolumn,aps,prx,superscriptaddress, 10pt]{revtex4-2}

\usepackage[english]{babel}
\usepackage{graphicx}
\usepackage[dvipsnames]{xcolor}
\usepackage{braket}
\usepackage{amsmath}
\usepackage{hyperref}
\usepackage{amsfonts}
\usepackage{amssymb}
\usepackage{dsfont}
\usepackage{ragged2e}
\usepackage{orcidlink}

\newtheorem{result}{Result}

\newcommand{\id}[0]{{\mathds{I}}}
\newcommand{\idmat}[0]{{\boldsymbol{I}}}
\newcommand{\dd}[0]{{\mathrm{d}}}
\newcommand{\tr}[1]{{\text{Tr}\left( #1 \right)}}
\newcommand{\trsquare}[1]{{ \text{Tr} \left[ #1 \right] }}
\newcommand{\trsys}[1]{{\operatorname{Tr}_{S}\!\left[ #1 \right]}}
\newcommand{\trenv}[1]{{\operatorname{Tr}_{E}\!\left[ #1 \right]}}
\newcommand{\myexp}[1]{\mathrm{e}^{#1}}
\newcommand{\norm}[1]{\left\lVert#1\right\rVert}

\begin{document}

\title{Global Precision Bounds and Success-Probability Guarantees in Quantum
  Parameter Learning}

\author{Federico Belliardo\,\orcidlink{0000-0002-1466-396X}}
\email{federico.belliardo@gmail.com}
\affiliation{Chicago Quantum Institute and Pritzker School of Molecular
  Engineering, University of Chicago, Chicago, Illinois 60637, USA}

\author{James W. Gardner\,\orcidlink{0000-0002-8592-1452}}
\affiliation{Chicago Quantum Institute and Pritzker School of Molecular
  Engineering, University of Chicago, Chicago, Illinois 60637, USA}

\author{Liang Jiang\,\orcidlink{0000-0002-0000-9342}\,}
\email{liangjiang@uchicago.edu}
\affiliation{Chicago Quantum Institute and Pritzker School of Molecular
  Engineering, University of Chicago, Chicago, Illinois 60637, USA}

\author{Aashish A. Clerk\,\orcidlink{0000-0001-7297-9068}\,}
\email{aaclerk@uchicago.edu}
\affiliation{Chicago Quantum Institute and Pritzker School of Molecular
  Engineering, University of Chicago, Chicago, Illinois 60637, USA}

\begin{abstract}

Quantum metrology offers the possibility of quantum enhancements of the
  precision of various sensing tasks. In this manuscript, we tackle two open
  problems in the theory of single-shot quantum parameter learning, going beyond
  the usual setting of local parameter estimation via repeated measurements. The
  first concerns the construction of global upper bounds on the learning
  precision. The second concerns rigorous guarantees on the success probability
  of parameter learning, namely, lower bounds on the probability of learning a
  parameter with a certain precision, given the constraints on the resources
  used for the quantum metrology task. We provide rigorous, practical, and
  global upper bounds and success-probability guarantees for quantum parameter
  learning. Most importantly, we establish a fidelity-based learning guarantee
  for generic mixed-state models that can be viewed as the achievability-side
  analogue of the quantum Cram\'er--Rao bound. Whereas the latter provides a
  no-go constraint, based on the local curvature of the fidelities, our bound
  uses only pairwise fidelities between parameter-encoded states to certify that
  a prescribed precision is attainable with a guaranteed success probability. We
  demonstrate the versatility of the new bounds in a Rabi-frequency-learning
  example involving a driven qubit coupled to a bosonic environment and a
  collective-spin Hamiltonian learning problem.
Together, the new global bounds and success-probability guarantees allow us to
  rule out unattainable precision and to certify attainable precision beyond
  what is possible via standard Fisher-information analysis or binary hypothesis
  testing bounds. They also allow one to tractably characterize the performance
  of various learning schemes, without the overhead of an explicit simulation.

\end{abstract}

\maketitle

\begin{figure*}[htbp!]
  \centering
  \includegraphics[width=\textwidth]{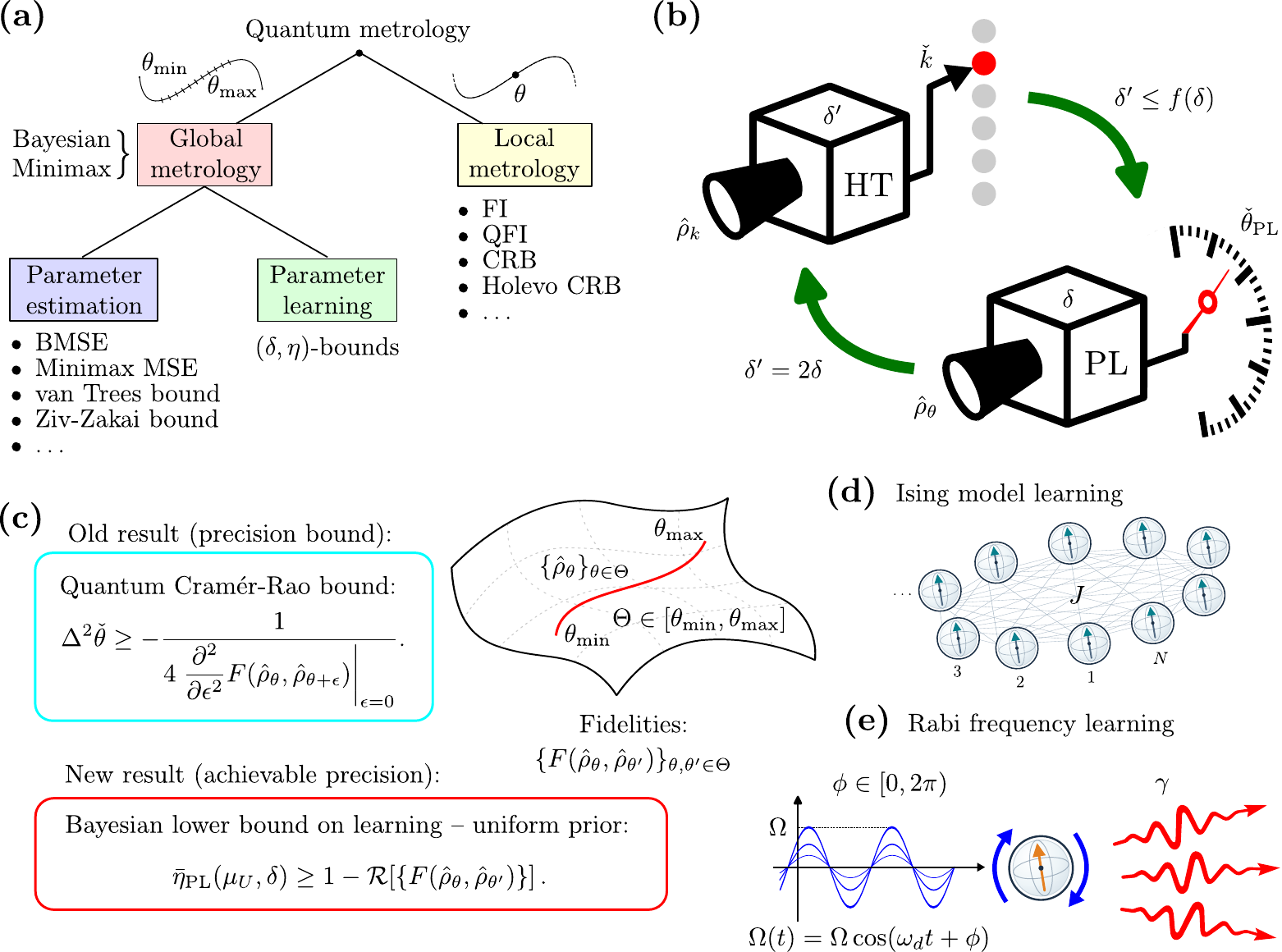}
  \caption{\justifying \textbf{(a)} Taxonomy of quantum metrology illustrating
    two distinctions: that between local and global metrology, determined by the
    prior information available about the parameter, and that between estimation
    and learning, determined by the choice of figure of merit. More details on
    this taxonomy are presented in
    Appendix~\ref{app:classification_metrology_tasks}. \textbf{(b)} An algorithm
    that solves parameter learning with precision $\delta$ can be used to
    construct a multi-hypothesis-testing algorithm with hypothesis spacing
    $\delta'=2\delta$. Conversely, an algorithm that solves multi-hypothesis
    testing with hypothesis spacing $0<\delta'\le f(\delta) \equiv \min \lbrace
    2\delta, 0.1/L \rbrace$ can be used to construct a parameter-learning
    algorithm with precision $\delta$. See Sec.~\ref{subsec:equivalence} for
    both results. \textbf{(c)} Schematic comparison of the main lower bound on
    the uniform-prior Bayesian success probability obtained in this work with
    the quantum Cram\'er--Rao bound. We consider a one-dimensional quantum
    statistical model $\{\hat{\rho}_{\theta}\}_{\theta\in\Theta}$ and the
    pairwise fidelities between its states. The quantum Cram\'er--Rao bound uses
    the local curvature of the fidelity to lower-bound the variance of a locally
    unbiased estimator. As such, it is a no-go result. By contrast, our result
    provides an informative lower bound on the uniform-prior Bayesian success
    probability, computed solely from pairwise fidelities, for learning $\theta$
    with prescribed finite precision $\delta$. In the notation shown,
    $\mathcal{R}$ is the function of the pairwise fidelities specified by
    Eq.~\eqref{eq:achievable_success_mixed_states}; see
    Sec.~\ref{subsubsec:achievable_success_mixed}. We show that this bound is
    informative for learning the Rabi frequency. \textbf{(d)} Single-shot
    learning of the all-to-all Ising coupling $J$ in a fully connected model of
    $N$ two-level systems at a fixed evolution time; see
    Sec.~\ref{subsec:hamiltonian_learning_ising}. \textbf{(e)} Learning the Rabi
    frequency $\Omega$ induced when a coherent drive of known carrier frequency
    and unknown field strength acts on a spin coupled to a dissipative
    environment. We obtain a rigorous lower bound on the uniform-prior Bayesian
    success probability even when the initial drive phase is unknown and
    uniformly distributed; see
    Sec.~\ref{subsec:dissipative_frequency_learning}.}
  \label{fig:figure_1}
\end{figure*}

\section{Introduction}

Quantum metrology seeks to employ non-classical states and other resources to
  provide advantages in determining one or more unknown parameters through
  quantum measurements. Its best-known form corresponds to local parameter
  estimation, where the target parameter is treated as an infinitesimal
  deviation from a known reference value. The experimentalist can therefore
  choose state preparation, measurement, and data processing so as to maximize
  information extraction at that reference
  value~\cite{giovannettiQuantumMetrology2006,
    giovannettiAdvancesQuantumMetrology2011,
    parisQuantumEstimationQuantum2009}. Success is usually quantified by the
      mean-squared estimation error, and rigorous bounds based on the
      Cram\'er--Rao (CR) bound and quantum Fisher information (QFI) are well
      known~\cite{parisQuantumEstimationQuantum2009} (though they are often only
      achievable in the limit of many repeated identical measurements). See
      Fig.~\ref{fig:figure_1} for a classification of quantum metrology tasks.

Despite its power, in many practical settings one needs to go beyond the
  framework of local estimation because the initial uncertainty in the target
  parameter is large. Further, one is often interested in the performance of a
  single experiment. In this more challenging global, single-shot setting, it is
  more useful to go beyond the mean-squared error and instead take a
  learning-theoretic approach. The goal is now to design a single-shot sensing
  protocol in which the estimator recovers the unknown parameter to within a
  prescribed target resolution with high probability. This framework is known as
  \textit{parameter learning}, also called \textit{PAC-metrology}
  (probably-approximately-correct metrology)~\cite{valiantTheoryLearnable1984,
    kearnsIntroductionComputationalLearning1994,
    meyerQuantumMetrologyFiniteSample2025}. This framework is also relevant
      beyond the strictly single-shot regime when the experimentalist has access
      to only a limited number of copies of the encoded probe state. In this
      finite-sample regime, it is therefore natural to characterize performance
      by both a target precision and a failure probability, defined as the
      probability that the estimate lies outside the prescribed
      tolerance~\cite{belliardoAchievingHeisenbergScaling2020}. An analogous
      separation is standard in classical statistics, where confidence-interval
      width and coverage probability play roles corresponding to precision and
      success probability, respectively~\cite{casellaStatisticalInference2002}.

Among its many virtues, parameter-learning metrics obey relatively strong
  continuity properties, making them robust against state-preparation and
  measurement (SPAM) errors. This is in contrast to the quantum Fisher
  information, where continuity is not controlled by trace-distance closeness
  alone\footnote{The quantum Fisher information is not controlled by
  trace-distance closeness alone: continuity additionally requires control of
  the quantum state derivatives, and discontinuities can occur when the support
  of the outcome distribution or the rank of the quantum statistical model
  changes.}~\cite{rezakhaniContinuityQuantumFisher2019,
    sevesoDiscontinuityQuantumFisher2020}. This robustness makes the PAC
      framework especially relevant to
recent quantum-computational quantum-sensing protocols whose implementations
  will invariably be subject to SPAM
  errors~\cite{khanQuantumComputationalSensing2025,
    khanQuantumComputationalSensingAdvantage2025}.

Given the importance and wide applicability of the learning approach to quantum
  metrology, it is crucial to understand the ultimate limits of such protocols
  and derive rigorous performance bounds. Previous work related
  finite-resolution global metrology to multi-hypothesis testing and derived a
  finite-sample Cram\'er--Rao-type bound based on the quantum Fisher
  information~\cite{meyerQuantumMetrologyFiniteSample2025}. However, one might
  expect that such approaches could miss constraints that are truly global in
  nature (e.g.,~involving the properties of the sensor's state throughout the
  entire parameter domain). In this work, we address this important open issue
  by introducing a family of \textit{global} upper bounds on the success
  probability of parameter learning. We emphasize that, unlike the finite-sample
  Cram\'er--Rao-type bound of Ref.~\cite{meyerQuantumMetrologyFiniteSample2025},
  our bounds retain the joint distinguishability of all encoded probe states
  across the allowed parameter range rather than reducing the problem to a
  collection of binary tests. In particular, we will see in the Applications
  section how the global bound is more informative than the quantum Fisher
  information.

We show that our global bounds can yield crucial insights through examples
  involving the quantum phase estimation algorithm and parameter learning in an
  all-to-all Ising model. They can certify the impossibility of successful
  learning or of achieving certain precision scalings in regimes where
  conventional bounds based on the Fisher information or binary hypothesis
  testing do not identify an obstruction. We use the term
  \emph{multi-hypothesis} to emphasize that these bounds account for the joint
  distinguishability of all possible hypotheses. In the same spirit, we also use
  the term \emph{global} for these bounds, emphasizing that they assess
  distinguishability over the full parameter range, rather than only over pairs
  of parameter values.

Crucially, our work also establishes a set of rigorous \textit{lower} bounds on
  the success probability of quantum parameter learning, namely,
  success-probability guarantees at a prescribed precision. We present in the
  main text an informative lower bound on the uniform-prior Bayesian success
  probability that applies to generic mixed-state models and requires only
  pairwise fidelities between parameter-encoded sensor states. We apply the
  mixed-state bound to a single interrogation in a continuous metrology
  protocol~\cite{albarelliPedagogicalIntroductionContinuously2024,
    tsangQuantumMetrologyOpen2013,
    gammelmarkFisherInformationQuantum2014}, where the goal is to use the
      continuously emitted field from a qubit sensor to estimate the amplitude
      of a Rabi drive (in a setting where the drive frequency is known, but the
      phase is completely uncertain).

Given a prescribed target precision for the amplitude estimate, we determine an
  integration time beyond which we can rigorously certify the existence of a
  joint measurement on the spin and the environment that achieves this
  precision.

Both the global upper bounds and the success-probability guarantees we derive
  rely on a rigorous connection between quantum parameter learning and quantum
  multi-hypothesis discrimination among states. The two frameworks are
  intuitively related: discriminating a sufficiently fine set of hypotheses is,
  in effect, a parameter-learning task. To obtain quantitative performance
  bounds, however, one needs a precise connection between the two tasks. We
  establish this by showing that a sufficiently fine hypothesis-discrimination
  task yields rigorous guarantees for parameter learning. We emphasize that the
  multi-hypothesis success-probability guarantee we derive in this manuscript
  for mixed states is expressible solely in terms of pairwise fidelities. It is
  thus dramatically more tractable to compute than bounds that require a direct
  optimization over the set of all possible measurements on the system.

The results of this manuscript provide a toolkit of widely applicable and
  computable success-probability guarantees and no-go results for global quantum
  metrological problems. Crucially, these tools go beyond the quantum Fisher
  information and bounds based only on binary hypothesis testing, and we expect
  them to be useful in future research on global quantum metrology. Through the
  Rabi-frequency-learning example, we have shown that our tools also apply to
  the highly relevant but challenging domain of continuous quantum metrology.

\subsection{Outline of the manuscript}
\label{subsec:overview_manuscript}
The remainder of this paper is structured as follows.
  Section~\ref{sec:background} establishes the background needed for our
  results, focusing on the definitions of parameter learning and hypothesis
  testing and on the connection between the two. Section~\ref{sec:results}
  contains the main results of the paper; specifically, in
  Section~\ref{sec:upper_bound_learning_success}, we present the global upper
  bounds on the success probability of learning, followed by the corresponding
  success-probability guarantees in Section~\ref{subsec:lower_bounds_success}.
  Section~\ref{sec:example} first applies the global upper bounds to learning
  the interaction strength of an all-to-all Ising model and then analyzes
  single-shot learning of the Rabi frequency of a driven spin coupled to a
  Markovian bosonic environment with and without a phase reference.

\section{Background}
\label{sec:background}
In this section, we introduce the background needed to state the main results of
  the paper and discuss the corresponding applications. We clarify the context
  of our work, introduce the basic definitions, and fix the notation.

\subsection{Quantum metrology}
\label{subsec:intro_quantum_metrology}
At the heart of quantum metrology lies the task of estimating an unknown
  physical parameter, $\theta$, encoded in a quantum system. We focus here on
  single-parameter estimation. We consider the quantum state of a \textit{probe}
  or \textit{sensor} encoded with a parameter $\theta$, i.e.,
  $\hat{\rho}_\theta$, which we measure to learn the parameter itself. The
  parameter $\theta$ takes values in a set $\Theta$, so that the statistical
  model on which we operate is $\lbrace \hat{\rho}_\theta \rbrace_{\theta \in
  \Theta}$. Throughout this manuscript, we assume that $\Theta \equiv
  [\theta_{\min}, \theta_{\max}]$ is an interval. We also define $|\Theta| =
  |\theta_{\max} - \theta_{\min}|$ as the range of the parameter. To retrieve
  $\theta$, we measure the state with a general positive operator-valued measure
  (POVM) $\mathcal{M}$, represented by the POVM density $\hat M_{\check\theta}$,
  which returns the estimator
  $\check{\theta} \in \Theta$. The corresponding probability density is given by
  the Born rule:

\begin{equation}
    p(\check{\theta}|\hat{\rho}_\theta, \mathcal{M}) \equiv
      \trsquare{\hat{\rho}_\theta \hat{M}_{\check{\theta}}} \,.
    \label{eq:born_rule_learning}
\end{equation}

Extracting the true value of $\theta$ from $\hat{\rho}_\theta$ requires an
  optimal choice of measurement $\mathcal{M}$, which we assume already contains
  the classical data-processing step. The standard scenario in quantum metrology
  is local estimation around a specific, known reference value $\theta$, where
  the task is to minimize the mean-squared error. For a fixed value of the
  parameter, the mean-squared error of a measurement-estimator strategy is
\begin{equation}
    \operatorname{MSE}_{\theta}(\mathcal{M})
    \equiv
    \int_{\Theta}
    (\check{\theta}-\theta)^2
    p(\check{\theta}|\hat{\rho}_\theta,\mathcal{M})\,\dd\check{\theta} \, .
    \label{eq:mse_definition}
\end{equation}
A review of the classification of quantum metrology tasks
  (Fig.~\ref{fig:figure_1}) is provided in
  Appendix~\ref{app:classification_metrology_tasks}. In short, quantum metrology
  tasks can be classified into \textit{local} or \textit{global} problems
  according to the prior information available about the parameters, and into
  \textit{estimation} or \textit{learning} problems according to the chosen
  figure of merit. In this manuscript, we focus on global learning, where
  $\theta$ may lie anywhere in $\Theta$, and we study the probability that the
  estimate lies within a prescribed distance of the true parameter, as described
  in the following section.

\subsection{Quantum parameter learning}
\label{subsec:quantum_parameter_learning}
We now define quantum parameter learning for a single parameter encoded in the
  quantum state of a system.

\subsubsection{Success probability of parameter learning}
In parameter learning, rather than evaluating the average error of an estimator,
  we seek probabilistic guarantees on its precision. Specifically, we seek
  measurement-and-estimation strategies that guarantee, with probability at
  least $\eta$, that the estimate lies within a distance $\delta$ of the true
  parameter. This is called a $(\delta, \eta)$-bound. We now formally define the
  Bayesian and minimax success probabilities for quantum parameter learning.

We first define the window function:
\begin{equation}
    w_\delta(\check{\theta} - \theta) \equiv
    \begin{cases}
      1 & \text{if } |\check{\theta} - \theta| \le \delta, \\
      0 & \text{otherwise}
    \end{cases} \,,
    \label{eq:window_function}
\end{equation}
which identifies the successful estimates for a given true parameter value. We
  can write the success probability of parameter learning for a fixed parameter
  $\theta$, precision $\delta$, and measurement $\mathcal{M}$ as
\begin{equation}
    \mathsf{P}_{\text{PL}} (\theta, \delta; \mathcal{M}) \equiv \int_{\Theta}
      w_\delta (\check{\theta} - \theta) p(\check{\theta}|\hat{\rho}_\theta,
      \mathcal{M})\,\dd\check{\theta} \,.
    \label{eq:parameter_learning_eta}
\end{equation}
The minimax success probability is
\begin{equation}
    \eta_{\text{PL}} (\delta) \equiv \sup_{\mathcal{M}} \inf_\theta
      \mathsf{P}_{\text{PL}} (\theta, \delta; \mathcal{M}) \,,
    \label{eq:parameter_learning_eta_minimax}
\end{equation}
which selects the POVM that optimizes parameter learning for the worst-case
  point $\theta \in \Theta$. We also define the Bayesian setting for parameter
  learning by introducing the Bayesian success probability:
\begin{equation}
    \bar{\mathsf{P}}_{\text{PL}} (\mu, \delta; \mathcal{M}) \equiv
      \int_{\theta_{\min}}^{\theta_{\max}} \dd \mu (\theta) \,
      \mathsf{P}_{\text{PL}} (\theta, \delta; \mathcal{M})\,,
    \label{eq:parameter_learning_eta_bayesian_povm}
\end{equation}
which represents the average success probability according to a Bayesian
  probability measure $\mu$ on $\Theta$.
Optimizing over the POVM gives
\begin{equation}
    \bar{\eta}_{\text{PL}} (\mu, \delta) \equiv \sup_{\mathcal{M}}
      \bar{\mathsf{P}}_{\text{PL}} (\mu, \delta; \mathcal{M}) \; .
    \label{eq:parameter_learning_eta_bayesian}
\end{equation}
In classical information theory, the minimax
  theorem~\cite{hollanderMathematicalStatisticsDecision1968} relates the
  Bayesian and minimax success probabilities. Under the regularity assumptions
  detailed in Appendix~\ref{app:application_minimax_theorem_pl}, the quantum
  minimax theorem~\cite{tanakaQuantumMinimaxTheorem2014} gives
\begin{equation}
    \eta_{\text{PL}} (\delta) = \inf_{\mu} \bar{\eta}_{\text{PL}} (\mu, \delta)
      \,,
    \label{eq:pl_minimax_theorem}
\end{equation}
where the infimum is computed over all probability measures on $\Theta$. If the
  infimum is attained, its minimizer is the least favorable prior. See
  Appendix~\ref{app:application_minimax_theorem_pl} for a discussion of the
  hypotheses required to apply the quantum minimax theorem. We define a learning
  task as successfully solved if the success probability is at least $3/4$. This
  threshold is chosen arbitrarily to maintain a significant gap from a $50\%$
  success probability, so that standard probability amplification using
  $O(\log(1/\varepsilon))$ independent repetitions can be used to reach any
  desired $1-\varepsilon$ success
  probability~\cite{schapireStrengthWeakLearnability1990}.

To summarize, an algorithm that solves the learning task seeks to maximize the
  Bayesian
($\bar{\eta}_{\rm PL}(\mu,\delta)$)
or the minimax ($\eta_{\rm PL}(\delta)$) success probability for a given
  precision $\delta$. Below, we show how connecting learning to hypothesis
  testing allows us to upper- and lower-bound the optimal success probability
  for a given precision, in both the Bayesian and minimax settings. An upper
  bound below $3/4$ certifies impossibility, whereas a lower bound of at least
  $3/4$ guarantees successful learning.

Finally, if $\hat{\rho}_\theta$ is a family of mutually commuting quantum
  states, then learning a parameter from a quantum state reduces to learning a
  parameter from a classical probability distribution. The relevant classical
  distribution is produced by measuring in the common eigenbasis that
  diagonalizes $\hat{\rho}_\theta$ for all $\theta$, which is also the optimal
  measurement that realizes the supremum in
  Eqs.~\eqref{eq:parameter_learning_eta_minimax}
  and~\eqref{eq:parameter_learning_eta_bayesian}.

\subsubsection{Nuisance parameters}
\label{subsubsec:nuisance_parameters}
The single-parameter formulation above assumes that the probe state depends only
  on the parameter of interest. In many metrological settings, however, the
  state also depends on additional unknown quantities that are not themselves
  the target of the learning task. We denote such nuisance parameters by $\xi$
  and write the corresponding family as $\ket{\psi_{\theta,\xi}}$, where the
  goal is to learn $\theta$ with precision $\delta$ without imposing any
  accuracy requirement on $\xi$. If the figure of merit is Bayesian with respect
  to the nuisance parameter, so that the success probability is averaged over a
  prior $p(\xi)$, then a single use of the experiment is equivalent, at the
  level of all measurement statistics for $\theta$, to learning from the
  averaged state
\begin{equation}
    \hat{\rho}_\theta \equiv \int \dd \xi \, p(\xi) \ket{\psi_{\theta, \xi}} \!
      \bra{\psi_{\theta, \xi}} \, ,
    \label{eq:mixed_state_nuisance}
\end{equation}
because both descriptions give the same probabilities for every POVM. A complete
  treatment of learning in the presence of nuisance parameters requires a theory
  of multiparameter quantum learning, which we leave to future work.

\subsection{Quantum multi-hypothesis testing}
\label{sec:quantum_multi_hypothesis_testings}

In this section, we review hypothesis testing between multiple quantum states
  and introduce a quasi-optimal solution to this problem: the pretty good
  measurement (PGM)~\cite{hausladenPrettyGoodMeasurement1994}.

As summarized in Table~\ref{table:pl_ht_comparison}, the definitions of the
  parameter-learning and hypothesis-testing success probabilities have closely
  parallel forms: the integral over the continuous parameter in the former is
  replaced by a sum over discrete hypotheses in the latter. Relating their
  operational performance, however, requires more than replacing an integral by
  a sum. In Sec.~\ref{subsec:equivalence}, we establish reductions in both
  directions, specifying the required discretization scales and the continuity
  corrections that arise when passing from hypothesis testing to parameter
  learning.

\subsubsection{Success probability of hypothesis testing}
We are given a state $\hat\rho$ that is guaranteed to be one of $\lbrace
  \hat{\rho}_k \rbrace_{k=1}^K$, say $\hat{\rho}_{k^\star}$, where $K$ is the
  total number of states (hypotheses). The task is to identify the index
  $k^\star$ with high success probability. As in the parameter-learning task, we
  use a POVM
\begin{equation}
    \mathcal{M} \equiv \lbrace \hat{M}_{\check{k}} \rbrace_{\check{k} \in
      \lbrace 1, \ldots, K\rbrace } \,,
\end{equation}
with a finite number of operators, including any classical processing of the
  measurement outcomes. The probability distribution for the estimator, given
  the true index $k^\star$ and the measurement $\mathcal{M}$, is computed via
  the Born rule:
\begin{equation}
    p(\check{k}|k^\star, \mathcal{M}) \equiv \trsquare{\hat{\rho}_{k^\star}
      \hat{M}_{\check{k}}} \,.
    \label{eq:born_rule_ht}
\end{equation}
The success probability for identifying the correct hypothesis is
\begin{equation}
    \mathsf{P}_{\text{HT}} (k^\star; \mathcal{M}) \equiv p(k^\star|k^\star,
      \mathcal{M}) \,.
\end{equation}
Analogously to parameter learning, we can define the minimax and Bayesian
  success probabilities. The minimax success probability is defined as
\begin{equation}
    \eta_{\text{HT}} \equiv \sup_{\mathcal{M}} \inf_{k^\star}
      \mathsf{P}_{\text{HT}} (k^\star; \mathcal{M}) \,.
    \label{eq:hypotheses_testing_eta_minimax}
\end{equation}
Unlike Eq.~\eqref{eq:parameter_learning_eta_minimax}, this optimal minimax
  success probability does not depend on a precision parameter, because success
  requires exact identification of the hypothesis. Given a discrete prior
  distribution $\mu \equiv \lbrace \mu_k \rbrace_{k=1}^K$, we define the
  Bayesian success probability
\begin{equation}
    \bar{\mathsf{P}}_{\text{HT}} (\mu; \mathcal{M}) \equiv \sum_{k=1}^K \mu_k \,
      \mathsf{P}_{\text{HT}} (k; \mathcal{M}) \,.
    \label{eq:hypotheses_testing_eta_bayesian_povm}
\end{equation}
Optimizing this quantity over the POVM gives
\begin{equation}
    \bar{\eta}_{\text{HT}} (\mu) \equiv \sup_{\mathcal{M}} \sum_{k=1}^K \mu_k \,
      \mathsf{P}_{\text{HT}} (k; \mathcal{M}) \,.
    \label{eq:hypotheses_testing_eta_bayesian}
\end{equation}
As for parameter learning, we invoke the quantum minimax
  theorem~\cite{tanakaQuantumMinimaxTheorem2014}, which states that
\begin{equation}
    \eta_{\text{HT}} = \inf_{\mu} \bar{\eta}_{\text{HT}} (\mu) \,.
    \label{eq:ht_quantum_minimax}
\end{equation}
See Appendix~\ref{app:application_minimax_theorem_ht} for a discussion of the
  applicability of the quantum minimax theorem in this scenario, including why
  the hypotheses of the theorem are satisfied.
In the Bayesian scenario, both the true hypothesis $k$ and the estimator
  $\check{k}$ are random variables, distributed according to the prior $\mu_k$
  and the marginal distribution
\begin{equation}
    p_\mu(\check{k}| \mathcal{M}) \equiv \sum_{k=1}^K \mu_k p(\check{k}|k,
      \mathcal{M}) \,,
    \label{eq:prob_estimator_ht}
\end{equation}
respectively.

\begin{table*}[htbp]
\centering
\renewcommand{\arraystretch}{2.2}
\begin{ruledtabular}
\begin{tabular}{ccc}
\textbf{Setting} & \textbf{Parameter learning} & \textbf{Hypothesis testing} \\
\hline
Minimax &
$\displaystyle \eta_{\text{PL}}(\delta) \equiv \sup_{\mathcal{M}} \inf_{\theta}
  \mathsf{P}_{\text{PL}}(\theta,\delta;\mathcal{M})$,
  Eq.~\eqref{eq:parameter_learning_eta_minimax} &
$\displaystyle \eta_{\text{HT}} \equiv \sup_{\mathcal{M}} \inf_{k^\star}
  \mathsf{P}_{\text{HT}}(k^\star;\mathcal{M})$,
  Eq.~\eqref{eq:hypotheses_testing_eta_minimax} \\
Bayesian (optimal POVM) &
$\displaystyle \bar{\eta}_{\text{PL}}(\mu,\delta) \equiv \sup_{\mathcal{M}}
  \int_{\theta_{\min}}^{\theta_{\max}} \!\! \dd\mu(\theta)\,
  \mathsf{P}_{\text{PL}}(\theta,\delta;\mathcal{M})$,
  Eq.~\eqref{eq:parameter_learning_eta_bayesian} &
$\bar{\eta}_{\text{HT}} (\mu) \equiv \sup_{\mathcal{M}} \sum_{k=1}^K \mu_k \,
  \mathsf{P}_{\text{HT}} (k; \mathcal{M})$,
  Eq.~\eqref{eq:hypotheses_testing_eta_bayesian}\\
\end{tabular}
\end{ruledtabular}
\caption{\justifying Comparison of the Bayesian (optimal measurement) and
  minimax success probabilities for parameter learning and hypothesis testing.
  The integral over the continuous parameter $\theta$ with prior measure $\mu$
  in parameter learning is replaced by a sum over the discrete hypotheses with
  prior $\mu_k$ in hypothesis testing, and the parameter-learning quantities
  additionally depend on the precision $\delta$.}
\label{table:pl_ht_comparison}
\end{table*}

\subsubsection{The pretty good measurement}
\label{subsec:pgm}
We introduce the pretty good
  measurement~\cite{hausladenPrettyGoodMeasurement1994} (PGM) as a quasi-optimal
  solution to multi-hypothesis testing on quantum states. We formulate the PGM
  in the Bayesian framework and therefore suppose that the unknown state belongs
  to the ensemble $\lbrace (\mu_k, \hat{\rho}_k)\rbrace_{k=1}^K$, where $\mu_k$
  is the prior probability assigned to the $k$th state. We define the average
  ensemble state as $\bar{\rho} = \sum_{k=1}^{K} \mu_k \hat{\rho}_k$. The
  measurement operators of the PGM are then
\begin{equation}
    \hat{M}_k = {\bar\rho}^{-1/2} (\mu_k \hat{\rho}_k) {\bar\rho}^{-1/2} \, .
    \label{eq:effect_PGM}
\end{equation}
The PGM construction remains well defined on infinite-dimensional Hilbert
  spaces~\cite{mishraNearoptimalPerformanceSquareroot2025}. The measurement
  operators $\hat{M}_k$ sum to the projector onto the support of $\bar\rho$, and
  the POVM can be completed arbitrarily on the orthogonal complement. We denote
  this POVM by $\mathcal{M}_{\text{PG}} \equiv \lbrace \hat{M}_k
  \rbrace_{k=1}^K$. Interestingly, if all $\hat{\rho}_k$ commute, the optimal
  measurement for hypothesis testing (measuring in the shared eigenbasis) does
  not, in general, coincide with the PGM.

\subsubsection{Quasi-optimality of the PGM (Barnum--Knill theorem)}
Let $\bar{\mathsf{P}}_{\text{HT}} (\mu; \mathcal{M}_{\text{PG}})$ denote the
  success probability of the PGM, as defined in
  Eq.~\eqref{eq:hypotheses_testing_eta_bayesian_povm}. The Barnum--Knill theorem
  gives the following inequalities relating it to the optimal Bayesian success
  probability~\cite{barnumReversingQuantumDynamics2000,
    barnumErratumReversingQuantumDynamics2026,
    mishraNearoptimalPerformanceSquareroot2025}:
\begin{equation}
    \bar{\eta}_{\text{HT}} (\mu)^2 \le \bar{\mathsf{P}}_{\text{HT}} (\mu;
      \mathcal{M}_{\text{PG}}) \le \bar{\eta}_{\text{HT}} (\mu) \,.
    \label{eq:barnum_knill}
\end{equation}
Equation~\eqref{eq:barnum_knill} holds separately for each prior. Taking the
  infimum over priors connects the minimax success probability to the infimum of
  the PGM success probability evaluated using the PGM associated with each
  prior.

\subsubsection{PGM success probability for pure states}
For pure hypothesis states of the form $\hat{\rho}_k = \ket{\psi_k} \!
  \bra{\psi_k}$, the PGM success probability admits a closed-form expression. We
  define the Gram matrix for the unnormalized states $\ket{\tilde{\psi}_k}
  \equiv \sqrt{\mu_k} \ket{\psi_k}$:
\begin{equation}
    \boldsymbol{G}_{k k'}\equiv\sqrt{\mu_k \mu_{k'}} \braket{\psi_k|\psi_{k'}}
      \,.
    \label{eq:gram_matrix}
\end{equation}
The success probability~\cite{wildeQuantumInformationTheory2013} is then given
  by
\begin{equation}
    \bar{\mathsf{P}}_{\text{HT}} (\mu; \mathcal{M}_{\text{PG}}) = \sum_{k=1}^K
      \big( \sqrt{\boldsymbol{G}}_{kk} \big)^2 \,.
    \label{eq:pgm_success}
\end{equation}
From the Barnum--Knill theorem in Eq.~\eqref{eq:barnum_knill}, we obtain the
  following upper bound for the optimal Bayesian success probability in quantum
  hypothesis testing:
\begin{equation}
    \bar{\eta}_{\text{HT}} (\mu) \le \sqrt{\sum_{k=1}^K \big(
      \sqrt{\boldsymbol{G}}_{kk} \big)^2} \, .
    \label{eq:pgm_upper_bound_success}
\end{equation}
We use this result to upper bound the success probability of quantum parameter
  learning. In Appendix~\ref{app:success_pgm}, we prove a generalized version of
  this result in which hypotheses are accepted within a tolerance band around
  the true hypothesis. This tolerance is essential for deriving the
  success-probability guarantees for parameter learning in
  Sec.~\ref{subsec:lower_bounds_success}. In the zero-tolerance limit, the
  generalized expression reduces to Eq.~\eqref{eq:pgm_success}.

\subsection{Equivalence between parameter learning and hypothesis testing}
\label{subsec:equivalence}

Reference~\cite{meyerQuantumMetrologyFiniteSample2025} established a formal
  connection between parameter learning and quantum multi-hypothesis testing.
  Intuitively, this connection is obtained by discretizing the one-dimensional
  parameter space $\Theta$ and hence the model for the probe state $\lbrace
  \hat\rho_\theta \rbrace_{\theta \in \Theta}$. Let $\delta$ be the precision of
  the parameter-learning task, as in Eq.~\eqref{eq:window_function}, and let
  $\delta'$ denote the resolution of the parameter discretization used to define
  the hypothesis-testing procedure. We realize this discretization through the
  $\delta'$-net
\begin{equation}
    \Omega \equiv \lbrace \theta_k \rbrace_{k=1}^{K(\delta')} \quad \text{with}
      \quad  \theta_k \equiv \theta_{\min} + \left(k - \tfrac{1}{2}\right)
      \delta'\,,
    \label{eq:set_hypothesis_definition}
\end{equation}
where
\begin{equation}
    K(\delta') \equiv {|\Theta|}/{\delta'} \; .
    \label{eq:definition_K_delta}
\end{equation}

For simplicity, we assume that $K(\delta') \in \mathbb{N}$, so that the
  intervals centered at $\theta_k \in \Omega$ exactly partition the original
  interval $\Theta$; otherwise, we slightly trim $\Theta$ so that this condition
  holds. For uniform-prior Bayesian bounds, trimming introduces corrections of
  order $\mathcal{O}(\delta'/|\Theta|)$; for minimax upper bounds, it may weaken
  the bound, whereas for minimax lower bounds, we instead should make sure that
  $\delta'=|\Theta|/K$, with $K\in\mathbb{N}$, so that $\Theta$ remains
  unchanged. Every point $\theta \in \Theta$ is then within a distance
  $\delta'/2$ of at least one element of $\Omega$. The grid points $\theta_k \in
  \Omega$ correspond to the hypothesis states $\lbrace \hat\rho_k \equiv
  \hat\rho_{\theta_k} \rbrace_{k=1}^{K(\delta')}$ that we aim to discriminate,
  with $K(\delta')$ defined in Eq.~\eqref{eq:definition_K_delta}. As illustrated
  in Fig.~\ref{fig:figure_1}, the parameter-learning and hypothesis-testing
  algorithms may be viewed as black boxes that perform tasks characterized by a
  learning precision $\delta$ and a hypothesis-testing resolution $\delta'$,
  respectively.

For a given spacing between discretized hypotheses, an algorithm that performs
  parameter learning with high success probability can be used to construct
  another algorithm that performs hypothesis testing over a discrete set of
  parameters, also with high success
  probability~\cite{meyerQuantumMetrologyFiniteSample2025} (see
  Fig.~\ref{fig:figure_1}). Conversely, an algorithm that performs hypothesis
  testing at sufficiently fine resolution on a discrete set of states can
  simulate parameter learning with a larger error and comparable success
  probability. This second implication requires the state family to be Lipschitz
  continuous in trace distance.

Together, these two results imply a formal equivalence between quantum parameter
  learning and discrete hypothesis testing (see Fig.~\ref{fig:figure_1}): when
  operated at sufficient resolution, either black-box algorithm can simulate the
  other with controlled success probability.

We summarize the equivalence between parameter learning and hypothesis testing
  in the following two statements.

\subsubsection{From parameter learning to hypothesis testing}
Given a black box that performs parameter learning with precision $\delta$ for a
  parameter $\theta \in \Theta$ encoded in a state $\hat{\rho}_\theta$, one can
  simulate a black box that performs hypothesis testing on any fixed finite set
  of states $\lbrace \hat{\rho}_{\theta_k} \rbrace_{k=1}^{K}$ whose parameter
  values $\theta_k\in\Theta$ are equally spaced by
  $\delta'>2\delta$~\cite{meyerQuantumMetrologyFiniteSample2025}. This simulated
  hypothesis-testing procedure has success probability at least as large as that
  of the learning procedure. Thus, the minimax success probability of parameter
  learning, $\eta_{\text{PL}}(\delta)$, is bounded above by that of hypothesis
  testing:

\begin{equation}
    \eta_{\text{PL}}(\delta) \le \eta_{\text{HT}} \,.
    \label{eq:theorem_statement_learning_to_testing}
\end{equation}
This implication is proved in Appendix~\ref{app:from_pl_to_ht}. In the same
  appendix, we derive an upper bound on the Bayesian learning success
  probability for a uniform prior. Let $\bar{\eta}_{\mathrm{HT}}^r(\mu_U)$
  denote the optimal Bayesian hypothesis-testing success probability for the
  uniform prior on the equally spaced grid with offset $r \in [0,\delta')$ from
  $\theta_{\min}$, meaning that the grid is defined by $\theta_k^r \equiv
  \theta_{\min}+r+(k-1)\delta'$. In Appendix~\ref{app:from_pl_to_ht}, we show
  that
\begin{equation}
    \bar{\eta}_{\mathrm{PL}}(\mu_U,\delta)
    \le
    \frac{1}{\delta'}\int_0^{\delta'} \dd r\,
    \bar{\eta}_{\mathrm{HT}}^r(\mu_U) \,.
    \label{eq:theorem_statement_learning_to_testing_bayesian_upper}
\end{equation}
Operationally, revealing the offset $r$ before the measurement can only make the
  learning problem easier. Conditioned on $r$, successful learning identifies
  the correct hypothesis because the grid spacing satisfies $\delta'>2\delta$;
  the resulting conditional learning success probability is therefore
  upper-bounded by the corresponding hypothesis-testing success probability.
  Equation~\eqref{eq:theorem_statement_learning_to_testing_bayesian_upper}
  follows by averaging this conditional bound over $r$, which is uniform because
  the prior on $\theta$ is uniform. If the hypothesis-testing success
  probability is invariant under shifts of the grid, the integrand is
  independent of $r$, and the bound reduces to
\begin{equation}
    \bar{\eta}_{\mathrm{PL}}(\mu_U,\delta)
    \le
    \bar{\eta}_{\mathrm{HT}}(\mu_U) \,.
    \label{eq:theorem_statement_learning_to_testing_bayesian_shift_invariant}
\end{equation}
Here, $\mu_U$ denotes the uniform prior on $\Theta$ for parameter learning and
  the uniform prior over the hypotheses for hypothesis testing. In the
  applications of the theorem, we take the limit $\delta'\to2\delta^+$; the
  notation $\delta'=2\delta$ refers to this limiting value.

\subsubsection{From hypothesis testing to parameter learning}
Given a black box that performs hypothesis testing on a set of hypotheses
  $\lbrace \hat{\rho}_{\theta_k} \rbrace_{k=1}^{K(\delta')}$ with $\theta_k \in
  \Omega$, as defined in Eq.~\eqref{eq:set_hypothesis_definition}, and
  $K(\delta')$ as defined in Eq.~\eqref{eq:definition_K_delta}, choose the
  discretization step $\delta'$ such that $\delta'\le 2\delta$. Let $L$ be the
  Lipschitz constant of the family of probe states $\lbrace \hat{\rho}_\theta
  \rbrace_{\theta \in \Theta}$ with respect to the trace distance $D(\hat{\rho},
  \hat{\sigma}) \equiv \tfrac{1}{2} \|\hat{\rho} - \hat{\sigma}\|_1$, i.e.,
\begin{equation}
     D(\hat{\rho}_{\theta}, \hat{\rho}_{\theta'}) =
       \frac{1}{2}\|\hat{\rho}_{\theta} - \hat{\rho}_{\theta'}\|_1 \le L |\theta
       - \theta'| \,.
\end{equation}
One can then implement an algorithm for parameter learning with precision
  $\delta$ whose minimax success probability is guaranteed to satisfy
\begin{equation}
    \eta_{\text{PL}} (\delta) \ge \inf_{\mu} \sup_{\mathcal{M}}
      \sum_{k=1}^{K(\delta')} \sum_{\check{k}\,:\,|k-\check{k}|\le a} \mu_k
      \trsquare{\hat{\rho}_{k} \hat{M}_{\check{k}}} - L\delta'/2 \,,
    \label{eq:theorem_statement_testing_to_learning}
\end{equation}
where
\begin{equation}
    a \equiv \left\lfloor \frac{\delta}{\delta'} - \frac{1}{2} \right\rfloor \,.
    \label{eq:expression_a_achievability}
\end{equation}
The integer $a$ is roughly the number of hypotheses that fit within a window of
  acceptance for the learning task. Here, the infimum is over all discrete prior
  probability distributions $\mu$, and the supremum is over all discrete POVMs
  $\mathcal{M}$ taking values in $\Omega$. A convenient choice of hypothesis
  spacing is
\begin{equation}
    \delta'_0 \equiv \min\left\lbrace 2\delta,\frac{0.1}{L}\right\rbrace .
    \label{eq:delta_prime_bayesian_convenient}
\end{equation}
With $\delta'=\delta'_0$, the correction $L \delta'/2$ reduces the
  success-probability guarantee by at most five percentage points. The precise
  numerical constant $0.1$ is not essential: it should be small enough not to
  substantially affect the guarantee, yet not so small that $\delta'$ becomes
  much finer than necessary.

Evaluating the minimax success-probability guarantee requires computing the
  infimum over priors, which cannot be avoided and is often infeasible. For this
  reason, we also prove a lower bound on the Bayesian success probability under
  a uniform prior $\mu_U(\theta)$ on the parameter. For every $\delta'\le
  2\delta$, this probability satisfies

\begin{equation}
    \begin{aligned}
    \bar{\eta}_{\text{PL}} & (\mu_U, \delta)
    \ge \frac{1}{K(\delta')} \sup_{\mathcal{M}}
    \sum_{k=1}^{K(\delta')}
    \sum_{\check{k}\,:\,|k-\check{k}|\le a}
    \trsquare{\hat{\rho}_{k} \hat{M}_{\check{k}}} \\
    &\quad - \frac{1}{|\Theta|}
    \sum_{k=1}^{K(\delta')}
    \int_{\theta_k-\delta'/2}^{\theta_k+\delta'/2}
    \dd\theta\,
    \sqrt{1-F^2\!\left(\hat{\rho}_{\theta},\hat{\rho}_{\theta_k}\right)} \, .
    \end{aligned}
    \label{eq:theorem_statement_testing_to_learning_bayesian}
\end{equation}
Here, $F(\hat{\rho},\hat{\sigma}) \equiv
  \tr{\sqrt{\sqrt{\hat{\rho}}\hat{\sigma}\sqrt{\hat{\rho}}}}$ is the fidelity.
  The displayed fidelity integral controls the continuity correction and is more
  precise than the term based on the Lipschitz constant used for the minimax
  result.

Maximizing the complete right-hand side over $\delta'\le 2\delta$ gives

\begin{equation}
    \begin{aligned}
    \bar{\eta}_{\text{PL}} & (\mu_U, \delta)
    \ge
    \sup_{\substack{\delta'\le 2\delta}}
    \Bigg\{
    \frac{1}{K(\delta')} \sup_{\mathcal{M}}
    \sum_{k=1}^{K(\delta')}
    \sum_{\check{k}\,:\,|k-\check{k}|\le a}
    \trsquare{\hat{\rho}_{k} \hat{M}_{\check{k}}} \\
    &\qquad - \frac{1}{|\Theta|}
    \sum_{k=1}^{K(\delta')}
    \int_{\theta_k-\delta'/2}^{\theta_k+\delta'/2}
    \dd\theta\,
    \sqrt{1-F^2\!\left(\hat{\rho}_{\theta},\hat{\rho}_{\theta_k}\right)}
    \Bigg\} \, .
    \end{aligned}
    \label{eq:theorem_statement_testing_to_learning_bayesian_optimized}
\end{equation}
Equation~\eqref{eq:theorem_statement_testing_to_learning_bayesian_optimized} can
  yield a stronger lower bound because it optimizes the full right-hand side
  over the grid rather than fixing $\delta'=\delta_0'$. The reference spacing in
  Eq.~\eqref{eq:delta_prime_bayesian_convenient} is simply one valid point in
  the optimization.

The implications above, including the Lipschitz and fidelity-integral continuity
  corrections, are proved in Appendices~\ref{app:from_ht_to_pl}
  and~\ref{app:from_pl_to_ht}, which together establish both directions of the
  equivalence between hypothesis testing and parameter learning. Although the
  precise testing-to-learning reduction is presented here for the first time, we
  include it in the Background section because of its conceptual simplicity and
  its close connection to Ref.~\cite{meyerQuantumMetrologyFiniteSample2025}.

\subsection{Binary hypothesis testing bounds on parameter learning}
\label{subsec:pairwise_local_bound}

We now review the minimax upper bound on the success probability of learning
  based on binary hypothesis testing, as used in
  Refs.~\cite{meyerQuantumMetrologyFiniteSample2025,
    huangQueryComplexitiesQuantum2026}. This bound reads
\begin{equation}
\begin{aligned}
    \eta_{\text{PL}} (\delta)
    &\le \inf_{\substack{\theta, \theta' \in \Omega \\ \theta \neq \theta'}}
      \left[ \frac{1}{2} + \frac{1}{4} \| \hat\rho_\theta - \hat\rho_{\theta'}
      \|_1 \right] \\
    &\le \inf_{\substack{\theta, \theta' \in \Omega \\ \theta \neq \theta'}}
      \left[ \frac{1}{2} + \frac{1}{2} \sqrt{1 - F^2(\hat\rho_\theta,
      \hat\rho_{\theta'})} \right] \,.
\end{aligned}
\label{eq:pairwise_success_probability}
\end{equation}
This upper bound follows from
  Eqs.~\eqref{eq:theorem_statement_learning_to_testing}
  and~\eqref{eq:ht_quantum_minimax}: we restrict the prior in the minimax
  representation to balanced binary distributions supported on two distinct
  hypotheses in the grid $\Omega$ defined in
  Eq.~\eqref{eq:set_hypothesis_definition}, evaluate the resulting binary
  success probability using the Helstrom bound, and minimize over the pair. It
  is a quantum generalization of Le Cam's two-point method for reducing
  classical parameter learning to binary hypothesis testing between classical
  probability
  distributions~\cite{tsybakovIntroductionNonparametricEstimation2009,
    camAsymptoticMethodsStatistical1986,
    yuAssouadFanoCam1997}. The fidelity expression follows from the Fuchs--van
      de Graaf inequality.

\subsection{Insufficiency of the binary hypothesis-testing bound}
\label{subsec:quantum_phase_estimation}
We present a phase-learning problem for which
  Eq.~\eqref{eq:pairwise_success_probability} is very loose: it does not rule
  out learning even in regimes excluded by a simple counting argument. The
  calculation is based on a variation of the quantum phase-estimation algorithm
  and motivates the introduction of tighter global upper bounds on the success
  probability of learning.

In the standard phase-estimation algorithm, we are given a state $\ket{\psi}$
  that is an eigenvector of a unitary operator $\hat{U}$, and our goal is to
  learn the parameter $\theta$ in $\hat{U}\ket{\psi} = \mathrm{e}^{2 \pi
  \mathrm{i} \theta} \ket{\psi}$ up to a target precision $\delta \simeq
  2^{-n}$. This is achieved using $n$ ancilla qubits initialized in the state
  $\hat{H}^{\otimes n} \ket{0}^{\otimes n}$, where $\hat{H}$ denotes the
  Hadamard gate. By applying successive controlled unitary operations
  $\text{C-}\hat{U}^{2^{j-1}}$ for $j=1, \dots, n$ on the register, we generate
  the encoded state
\begin{equation}
    \ket{\psi_\theta} \equiv \bigotimes_{j=1}^{n} \frac{1}{\sqrt{2}} \left(
      |0\rangle + \mathrm{e}^{2\pi \mathrm{i} \, 2^{j-1} \theta} |1\rangle
      \right) \,.
\end{equation}
Applying the inverse quantum Fourier transform and measuring yields an $n$-bit
  estimate of $\theta$.

We now consider a variation of this algorithm in which the controlled operations
  are instead $\text{C-}\hat{U}^{3^{j-1}}$ for $j=1, \dots, n$. Let $\theta =
  0.\theta_1 \theta_2 \theta_3 \dots$ be the infinite ternary expansion (base 3)
  of the parameter. The resulting encoded state $\ket{\psi_\theta}$ is
\begin{equation}
\begin{split}
    \ket{\psi_\theta} = {} & \frac{1}{\sqrt{2}} \left( |0\rangle +
      \mathrm{e}^{2\pi \mathrm{i} \, 0.\theta_1 \theta_2 \dots } |1\rangle
      \right) \\
    & \otimes \frac{1}{\sqrt{2}} \left( |0\rangle + \mathrm{e}^{2\pi \mathrm{i}
      \, 0.\theta_2 \theta_3 \dots } |1\rangle \right) \\
    & \otimes \dots \otimes \frac{1}{\sqrt{2}} \left( |0\rangle +
      \mathrm{e}^{2\pi \mathrm{i} \, 0.\theta_n \theta_{n+1} \dots } |1\rangle
      \right) \,.
\end{split}
\label{eq:initial_state_phase_estimation_ternary}
\end{equation}
We target a learning precision of $\delta = 3^{-n}/2$, which implies a
  hypothesis-testing resolution of $\delta' = 3^{-n}$ according to
  Eq.~\eqref{eq:theorem_statement_learning_to_testing}. This resolution
  corresponds to truncating $\theta$ to its first $n$ ternary digits, yielding
  the following set of hypotheses:
\begin{equation}
\begin{split}
    \ket{\psi_{0.\theta_1 \theta_2 \dots \theta_n}} = {} & \frac{1}{\sqrt{2}}
      \left( |0\rangle + \mathrm{e}^{2\pi \mathrm{i} \, 0.\theta_1 \theta_2
      \dots \theta_n} |1\rangle \right) \\
    & \otimes \frac{1}{\sqrt{2}} \left( |0\rangle + \mathrm{e}^{2\pi \mathrm{i}
      \, 0.\theta_2 \dots \theta_n} |1\rangle \right) \\
    & \otimes \dots \otimes \frac{1}{\sqrt{2}} \left( |0\rangle +
      \mathrm{e}^{2\pi \mathrm{i} \, 0.\theta_n} |1\rangle \right) \,.
\end{split}
\end{equation}

We now calculate the squared magnitude of the overlap between two distinct
  hypothesis states to evaluate the corresponding pairwise learning bound in
  Eq.~\eqref{eq:pairwise_success_probability}. This quantity is
\begin{equation}
    |\braket{\psi_{\theta} | \psi_{\theta'}}|^2 = \prod_{j=1}^n \cos^2 \left[
      \pi 3^{j-1} (\theta - \theta') \right] \,.
    \label{eq:overlap_hypotheses}
\end{equation}
Since each hypothesis is characterized by a finite number of digits, we can
  explicitly express this overlap in terms of the ternary expansions of $\theta$
  and $\theta'$:
\begin{equation}
\begin{aligned}
    |\braket{\psi_{\theta} | \psi_{\theta'}}|^2 =
    & \cos^2 \left[ \pi (0.\theta_1 \dots \theta_n - 0.\theta_1' \dots
      \theta_n') \right] \\
    & \cdot \cos^2 \left[ \pi (0.\theta_2 \dots \theta_n - 0.\theta_2' \dots
      \theta_n') \right] \\
    & \ \vdots \\
    & \cdot \cos^2 \left[ \pi (0.\theta_n - 0.\theta_n') \right] \,.
\end{aligned}
\label{eq:overlap_hypotheses_digits}
\end{equation}
Consider the final digits $\theta_n$ and $\theta_n'$. If they differ, the
  argument of the corresponding cosine term is $\pi/3$ or $2\pi/3$, in either
  case giving $\cos^2 = 1/4$. If they are identical, we examine the $(n-1)$-th
  digits. If these differ, the expression $\pi (0.\theta_{n-1}\theta_n -
  0.\theta_{n-1}'\theta_n')$ simplifies to $\pi/3$ or $2\pi/3$ because $\theta_n
  = \theta_n'$. Proceeding inductively, we find that, because the two hypotheses
  are distinct ($\theta \neq \theta'$), at least one pair of digits differs.
  Hence, the argument of at least one cosine term is $\pi/3$ or $2\pi/3$. This
  observation upper-bounds the squared overlap by $\cos^2 (\pi/3) = 1/4$, giving
  the following lower bound on the pairwise Helstrom upper-bound expression:
\begin{equation}
\begin{aligned}
    \inf_{\substack{\theta, \theta' \in \Omega \\ \theta \neq \theta'}}
    \left[ \frac{1}{2} + \frac{1}{2}\sqrt{1 - |\braket{\psi_\theta |
      \psi_{\theta'}}|^2} \right]  \ge \frac{1}{2} + \frac{\sqrt{3}}{4} \approx
      0.933 \,.
\end{aligned}
\label{eq:phase_pairwise_helstrom_bound}
\end{equation}
Thus, the pairwise bound in Eq.~\eqref{eq:pairwise_success_probability} does not
  flag the impossibility of learning.

We now consider a multi-hypothesis upper bound obtained from a counting
  argument. For a $\delta'$-net of $K(\delta')$ hypotheses whose supports span
  an effective Hilbert space of dimension $D_{\text{eff}}$, the average success
  probability for a uniform prior is at most $D_{\text{eff}}/K(\delta')$. Let
  $\hat{\Pi}_{\text{eff}}$ denote the projector onto the span of the hypothesis
  supports. For any POVM $\lbrace \hat{M}_k \rbrace_{k=1}^{K(\delta')}$,
\begin{equation}
    \begin{aligned}
    \frac{1}{K(\delta')} \sum_{k=1}^{K(\delta')} \trsquare{\hat{\rho}_k
      \hat{M}_k}
    &\le \frac{1}{K(\delta')} \sum_{k=1}^{K(\delta')}
      \trsquare{\hat{\Pi}_{\text{eff}} \hat{M}_k} \\
    &= \frac{D_{\text{eff}}}{K(\delta')} \,.
    \end{aligned}
\end{equation}
Combining this estimate with the learning-to-testing reduction gives the
  counting bound
\begin{equation}
    \eta_{\text{PL}}(\delta) \le \frac{D_{\text{eff}}}{K(\delta')} \,.
    \label{eq:counting_bound}
\end{equation}
The hypotheses are encoded in an $n$-qubit Hilbert space, so $D_{\text{eff}} \le
  2^n$. The target precision $\delta=3^{-n}/2$ implies $\delta'=3^{-n}$ and
  therefore $K(\delta') = 3^n$. Hence, Eq.~\eqref{eq:counting_bound} yields
\begin{equation}
    \eta_{\text{PL}}(\delta) \le \left( \frac{2}{3} \right)^n \,,
    \label{eq:phase_counting_bound}
\end{equation}
which is below the success threshold $3/4$ for $n \ge 1$ and vanishes as
  $n\rightarrow\infty$.

With only $n$ qubits, the system's Hilbert-space dimension restricts the number
  of orthogonal measurement outcomes to $2^n$, whereas the requested precision
  scales as $3^{-n}$. Consequently, there are too many hypotheses to
  discriminate successfully relative to the available dimensionality of the
  system. This phenomenon is analogous to what is observed in symmetric
  informationally complete positive operator-valued measures (SIC-POVMs):
  although any pair of hypotheses remains highly distinguishable, the
  impossibility of performing learning at the required precision stems from the
  limitations of the Hilbert space.

Crucially, this obstruction is not flagged by
  Eq.~\eqref{eq:pairwise_success_probability}, which motivates the
  multi-hypothesis upper bounds on the success probability introduced below.

\section{Results}
\label{sec:results}

In the background section, we introduced the quasi-optimal solution to the
  hypothesis-testing task and showed how to rigorously connect quantum
  multi-hypothesis testing with quantum parameter learning. In this section, we
  combine these results to obtain both upper and lower bounds on the learning
  success probability based on the pretty good measurement.

Alternative upper bounds on the success probability are collected in
  Appendix~\ref{app:alternative_upper_bounds} and summarized in
  Table~\ref{table:alternative_upper_bounds}.

\subsection{Multi-hypothesis upper bounds on the success probability}
\label{sec:upper_bound_learning_success}
In this section, we present the PGM-based multi-hypothesis upper bound on
  quantum parameter learning.

\subsubsection{Pure-state multi-hypothesis bound from the Barnum--Knill theorem}
\label{subsubsec:barnum_knills_theorem}
We first focus on the case where all hypothesis states are pure. The
  learning-to-testing simulation upper-bounds the minimax success probability of
  learning by the success probability of hypothesis testing with spacing
  $\delta'>2\delta$. Combining the Barnum--Knill theorem of
  Sec.~\ref{subsec:pgm} with
  Eq.~\eqref{eq:theorem_statement_learning_to_testing}, and using the definition
  of $K(\delta')$ in Eq.~\eqref{eq:definition_K_delta}, we obtain the upper bound

\begin{result}[Minimax upper bound---pure states]
\begin{equation}
    \eta_{\text{PL}} (\delta) \le \inf_{\mu} \sqrt{\sum_{k=1}^{K(\delta')} \big(
      \sqrt{\boldsymbol{G}}_{kk} \big)^2} \,.
    \label{eq:upper_bound_eta_pl_pgm}
\end{equation}
\end{result}
Here, $\eta_{\text{PL}}(\delta)$ is the minimax learning success probability at
  learning precision $\delta$, $\mu=\lbrace\mu_k\rbrace_{k=1}^{K(\delta')}$ is a
  discrete prior over the parameter hypotheses, $\delta'$ is the spacing between
  hypotheses, and $K(\delta')$ is the number of hypotheses, as defined in
  Eq.~\eqref{eq:definition_K_delta}. The matrix $\boldsymbol{G}$ is the Gram
  matrix of the ensemble $\lbrace (\mu_k,
  \ket{\psi_{\theta_k}})\rbrace_{\theta_k \in \Omega}$ (see
  Eq.~\eqref{eq:gram_matrix}). Choosing any prior on the hypotheses (e.g., a
  uniform prior) provides a valid bound on the success probability. This bound
  naturally goes beyond binary hypothesis testing and considers the
  distinguishability of the probe states across the whole range of parameters
  $\theta \in \Theta$. Although only diagonal elements of
  $\sqrt{\boldsymbol{G}}$ appear explicitly in
  Eq.~\eqref{eq:upper_bound_eta_pl_pgm}, the matrix square root is determined by
  the full complex Gram matrix; consequently, these diagonal elements generally
  depend on the complete set of overlaps, including their phases, which are
  invisible to pairwise fidelities but can affect joint distinguishability.

Two alternative upper bounds are also computable from the full Gram matrix for
  pure-state ensembles. Appendix~\ref{app:alternative_upper_bounds} derives the
  Fano-based entropy bound of Eq.~\eqref{eq:upper_bound_entropy_pure_states_2}
  and gives the pure-state Gram-matrix form of the Holevo--Curlander bound in
  Eq.~\eqref{eq:holevo_curlander_bound_pure_gram}. These bounds are independent
  of Eq.~\eqref{eq:upper_bound_eta_pl_pgm} and may be tighter in certain
  regimes; see Sec.~\ref{subsec:comparison_upper_bounds} for a comparison.

\subsection{Success-probability guarantees}
\label{subsec:lower_bounds_success}
The testing-to-learning reduction,
  Eq.~\eqref{eq:theorem_statement_testing_to_learning}, shows that a
  sufficiently fine multi-hypothesis-testing problem gives a rigorous lower
  bound on the success probability of parameter learning. Two features of this
  construction are important. First, one does not need to take a continuous
  limit of hypothesis testing: a finite $\delta'$-net of the parameter interval
  is sufficient, as defined in Eq.~\eqref{eq:set_hypothesis_definition}. Second,
  the reduction uses the tolerance band $|\check{k}-k|\le a$, where the integer
  $a$ is defined in Eq.~\eqref{eq:expression_a_achievability} and is roughly
  the number of hypotheses that fit within a window of acceptance for the
  learning task.

We now combine the preceding minimax and uniform-prior Bayesian
  testing-to-learning bounds with performance guarantees for the pretty good
  measurement, thereby obtaining explicit lower bounds on the corresponding
  learning success probabilities. For pure-state models, the exact PGM success
  probability extends directly to the tolerance setting, yielding
  Eqs.~\eqref{eq:achievable_probability}
  and~\eqref{eq:achievable_probability_bayesian}. For mixed-state models,
  extending PGM error bounds to the tolerance setting yields
  Eq.~\eqref{eq:achievable_success_mixed_states}, which depends only on pairwise
  fidelities.

\subsubsection{Pure states: success-probability guarantees from the PGM}
\label{subsubsec:achievable_bound_pgm}
We now give an explicit PGM-based lower bound on the minimax learning success
  probability for pure-state models. Combining the PGM success probability,
  defined with tolerance and derived in Appendix~\ref{app:success_pgm}, with the
  testing-to-learning reduction in
  Eq.~\eqref{eq:theorem_statement_testing_to_learning} yields

\begin{result}
{\bfseries\upshape
(Minimax success-probability\\
lower bound---pure states)}
  \begin{equation}
    \eta_{\text{PL}} (\delta) \ge \inf_{\mu} \sum_{k=1}^{K(\delta')}
      \sum_{\check{k}\,:\,|k-\check{k}| \le a} \left| (\sqrt{\boldsymbol{G}})_{k
      \check{k}} \right|^2 - L \delta'/2 \,.
    \label{eq:achievable_probability}
\end{equation}
\end{result}
Here, $\eta_{\text{PL}}(\delta)$ is the minimax learning success probability at
  learning precision $\delta$, $\mu=\lbrace\mu_k\rbrace_{k=1}^{K(\delta')}$ is a
  discrete prior over the parameter hypotheses, and $K(\delta')$ is the number
  of hypotheses in the $\delta'$-net, as defined in
  Eq.~\eqref{eq:definition_K_delta}. The indices $k$ and $\check{k}$ label the
  true and returned hypotheses, respectively. The quantity $L$ is the
  trace-distance Lipschitz constant satisfying
  $D(\hat{\rho}_{\theta},\hat{\rho}_{\theta'})\le L|\theta-\theta'|$, and we use
  the reference hypothesis spacing $\delta'=\delta'_0$ defined in
  Eq.~\eqref{eq:delta_prime_bayesian_convenient}. The matrix $\boldsymbol{G}$ is
  the prior-weighted Gram matrix defined in Eq.~\eqref{eq:gram_matrix}. The
  Gram-matrix sum in Eq.~\eqref{eq:achievable_probability} contains the main
  diagonal of $\sqrt{\boldsymbol{G}}$ and, when $a\ge1$, its first $a$ super-
  and subdiagonals, thereby accounting for PGM outcomes associated with
  neighboring hypotheses inside the learning-tolerance window.

For a uniform prior distribution on $\theta$, the corresponding lower bound on
  the Bayesian success probability follows by evaluating
  Eq.~\eqref{eq:theorem_statement_testing_to_learning_bayesian} for the PGM
  success probability with tolerance.

\begin{result}
{\bfseries\upshape
(Uniform-prior Bayesian\\
success-probability lower bound---pure states)}
\begin{equation}
    \begin{aligned}
    \bar{\eta}_{\text{PL}} &(\mu_U, \delta)
    \ge
    \sum_{k=1}^{K(\delta')}
    \sum_{\check{k}\,:\,|k-\check{k}| \le a}
    \left| (\sqrt{\boldsymbol{G}})_{k \check{k}} \right|^2 \\
    &\qquad
    -\frac{1}{|\Theta|}
    \sum_{k=1}^{K(\delta')}
    \int_{\theta_k-\delta'/2}^{\theta_k+\delta'/2}
    \dd\theta\,
    \sqrt{1-F^2\!\left(\hat{\rho}_{\theta},\hat{\rho}_{\theta_k}\right)}
    \,.
    \end{aligned}
    \label{eq:achievable_probability_bayesian}
\end{equation}
\end{result}
Here, $\bar{\eta}_{\text{PL}}(\mu_U,\delta)$ is the optimal Bayesian learning
  success probability at learning precision $\delta$ for the uniform prior
  $\mu_U$ on $\Theta$. The interval $\Theta=[\theta_{\min},\theta_{\max}]$ is
  the parameter space, and $|\Theta|=|\theta_{\max}-\theta_{\min}|$ is its
  length. The set
  $\Omega=\lbrace\theta_k\rbrace_{k=1}^{K(\delta')}\subset\Theta$ is the
  $\delta'$-net defined in Eq.~\eqref{eq:set_hypothesis_definition}, where
  $\delta'$ is the hypothesis spacing, $\theta_k$ are the hypothesis parameter
  values, and $K(\delta')$ is their number, as defined in
  Eq.~\eqref{eq:definition_K_delta}. The state $\hat{\rho}_{\theta}$ is the
  probe state encoded at parameter value $\theta$, and
  $F(\hat{\rho},\hat{\sigma}) \equiv
  \tr{\sqrt{\sqrt{\hat{\rho}}\hat{\sigma}\sqrt{\hat{\rho}}}}$ is the fidelity.
  The integer $a$ is determined by Eq.~\eqref{eq:expression_a_achievability},
  and the uniform-prior Gram matrix $\boldsymbol{G}$ has weights
  $\mu_k=1/K(\delta')$.

\subsubsection{Mixed states: lower bound on the uniform-prior Bayesian success
  probability from pairwise fidelities}
\label{subsubsec:achievable_success_mixed}

The mixed-state case requires an additional step. For pure states, the exact PGM
  success probability can be generalized directly by summing the PGM outcome
  probabilities inside the tolerance band $|\check{k}-k|\le a$; for generic
  mixed-state ensembles, the corresponding exact expression requires the full
  spectral decomposition of the mixed states entering an enlarged Gram matrix
  and is generally difficult to evaluate. We therefore lower-bound the success
  probability with tolerance by extending the PGM hypothesis-testing error bound
  of~\cite{montanaroPrettySimpleBounds}. This gives a success-probability
  guarantee expressed in terms of fidelities that can be practical to evaluate
  even when directly computing the performance of the PGM is infeasible.

\begin{figure}[htbp]
  \centering
  \includegraphics[width=0.45\textwidth]{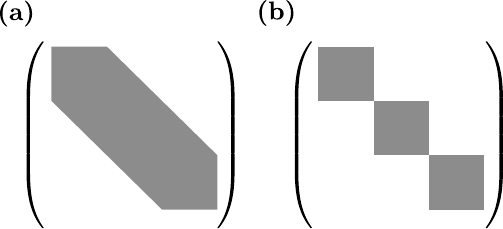}
  \caption{\justifying Representation of the matrix of outcome probabilities in
    Eq.~\eqref{eq:matrix_outcomes}. (a) Success probability with tolerance, as
    appearing in Eq.~\eqref{eq:theorem_statement_testing_to_learning}. (b) Block
    success probability, with blocks of at most $\ell=a+1$ elements, defined in
    Eq.~\eqref{eq:block_success_probability}.}
  \label{fig:matrix_representation}
\end{figure}

Let $I = \{1, 2, \dots, K(\delta')\}$ and define $\ell \equiv a+1$. We set $T
  \equiv \lceil K(\delta')/\ell\rceil$ and partition $I$ into the contiguous
  blocks
\begin{equation}
    I_j
    \equiv
    \{(j-1)\ell+1,\dots,\min\{j\ell,K(\delta')\}\} \, ,
    \label{eq:contiguous_hypothesis_blocks}
\end{equation}
with $j=1,\dots,T$, so that $I = \bigcup_{j=1}^{T} I_j$. Every block has at most
  $\ell$ elements. Hence, any two indices in the same block differ by at most
  $\ell-1=a$. In this setting, the hypothesis-testing task is considered
  successful if the algorithm returns an estimator belonging to the same subset
  $I_j$ as the true hypothesis. This defines a coarser hypothesis-testing
  problem. Success in the coarse problem implies success in the
  hypothesis-testing problem with a tolerance band, so for a given prior $\mu$
  and POVM $\mathcal{M}$, we define the block success probability as
\begin{equation}
    \bar{\mathsf{P}}_{\text{Block}} (\mu; \mathcal{M}) \equiv \sum_{j=1}^T
      \sum_{k, \check{k} \in I_j} \mu_k \trsquare{\hat{\rho}_{k}
      \hat{M}_{\check{k}}} \,.
    \label{eq:block_success_probability}
\end{equation}
Figure~\ref{fig:matrix_representation} illustrates the outcome probability
  matrix for a given POVM and prior:
\begin{equation}
    [\boldsymbol{M}(\mu, \mathcal{M})]_{k, \check{k}} \equiv \mu_k
      \trsquare{\hat{\rho}_{k} \hat{M}_{\check{k}}} \,.
    \label{eq:matrix_outcomes}
\end{equation}
The highlighted regions indicate the matrix elements that must be summed to
  obtain either the success probability with tolerance (left) or the block
  success probability of Eq.~\eqref{eq:block_success_probability} (right).

Because the blocks lie inside the tolerance band around the diagonal, success in
  the block problem implies success in the original hypothesis-testing problem
  with a full tolerance band around the true hypothesis. Therefore, the success
  probability with a tolerance band is lower-bounded by the block success
  probability:
\begin{equation}
    \sum_{k=1}^{K(\delta')} \sum_{\check{k}\,:\,|k-\check{k}|\le a} \mu_k
      \trsquare{\hat{\rho}_{k} \hat{M}_{\check{k}}} \ge
      \bar{\mathsf{P}}_{\text{Block}} (\mu; \mathcal{M}) \; .
    \label{eq:block_success_lower_bounds_tolerance}
\end{equation}
We focus on a uniform prior over the parameter $\theta$. From
  Eq.~\eqref{eq:theorem_statement_testing_to_learning_bayesian_optimized}, the
  optimal Bayesian learning probability with a uniform prior distribution is
  lower-bounded by
\begin{align}
    \bar{\eta}_{\text{PL}} &(\mu_U, \delta)
    \ge
    \sup_{\substack{\delta'\le 2\delta}}
    \Bigg\{
    \sup_{\mathcal{M}}
    \bar{\mathsf{P}}_{\text{Block}}(\mu_U;\mathcal{M})
    \notag\\
    &\qquad
    -\frac{1}{|\Theta|}
    \sum_{k=1}^{K(\delta')}
    \int_{\theta_k-\delta'/2}^{\theta_k+\delta'/2}
    \dd\theta\,
    \sqrt{1-F^2\!\left(\hat{\rho}_{\theta},\hat{\rho}_{\theta_k}\right)}
    \Bigg\} \,.
    \label{eq:theorem_statement_testing_to_learning_block_uniform_bayesian}
\end{align}
In Appendix~\ref{app:block_bounds}, we lower-bound
  $\bar{\mathsf{P}}_{\text{Block}}(\mu; \mathcal{M}_{\text{PG}})$ in terms of
  pairwise fidelities. Evaluating the grid supremum in
  Eq.~\eqref{eq:theorem_statement_testing_to_learning_block_uniform_bayesian} at
  the reference spacing $\delta'=\delta'_0$ from
  Eq.~\eqref{eq:delta_prime_bayesian_convenient} yields the following bound for
  the uniform prior:

\begin{result}
{\bfseries\upshape
(Uniform-prior Bayesian\\
success-probability lower bound---mixed states)}
\begin{equation}
    \begin{aligned}
    \bar{\eta}_{\text{PL}} &(\mu_U, \delta)
    \ge
    1 - \frac{1}{2K(\delta')}
    \sum_{\substack{t,t'=1\\t\neq t'}}^T
    \sum_{\substack{i \in I_t \\ j \in I_{t'}}}
    F(\hat{\rho}_{\theta_i}, \hat{\rho}_{\theta_j}) \\
    &\qquad
    -\frac{1}{|\Theta|}
    \sum_{k=1}^{K(\delta')}
    \int_{\theta_k-\delta'/2}^{\theta_k+\delta'/2}
    \dd\theta\,
    \sqrt{1-F^2\!\left(\hat{\rho}_{\theta},\hat{\rho}_{\theta_k}\right)}
    \,.
    \end{aligned}
    \label{eq:achievable_success_mixed_states}
\end{equation}
\end{result}

Recall that the tolerance radius $a$ is fixed by
  Eq.~\eqref{eq:expression_a_achievability}, so $a$, $\ell$, $T$, and the blocks
  $I_t$ in Eq.~\eqref{eq:achievable_success_mixed_states} are all evaluated on
  the reference grid specified above.

We conclude by commenting on the applicability of these success-probability
  guarantees. Equation~\eqref{eq:achievable_success_mixed_states} is a one-shot,
  pairwise lower bound on the uniform-prior Bayesian success probability for
  learning a single parameter from a mixed-state model, and it requires only the
  ability to compute fidelities between sensor states encoded with different
  parameter values. The fidelity integral controls the passage from the discrete
  hypotheses to the continuous uniform-prior problem. In general, one cannot
  obtain a valid fidelity-only formula by replacing the disjoint-block penalty
  with a sum restricted to pairs outside the tolerance band, as long as we use
  the PGM to certify achievable performance.
  Appendix~\ref{subsec:out_of_band_fidelity_penalty} gives an explicit
  pure-state counterexample.

For a mixed-state model, computing the fidelities can still be difficult, but
  there are important tractable cases. For example, the fidelity of bosonic or
  fermionic Gaussian states admits a closed-form
  expression~\cite{banchiQuantumFidelityArbitrary2015,
    swingleRecoveryMapFermionic2019}. Another example is a sensor state that
      factorizes into smaller mixed states, $\hat{\rho}_\theta =
      \bigotimes_{b=1}^B \rho_\theta^b$, for which the fidelity factorizes. In
      Section~\ref{subsec:dissipative_frequency_learning}, we discuss a further
      example in which the fidelities between matrix-product states of a
      Markovian environment~\cite{ciccarelloQuantumCollisionModels2022} can be
      evaluated exactly using the two-sided master-equation
      formalism~\cite{molmerHypothesisTestingOpen2015,
        khanTensorNetworkApproach2025}.
      In that example, a generically infinite-rank, phase-averaged joint
      spin--field state has an elementary closed-form fidelity. We conclude by
      saying that Eq.~\eqref{eq:achievable_success_mixed_states} can therefore
      provide a computable, rigorous, and informative lower bound on the
      uniform-prior Bayesian learning success probability for a broad class of
      high-dimensional mixed-state models.

\subsection{Comparison with prior work}
\label{subsec:related_works_comparison}

To provide context for our results, we briefly review relevant recent work on
  quantum parameter learning.
In Ref.~\cite{meyerQuantumMetrologyFiniteSample2025}, the authors establish a 
  precision bound for learning a single quantum parameter by discretizing the
  parameter space and performing binary hypothesis testing between adjacent
  pairs of hypotheses. This yields a practically computable bound expressed in
  terms of the quantum Fisher information (QFI), making transparent contact with
  Cram\'er--Rao (CR) bounds. However, this bound does not generally capture
  constraints arising from the joint distinguishability of all possible encoded
  states. In a closely related work, Huang~\textit{et
  al.}~\cite{huangQueryComplexitiesQuantum2026} establish sample-complexity
  lower bounds for channel discrimination in both parallel and sequential
  (adaptive) schemes, which are also rooted in binary hypothesis testing and
  therefore do not capture the global nature of parameter learning.

In earlier work~\cite{walterLowerBoundsQuantum2014}, lower bounds on the volume
  of the confidence region in quantum parameter learning were derived using the
  hypothesis-testing relative entropy. This approach is complementary to ours,
  since we compute upper bounds on the probability of success for a fixed
  confidence-interval size. The bounds of
  Ref.~\cite{walterLowerBoundsQuantum2014} are potentially complicated to
  evaluate because they involve semidefinite optimization over the space of
  measurement operators acting on the probe state.

A recent work~\cite{kwonUniversalSampleComplexity2026} studying quantum
  parameter learning explicitly connects its sample complexity to the Fisher
  information. In this work, the authors analyze the sample complexity of the
  maximum likelihood estimator (MLE), i.e., the number of identical copies of
  the probe state necessary for the MLE to achieve a certain precision, showing
  that it is determined by the diagonal elements of the Fisher information
  matrix in the asymptotic limit of small error. This work assumes the use of
  multiple copies, whereas we provide single-shot upper bounds and
  success-probability guarantees.

In~\cite{chenInstanceoptimalHighprecisionShadow2026}, the authors study the
  sample complexity of estimating the expectation values of a known set of
  observables, or an unknown linear combination of them, using a two-stage
  procedure based on state tomography followed by local estimation. This sample
  complexity is characterized by a quantity computable from the Fisher
  information of an appropriately optimized single-copy measurement on the
  unknown state. This is again a work based on a multi-copy setting, while we
  study a single-shot scenario.

\section{Applications}
\label{sec:example}
We now apply the global bounds to two different settings. The first is
  single-shot learning of the coupling parameter from the fixed-time unitary
  evolution generated by an all-to-all Ising Hamiltonian acting on a system of
  spins. This application illustrates how global multi-hypothesis bounds can
  reveal limitations missed by local or binary analyses in single-shot learning
  problems. The second setting concerns learning the Rabi frequency of a driven
  spin coupled to a bosonic environment using joint measurements of the spin and
  the environment. We consider two versions of this application: in the first,
  the drive phase is known; in the second, the drive phase is unknown and
  treated as a nuisance parameter.

\subsection{Learning an all-to-all Ising interaction}
\label{subsec:hamiltonian_learning_ising}
We first consider learning a global Ising interaction. We compare the PGM-based
  multi-hypothesis upper bound in Eq.~\eqref{eq:upper_bound_eta_pl_pgm} with the
  binary hypothesis-testing upper bound in
  Eq.~\eqref{eq:pairwise_success_probability}, and we evaluate the PGM-based
  achievability result in Eq.~\eqref{eq:achievable_probability_bayesian} for the
  uniform-prior Bayesian success probability. Together, these results constrain
  the attainable precision of the prescribed single-shot strategy and reveal
  features missed by local and binary analyses. Our aim is not to provide an
  extensive analysis of Hamiltonian
  learning~\cite{huangLearningManyBodyHamiltonians2023,
    huAnsatzFreeHamiltonianLearning2025} for Ising Hamiltonians; rather, we use
      this example to illustrate how global and local bounds can be dramatically
      different for a learning scenario in a many-body system.

Consider $N$ spin-$1/2$ particles subjected to an all-to-all Ising interaction
  with a uniform coupling parameter $J\in[0,J_{\max}]$, where $J_{\max}>0$ sets
  the coupling scale. The Hamiltonian is
\begin{equation}
    \hat{H}(J)
    \equiv
    J\sum_{1\le i<j\le N}\hat{\sigma}^i_z \hat{\sigma}^j_z \;.
    \label{eq:ising_hamiltonian_main}
\end{equation}
We assume the product input $\ket{\psi_0}\equiv\ket{+}^{\otimes N}$, polarized
  along the $x$ axis, and evolve it for a fixed duration $t$. The task is to
  learn $J$ from the resulting state using an arbitrary collective POVM on the
  $N$-spin output. We study the precision as a function of the number of spins
  $N$; note that $J_{\max}$ and $t$ do not scale with $N$. All conclusions in
  this subsection pertain to the single-shot, fixed-input, fixed-time strategy.
  The resulting family of sensor states is
\begin{equation}
    \ket{\psi(J)} = \mathrm{e}^{-\mathrm{i}t\hat{H}(J)}\ket{+}^{\otimes N}.
    \label{eq:ising_state_main}
\end{equation}

Before computing the multi-hypothesis and binary hypothesis-testing bounds, we
  recall the QFI of the state $\ket{\psi(J)}$, which provides the corresponding
  local benchmark. As derived in Appendix~\ref{app:ising_qfi}, this QFI is
\begin{equation}
    \mathcal{F}(J) = 2 \, N(N-1) \, t^2 \,.
    \label{eq:ising_qfi}
\end{equation}
It is independent of $J$ and scales as $\mathcal{F}\sim 2N^2t^2$ for large $N$.
  Its quadratic scaling is compatible with a local precision $\delta J\propto
  N^{-1}$~\cite{parisQuantumEstimationQuantum2009,
    boixoQuantumMetrologyDynamics2008}, but it does not establish that this
      precision is achievable in the present single-shot global learning task.

\begin{figure*}[htbp]
  \centering
  \includegraphics[width=\textwidth]{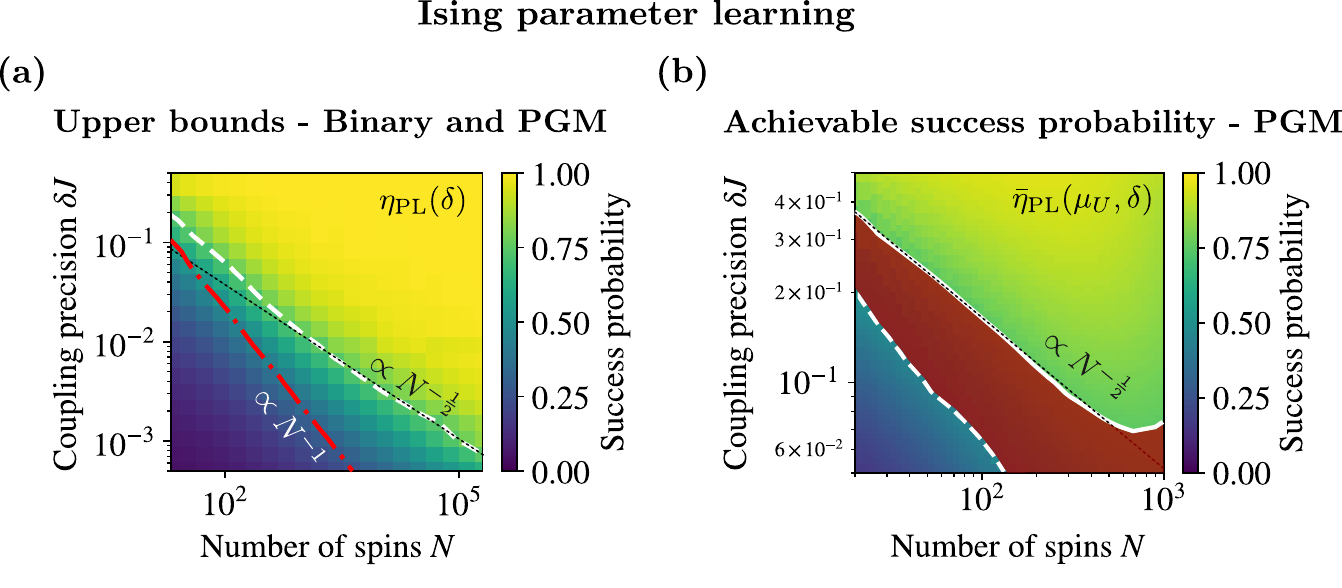}
  \caption{\justifying Success-probability bounds for single-shot learning of
    the coupling $J$ in Eq.~\eqref{eq:ising_hamiltonian_main}, for the input
    $\ket{+}^{\otimes N}$ and evolution time $t=0.1/J_{\max}$. \textbf{(a)} The
    heat map shows the PGM-based multi-hypothesis upper bound in
    Eq.~\eqref{eq:upper_bound_eta_pl_pgm} on the minimax success probability,
    evaluated from the exact Gram matrix constructed from the overlaps in
    Eq.~\eqref{eq:exact_overlap_ising}. The dashed white curve is its $3/4$
    contour, and the red dot-dashed curve is the $3/4$ contour of the binary
    hypothesis-testing bound in Eq.~\eqref{eq:pairwise_success_probability}.
    Below either curve, the corresponding bound rules out successful minimax
    learning. The binary contour is compatible with $\delta J\propto N^{-1}$,
    whereas the PGM-based exclusion contour approximately follows the black
    dotted $\delta J\propto N^{-1/2}$ reference over the numerically accessible
    range, after an initial regime in which the scaling is intermediate between
    $\delta J\propto N^{-1}$ and $\delta J\propto N^{-1/2}$. \textbf{(b)} The
    heat map shows the PGM-based achievable success probability obtained from
    Eq.~\eqref{eq:achievable_probability_bayesian} for Bayesian learning under
    the uniform prior on $J$. The solid white curve is its $3/4$ contour, while
    the dashed white curve is the $3/4$ contour of the PGM-based upper bound on
    the optimal uniform-prior Bayesian success probability obtained from
    Eqs.~\eqref{eq:theorem_statement_learning_to_testing_%
bayesian_shift_invariant} and~\eqref{eq:pgm_upper_bound_success}; this is the
  same dashed curve as in panel~{(a)}. The red shaded region between the two
  contours contains the optimal precision $\delta J$ for which the uniform-prior
  Bayesian success probability reaches $3/4$: below the dashed curve, success is
  ruled out, whereas on and above the solid curve it is certified by the PGM.
  The achievable precision approximately follows the black dotted $\delta
  J\propto N^{-1/2}$ reference for smaller $N$ before bending toward larger
  $\delta J$ at larger $N$.}
  \label{fig:ising_probabilities}
\end{figure*}

We compare the binary hypothesis-testing bound of
  Eq.~\eqref{eq:pairwise_success_probability} with the PGM-based
  multi-hypothesis bound of Eq.~\eqref{eq:upper_bound_eta_pl_pgm}.
  Figure~\ref{fig:ising_probabilities}(a) summarizes the comparison. Because the
  states are pure, both bounds can be evaluated from the exact overlaps derived
  in Appendix~\ref{app:ising_effective_dimension_overlap}. The QFI in
  Eq.~\eqref{eq:ising_qfi} and the binary bound are compatible with $\delta
  J\propto N^{-1}$. These results could suggest the possibility of learning $J$
  with this precision scaling. However, the global PGM-based bound, which
  accounts for the requirement of joint discrimination, yields a substantially
  more restrictive precision constraint that definitely excludes the scaling
  suggested by the QFI and binary hypothesis-testing approaches, giving instead
  $\delta J\propto N^{-\frac{1}{2}}$.

For a uniform prior on $J$, we also evaluate the PGM-based lower bound in
  Eq.~\eqref{eq:achievable_probability_bayesian}, which provides an
  achievability guarantee for successful learning.
  Figure~\ref{fig:ising_probabilities}(b) reports both the PGM-based precision
  upper bound and the achievable precision, thereby identifying the viable
  region containing the exact optimal precision. While the achievable precision
  initially scales as $\delta J\propto N^{-\frac{1}{2}}$, it deteriorates for
  larger numbers of spins, likely because of the choice of $t$, which does not
  depend on $N$. Interestingly, the value of $N$ at which the one-shot learning
  guarantee starts to deteriorate coincides with an inflection point in the
  precision inferred from the PGM upper bound, beyond which this precision
  starts to scale as $\delta J\propto N^{-\frac{1}{2}}$. All of this indicates a
  relatively complex structure for learning from such a simple family of states.
  This structure can be explored using the tools introduced in this manuscript.

We conclude by noting that the information that we can guarantee to be
  extractable from a single copy of $\ket{\psi(J)}$ is relatively small, as can
  be seen from the scale of the axis in Figure~\ref{fig:ising_probabilities}(b).
  More information can be extracted from multiple copies of the state, and this
  approach would asymptotically attain the QFI limit. Nevertheless, our interest
  here is in studying the single-copy performance of the learning task, and this
  result shows that some information about the coupling can still be reliably
  retrieved from a single copy of the state.

\subsection{Dissipative Rabi-frequency learning}
\label{subsec:dissipative_frequency_learning}
In this application, the sensor is a driven spin coupled to a Markovian bosonic
  environment (e.g.,~a waveguide). A coherent drive at a known carrier
  frequency, resonant with the known spin transition, has an unknown amplitude
  that determines the Rabi frequency $\Omega$ to be learned using both the state
  of the qubit and the emitted waveguide field. We compare a setting in which
  the drive phase is known with a phase-agnostic setting in which no phase
  reference tied to the drive is available. In the spin's rotating frame, the
  drive phase specifies the direction of the effective static magnetic field.
  Uniformly averaging over the phase of the drive produces a mixed joint
  spin--field state, for which the lower bound on the uniform-prior Bayesian
  success probability becomes the relevant result.

This setting is an example of a continuous quantum metrology
  protocol~\cite{albarelliPedagogicalIntroductionContinuously2024,
    tsangQuantumMetrologyOpen2013,
    gammelmarkFisherInformationQuantum2014}. Bayesian inference over broad
      parameter ranges in this setting has been studied using stochastic
      simulations of continuously monitored
      records~\cite{gammelmarkBayesianParameterInference2013}, while recent work
      computes the QFI of continuously encoded quantum sensors and their
      environment~\cite{leeTimescalesSqueezingHeisenberg2025}. Proposed readout
      strategies include quantum decoders, coherent absorbers, and adaptive or
      pattern-counting measurements~\cite{yangEfficientInformationRetrieval2023,
    godleyAdaptiveMeasurementFilter2023,
    girottiEstimatingQuantumMarkov2025}. These techniques, however, do not by
      themselves provide a completely analytical, attainable one-shot precision
      guarantee for a joint measurement on the spin--field output in the
      Bayesian setting. The present framework fixes this issue by providing
      finite-time guarantees on the learning success probability at a certain
      precision without requiring the simulation of the experiment.

The spin is initially in its ground state, and the bosonic environment is
  initially in the vacuum state. We work in a frame rotating at the known drive
  frequency. Let $\hat b_\omega$ annihilate an environmental mode with detuning
  $\omega$ from the drive frequency. For a fixed offset $\phi$ between the drive
  and the rotating-frame phase origin, we assume that the joint
  spin--environment evolution is generated by
\begin{equation}
    \begin{aligned}
    \hat H_{\Omega,\phi}
    &\equiv
    \frac{\Omega}{2}
    \left(
    \myexp{-\mathrm{i}\phi}\hat\sigma_+
    +
    \myexp{\mathrm{i}\phi}\hat\sigma_-
    \right),
    \\
    \hat H^{SE}_{\Omega,\phi}
    &={}
    \hat H_{\Omega,\phi}
    +
    \int_{-\infty}^{\infty}
    \omega\,\hat b_{\omega}^{\dagger}\hat b_{\omega}\,\dd\omega
    \\
    &+
    \mathrm{i}\sqrt{\frac{\gamma}{2\pi}}
    \int_{-\infty}^{\infty}
    \left(
    \hat b_{\omega}^{\dagger}\hat\sigma_-
    -
    \hat\sigma_+\hat b_{\omega}
    \right)\dd\omega \; .
    \end{aligned}
    \label{eq:dissipative_joint_hamiltonian}
\end{equation}
The last term describes the interaction between the spin and the bosonic
  environment. This interaction is independent of $\Omega$ and produces the jump
  operator $\sqrt{\gamma}\hat\sigma_-$. The corresponding joint state of the
  spin and the field is
\begin{equation}
    \ket{\Psi_{\Omega,\phi}(t)}
    =
    \myexp{-\mathrm{i}t\hat H^{SE}_{\Omega,\phi}}
    \ket{g}\ket{\mathrm{vac}}_E \; .
    \label{eq:dissipative_joint_state}
\end{equation}
For an environment initially in the vacuum state, tracing out the field yields
  the reduced spin dynamics
\begin{equation}
    \begin{aligned}
    \frac{\dd\hat{\rho}}{\dd t}
    &= -\mathrm{i}\left[\hat{H}_{\Omega,\phi},\hat{\rho}\right] \\
    &\quad + \gamma\!\left(\hat{\sigma}_-\hat{\rho}\,\hat{\sigma}_+ -
      \tfrac{1}{2}\{\hat{\sigma}_+\hat{\sigma}_-,\hat{\rho}\}\right).
    \end{aligned}
    \label{eq:master_equation_qubit}
\end{equation}

\subsubsection{Known drive phase}
When the drive phase is known, we may set $\phi=0$ without loss of generality.
  We use the decay rate $\gamma$ as the reference frequency scale and take the
  Rabi frequency to lie in $[0,10\gamma]$. The joint spin--field output
  $\ket{\Psi_{\Omega,0}(t)}$ is pure. For two candidate Rabi frequencies
  $\Omega_i, \Omega_j$, let $\Delta\Omega_{ij}\equiv\Omega_i-\Omega_j$ and
  define
\begin{equation}
    \nu_{ij}
    \equiv
    \sqrt{
    \frac{\Delta\Omega_{ij}^2}{4}
    -
    \frac{\gamma^2}{16}
    }.
    \label{eq:dissipative_overlap_rate}
\end{equation}
As derived in Appendix~\ref{app:dissipative_fidelity} using the two-sided master equation~\cite{molmerHypothesisTestingOpen2015}, the exact overlap between two pure joint spin--field states is
  \begin{equation}
    \begin{aligned}
    \braket{\Psi_{\Omega_j,0}(t)|\Psi_{\Omega_i,0}(t)}
    &=
    \myexp{-\gamma t/4}
    \Biggl[
    \cos\left(\nu_{ij}t\right)
    \\
    &\qquad+
    \frac{\gamma}{4\nu_{ij}}
    \sin\left(\nu_{ij}t\right)
    \Biggr].
    \end{aligned}
    \label{eq:joint_overlap_dissipative_main}
\end{equation}
When $|\Delta\Omega_{ij}|<\gamma/2$,
  Eq.~\eqref{eq:joint_overlap_dissipative_main} is understood by analytic
  continuation, and the trigonometric functions become hyperbolic. The
  expression at $|\Delta\Omega_{ij}|=\gamma/2$ is defined by continuity.
  Equation~\eqref{eq:joint_overlap_dissipative_main} determines the full Gram
  matrix of any discretized hypothesis ensemble. Because the full Gram matrix
  depends only on pairwise frequency differences, the PGM hypothesis-testing
  success probability is invariant under a common shift of the $\delta'$-net.
  Combining Eqs.~\eqref{eq:theorem_statement_learning_to_testing_%
bayesian_shift_invariant} and~\eqref{eq:pgm_upper_bound_success} therefore gives
  a PGM-based upper bound on the uniform-prior Bayesian learning success
  probability. We also evaluate the Bayesian pure-state PGM achievable success
  probability for the uniform prior using
  Eq.~\eqref{eq:achievable_probability_bayesian}. These results are shown in
  Fig.~\ref{fig:dissipative_probabilities}(a).

The resulting bracket constrains both the asymptotic scaling and the optimal
  prefactor without simulating the physical measurement process or invoking a
  numerical inference procedure. The finite-time bounds resolve the
  pre-asymptotic regime and its crossover, on the timescale set by
  $\gamma^{-1}$, from transient coherent behavior to the standard-quantum-limit
  behavior expected at long times for this Markovian model. This tight result on
  the asymptotic performance of Rabi-frequency learning in Bayesian continuous
  metrology could not be obtained with existing
  techniques~\cite{gutaFisherInformationAsymptotic2011,
    catanaFisherInformationsLocal2015,
    gutaInformationGeometryLocal2017,
    yangEfficientInformationRetrieval2023,
    godleyAdaptiveMeasurementFilter2023,
    girottiEstimatingQuantumMarkov2025} without simulations of the measurement
      dynamics and the evolution of the estimator.

\subsubsection{Phase-agnostic drive}
We now suppose that no phase reference is available. As explained in
  Sec.~\ref{subsubsec:nuisance_parameters}, the operational joint spin--field
  state is then obtained by averaging over the unknown phase. In this setting,
  we restrict the Rabi frequency to the regime of overdamped oscillation,
  $[0,\gamma/2]$. This regime has the advantage that the fidelities between the
  phase-averaged states needed for our bound can be evaluated analytically, as
  shown in Appendix~\ref{app:phase_averaged_joint_positivity}.

The drive is already switched on when the spin is prepared and the interrogation
  begins, and its phase origin is not synchronized with the experimentalist's
  clock. We model complete uncertainty in this phase offset by a uniform prior
  on $[0,2\pi)$ for the initial drive phase $\phi$. We assume that the drive
  remains coherent, so that the corresponding rotating-frame phase offset is
  constant throughout the experiment. Averaging the single-interrogation state
  over the nuisance phase gives
\begin{equation}
    \hat{\rho}^{SE}_\Omega(t)
    \equiv
    \int_0^{2\pi}\frac{\dd\phi}{2\pi}\,
    \ket{\Psi_{\Omega,\phi}(t)}
    \!\bra{\Psi_{\Omega,\phi}(t)} \,.
    \label{eq:phase_averaged_joint_state_main}
\end{equation}

For two candidate Rabi frequencies, define
\begin{equation}
    \lambda_{ij}
    \equiv
    \sqrt{
    \frac{\gamma^2}{16} - \frac{\Delta\Omega_{ij}^2}{4}
    }.
    \label{eq:overdamped_fidelity_rate}
\end{equation}
For $0\le\Omega_i,\Omega_j\le\gamma/2$, one has $\lambda_{ij}\ge0$. On this
  domain, the fidelity between the mixed states can be evaluated analytically.
  As derived in Appendix~\ref{app:dissipative_fidelity}, the exact fidelity is
\begin{equation}
    \begin{aligned}
    F\!\left(
    \hat{\rho}^{SE}_{\Omega_i}(t),
     \hat{\rho}^{SE}_{\Omega_j}(t)
    \right) &= \myexp{-\gamma t/4}
    \Biggl[
    \cosh\left(\lambda_{ij}t\right)
    \\
    &\qquad+
    \frac{\gamma}{4\lambda_{ij}}
    \sinh\left(\lambda_{ij}t\right)
    \Biggr].
    \end{aligned}
    \label{eq:fidelity_dissipative_main}
\end{equation}

\begin{figure*}[htbp]
  \centering
  \includegraphics[width=\textwidth]{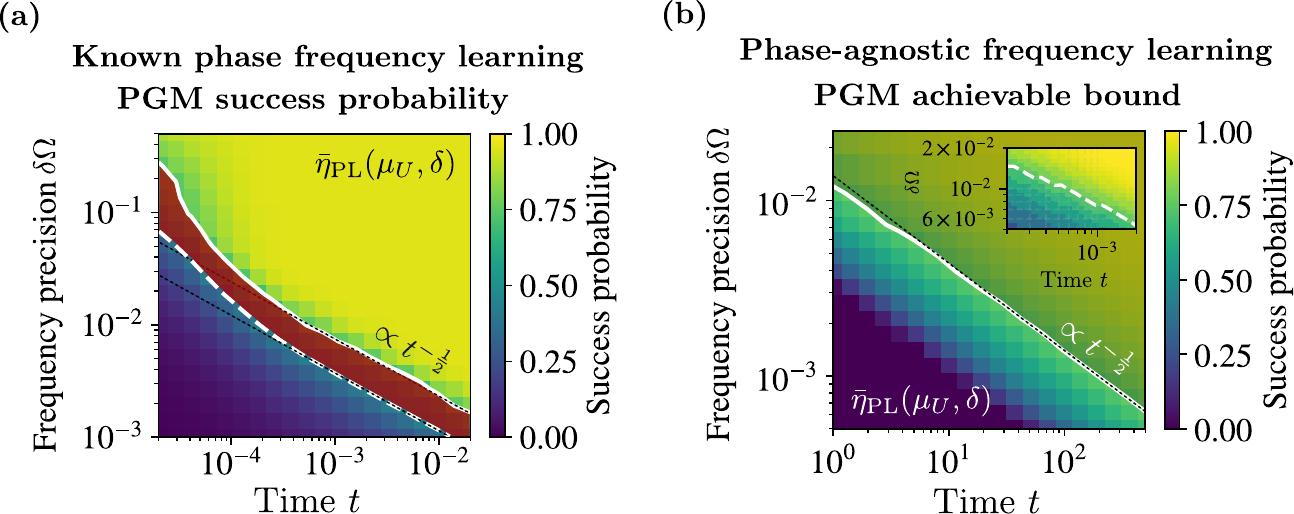}
  \caption{\justifying Uniform-prior Bayesian success-probability bounds for
    dissipative Rabi-frequency learning as functions of the observation time $t$
    and target precision $\delta\Omega$. In both main panels, the heat map shows
    a PGM-based lower bound (achievable probability), the solid white curve is
    its $3/4$ contour, and the black dotted line indicates the reference scaling
    $\delta\Omega\propto t^{-1/2}$. \textbf{(a)} Learning with a known drive
    phase for $\Omega\in[0,10\gamma]$, for which the joint spin--field output is
    pure. The lower bound is given by
    Eq.~\eqref{eq:achievable_probability_bayesian}. The dashed white curve is
    the $3/4$ contour of the PGM-based upper bound obtained from
    Eqs.~\eqref{eq:theorem_statement_learning_to_testing_%
bayesian_shift_invariant} and~\eqref{eq:pgm_upper_bound_success}, using the Gram
  matrix determined by Eq.~\eqref{eq:joint_overlap_dissipative_main}. The red
  shaded region between the two contours contains the precision $\delta\Omega$
  for which the optimal success probability reaches $3/4$: below the dashed
  curve, success is ruled out, whereas on and above the solid curve it is
  certified by the PGM. The contours cross over from transient behavior
  consistent with $\delta\Omega\propto t^{-1}$ to the long-time scaling
  $\delta\Omega\propto t^{-1/2}$; at long times, their values of $\delta\Omega$
  differ by approximately a factor of two. \textbf{(b)} Phase-agnostic learning
  for $\Omega\in[0,\gamma/2]$, for which phase averaging produces a mixed joint
  spin--field state. The lower bound in the main panel is given by
  Eq.~\eqref{eq:achievable_success_mixed_states}. The inset shows the exact PGM
  band-success probability in Eq.~\eqref{eq:phase_averaged_pgm_band_success},
  evaluated using the sector-resolved construction of
  Appendix~\ref{app:phase_averaged_joint_pgm}; its dashed white curve is the
  corresponding $3/4$ contour. The separation between this contour and the solid
  contour in the main panel directly demonstrates the looseness of the
  fidelity-only bound for this example. The gray shaded region above the solid
  white contour is excluded as the location of the optimal threshold, because
  the PGM construction already certifies successful learning at the smaller
  tolerance marked by the contour. The bound crosses $3/4$ at finite observation
  times for suitable tolerances, and its long-time contour follows the reference
  scaling $\delta\Omega\propto t^{-1/2}$. All displayed achievable probabilities
  are evaluated at the reference spacing $\delta'=\delta'_0$ defined in
  Eq.~\eqref{eq:delta_prime_bayesian_convenient}.}
  \label{fig:dissipative_probabilities}
\end{figure*}

In contrast to the known-phase setting, we do not evaluate a correspondingly
  tight upper bound on the success probability here. The main heat map in
  Fig.~\ref{fig:dissipative_probabilities}(b) retains the fidelity-only bound in
  Eq.~\eqref{eq:achievable_success_mixed_states} to illustrate the mixed-state
  result, whereas the inset shows the exact PGM band-success probability
  computed using the sector-resolved construction of
  Appendix~\ref{app:phase_averaged_joint_pgm}. This exact success probability is
  more challenging to compute and was evaluated only for a restricted range of
  parameters. The separation between their $3/4$ contours demonstrates that the
  fidelity-only bound is loose for this family and that substantially better
  performance is achievable with the PGM, provided we are able to evaluate its
  performance in the first place. The fidelity-only achievable bound on the
  success probability in Eq.~\eqref{eq:achievable_success_mixed_states} is not
  guaranteed, a priori, to be informative: for a given problem, its right-hand
  side could remain non-positive, and hence trivial, throughout the parameter
  regime. Yet, using this example with a nuisance parameter, we show that the
  bound provides a nontrivial certificate of successful learning from pairwise
  fidelities alone, thereby showing its utility. In particular, we see how this
  bound is tight enough to recover the standard quantum limit in time for this
  example. In other words, this bound misses the correct prefactor of the
  precision as a function of the observation time but recovers the correct
  scaling. Interestingly, in the inset, the exact computation of the PGM
  performance seems to give some information about the pre-asymptotic scaling of
  the precision, which is worse than $t^{-\frac{1}{2}}$, a feature that the main
  plot (the fidelity-based lower bound) seems to capture slightly.

We expect the achievable success probability of
  Eq.~\eqref{eq:achievable_success_mixed_states} to be applicable to and useful
  for other learning scenarios with mixed states.

Even when it is not tight, Eq.~\eqref{eq:achievable_success_mixed_states} can
  certify the finite-resolution learnability of a parameter or, within a fully
  developed multiparameter theory, of a prescribed group of parameters, and it
  can recover the scaling of the achievable precision with the available
  resources. Related questions of parameter identifiability under restricted
  access arise in quantum probe
  tomography~\cite{chenQuantumProbeTomography2025}.

\section{Discussion}
\label{sec:discussion}
Parameter learning remains a comparatively underexplored framework for quantum
  metrology~\cite{meyerQuantumMetrologyFiniteSample2025}, in comparison to
  conventional parameter estimation. In this work, we addressed two
  complementary questions. First, we derived computationally tractable upper
  bounds on the learning success probability that retain genuinely
  multi-hypothesis information for a given precision, and hence identify truly
  global constraints on learning a parameter over a finite interval, which are
  missed by the quantum Fisher information or an approach based on binary
  hypothesis testing. Second, we established rigorous lower bounds on the
  success probability and thus obtained explicit success-probability guarantees
  at a prescribed finite precision.

The phase-estimation example and the Ising application illustrate how global
  features of the learning problem can impede performance in a manner that is
  completely missed if one analyzes performance using local-estimation metrics
  like the quantum Fisher information or binary hypothesis testing constraints.
  We also used our new bounds to gain insight into a continuous metrology
  problem, namely, dissipative Rabi-frequency learning both when the drive phase
  is known and when it is unknown. In this last setting, the lower bound for the
  success probability of learning in
  Eq.~\eqref{eq:achievable_success_mixed_states}, valid for mixed-state
  models, was the key to certifying the learnability of the frequency in the
  high-precision limit.
It is instructive to compare this result to the role of the quantum
  Cram\'er--Rao bound in local parameter
  estimation~\cite{braunsteinStatisticalDistanceGeometry1994,
    parisQuantumEstimationQuantum2009}. Both constraints are derived from
      fidelity information, but they use it in fundamentally different ways. The
      quantum Cram\'er--Rao bound uses the local curvature of the fidelity,
      equivalently the quantum Fisher information, to lower-bound the variance
      of a locally unbiased estimator. It is therefore a no-go result. By
      contrast, Eq.~\eqref{eq:achievable_success_mixed_states} uses pairwise
      fidelities to lower-bound the success probability at a prescribed finite
      precision. It is consequently a one-shot lower bound on the uniform-prior
      Bayesian success probability, and therefore represents an achievable
      precision.

A central practical advantage of the mixed-state bound in
  Eq.~\eqref{eq:achievable_success_mixed_states} is that it depends only on
  pairwise fidelities between the original sensor states. In the phase-agnostic
  Rabi-frequency estimation example, these fidelities admit an elementary closed
  form even though the phase-averaged states are generically infinite rank. The
  success-probability guarantee can therefore be evaluated without constructing
  the phase-averaged density operators or optimizing over measurements. This
  simplification comes at the cost of the disjoint-block and pairwise
  relaxations, and the resulting bound need not be tight. Nevertheless, we have
  shown this bound to be informative for at least one example, in which not only
  can the identifiability of the parameter at every precision be certified,
  given sufficient observation time, but the correct scaling is also recovered.

Compared with other well-known Bayesian bounds, such as the van Trees and
  Ziv--Zakai bounds, we expect our bounds to be tighter for learning problems.
  While these Bayesian bounds concern estimation rather than learning, and are
  therefore not immediately comparable to ours, the van Trees bound remains
  based on Fisher information, whereas Ziv--Zakai bounds reduce estimation to
  binary hypothesis-testing problems~\cite{ClassicalDetectionEstimation2001,
    daurelioExperimentalInvestigationBayesian2022}.

Stepping back, our results provide a toolkit for deriving both no-go statements
  and success-probability guarantees in global quantum sensing. They may be
  particularly useful for quantum-computing-enhanced sensing
  protocols~\cite{allenQuantumComputingEnhanced2025,
    kannanExponentialQuantumAdvantage2026,
    khanQuantumComputationalSensing2025,
    khanQuantumComputationalSensingAdvantage2025,
    prabhuExponentialAdvantageQuantum2025}, where the framework of learning has
      an advantage over estimation because of its greater robustness to errors.

Regarding future research directions, all the results of this manuscript are
  one-shot, which means that they apply to a single copy of the sensor state. We
  leave for future work the derivation of multicopy asymptotic results on the
  rate of convergence of the success probability, as done in
  Ref.~\cite{meyerQuantumMetrologyFiniteSample2025}. This is not entirely
  trivial, as the success-probability guarantees have both a
  pretty-good-measurement component and a continuity correction, which must be
  accounted for in the multicopy asymptotic expansion.

At the level of achievability, the hypothesis-testing formulation and the PGM
  construction extend directly to multiple parameters by replacing the
  one-dimensional hypothesis net and tolerance band with a net over the
  multiparameter domain and an appropriate acceptance region. For a Bayesian
  nuisance prior, nuisance parameters can likewise be handled by grouping
  hypotheses according to the parameters of interest: an outcome is considered
  successful whenever those parameters lie in the prescribed acceptance region,
  while the success probability is summed over all nuisance-parameter labels.
  This grouping is operationally equivalent to averaging the sensor state over
  the nuisance parameters, as in Eq.~\eqref{eq:mixed_state_nuisance}. Although
  the grouped-hypothesis construction makes the extension of the achievability
  results straightforward, we leave a complete formulation of multiparameter
  quantum learning and its related applications to future work. A further
  challenge is to formulate analogous guarantees for waveform learning, thereby
  connecting parameter learning with existing quantum waveform-estimation
  theory~\cite{tsangFundamentalQuantumLimit2011,
    gardnerAchievingFundamentalQuantum2024,
    dingGaussianQuantumMetrology2026}.

Assouad's lemma provides a complementary route to minimax upper bounds on the
  success probability by embedding a hypercube of hypothesis points in the space
  of parameters and reducing the resulting estimation problem to a collection of
  binary tests~\cite{assouadDeuxRemarquesLestimation1983,yuAssouadFanoCam1997}.
  This construction is particularly natural for multiparameter models and ought
  to be explored in future work.

It is also worth exploring whether the coarse-graining approach that we
  introduced in Appendix~\ref{app:coarse_grained_reformulation} can lead to
  practical achievability results for mixed-state models.

\section*{AI disclosure}
Claude Opus 4.8 and ChatGPT 5.6 Sol were used in iterative exchanges directed by
  the authors to assist with mathematical derivations in
  Appendices~\ref{app:ising_qfi_lipschitz} and~\ref{app:dissipative_fidelity},
  the counterexample in Appendix~\ref{subsec:out_of_band_fidelity_penalty}, the
  upper-bound comparison in Appendix~\ref{subsec:comparison_upper_bounds}, and
  part of the coarse-grained reformulation in
  Appendix~\ref{app:coarse_grained_reformulation}. The authors independently
  checked every AI-assisted derivation for mathematical correctness. The same
  tools assisted in writing code implementing the formulas used to produce all
  plots. The authors verified that this code evaluates the intended mathematical
  expressions correctly. The AI-assisted code was based on an earlier
  implementation, written without AI assistance (all versions are available in
  the git repository). The AI tools were also used to improve the stylistic and
  grammatical consistency of the manuscript text and explanations. The resulting
  prose reflects entirely the authors' own scientific ideas and judgments.

\begin{acknowledgments}
F.B. thanks Dayou Yang, Johannes Jakob Meyer, Francesco Albarelli, Richard
  Allen, and Takuya Isogawa for useful discussions. This material is based upon
  work supported by the U.S. Department of Energy Office of Science National
  Quantum Information Science Research Centers as part of the Q-NEXT center.
  Q-NEXT provided primary funding support for F.B. during the completion of this
  research. We acknowledge support from the ARO (W911NF-23-1-0077), ARO MURI
  (W911NF-25-1-0261, W911NF-21-1-0325), AFOSR MURI (FA9550-21-1-0209,
  FA9550-23-1-0338), ONR MURI (N000142612102), DARPA (HR0011-24-9-0361), NSF
  (ERC-1941583, OMA-2137642, OSI-2326767, CCF-2312755, OSI-2426975). A.C.
  acknowledges support from the Simons Foundation through a Simons Investigator
  Award (Grant No. 669487). This material is based upon work supported by the
  U.S. Department of Energy, Office of Science, National Quantum Information
  Science Research Centers and Advanced Scientific Computing Research (ASCR)
  program under contract number DE-AC02-06CH11357 as part of the InterQnet
  quantum networking project. This work was completed with resources provided by
  the University of Chicago's Research Computing Center. This work was supported
  in part by a grant of access to OpenAI models through the ChatGPT for Academic
  Researchers program.

\end{acknowledgments}

\section*{Data Availability}
No separate numerical data set is associated with this work. All numerical
  results and figures can be reproduced using the authors' code, which will be
  made publicly available on GitLab and Zenodo upon publication. Repository
  links, version information, persistent identifiers, and the corresponding
  software citations will be included in the final version of the manuscript.

\appendix

\section{Classification of quantum metrology tasks}
\label{app:classification_metrology_tasks}
We classify quantum metrology tasks along the local--global and
  estimation--learning axes, establish quantitative connections between
  estimation and learning guarantees, and show that parameter-learning success
  probabilities are robust under trace-distance perturbations.

\subsection{Local versus global priors}
We begin by discussing the characterization of a metrological task through the
  amount of information we possess about the parameter before the measurement
  begins. In the regime of \textit{local metrology}, we assume that $\theta$ is
  already known to lie within a narrow neighborhood around a reference value,
  and the goal is to estimate infinitesimal deviations from this reference
  value. Most theory and practice in quantum sensing are primarily concerned
  with local estimation, tracking a small variation from a known parameter
  value. This scenario is the natural domain for applying the Fisher information
  and the quantum Fisher information, which appear in the quantum Cram\'er--Rao
  bound. In contrast, if $\theta$ is completely unknown or known only to follow
  a broad prior distribution over an extended region, local approximations break
  down, placing us in the regime of global metrology. Here, we must employ
  Bayesian statistics~\cite{albarelliMeasurementIncompatibilityBayesian2025} or
  global minimax strategies to bound the error across the entire available
  parameter space. In global estimation, the parameter belongs to an interval
  $[\theta_{\min}, \theta_{\max}]$, on which we might or might not have a prior
  distribution, according to the setting (Bayesian or minimax). According to the
  figure of merit used to evaluate the performance of the estimator, global
  metrology can be based on parameter estimation or on parameter learning. In
  the first scenario, some relevant tools are the Bayesian MSE and the minimax
  MSE~\cite{demkowicz-dobrzanskiMultiparameterEstimationQuantum2020a}, the van
  Trees bound~\cite{ClassicalDetectionEstimation2001}, and the Ziv--Zakai
  bound~\cite{daurelioExperimentalInvestigationBayesian2022}.
  See~\cite{shiGlobalBoundsLocal2026} for recent results on global estimation.

\subsection{The figure of merit: estimation versus learning}
The distinction between estimation and learning is in the definition of what
  constitutes a ``good'' estimator. In the quantum estimation scenario, we
  evaluate the error of the estimator $\check{\theta}$ using figures of merit
  that are \textit{averages} of a certain \textit{risk} function over the
  probability distribution of the estimator given the observed data. In the
  Bayesian case, we also average over the prior. The choice of risk or error
  function fundamentally dictates the optimal measurement strategy and can be
  informed by the features of the parameter space $\Theta$. The standard error
  for continuous, unbounded parameters (such as physical displacements or
  magnetic field strengths) is the mean squared error (MSE). When dealing with
  periodic parameters, such as an interferometric phase $\theta \in [-\pi,
  \pi)$, the standard MSE becomes ill-defined due to cyclic boundaries. In these
  scenarios, phase-appropriate metrics like the Holevo
  variance~\cite{holevoCovariantMeasurementsImprimitivity1984} are required to
  correctly measure statistical dispersion.

Both learning and estimation can be formulated within a Bayesian or minimax
  framework, depending on whether we average the error probability over a prior
  distribution for the parameter $\theta \in \Theta$ or we assess the worst-case
  error for the estimator $\check{\theta}$ across the entire parameter space
  $\Theta$. More precisely, given a figure of merit like the mean squared error
  (MSE), we can define the error in global estimation in two ways. First, we can
  take a Bayesian approach and define a Bayesian error, such as the Bayesian
  MSE, which averages the MSE over the prior distribution (see, for
  example,~\cite{vasilyevOptimalMultiparameterMetrology2024a,
    gardnerBayesianFrequencyEstimation2025}). The optimal estimator in this
      scenario is the mean of the posterior~\cite{TheoryPointEstimation1998}.
      The second approach is the minimax strategy, where we maximize the error
      over the parameter space and define the optimal measurement and estimator
      as those that minimize the worst-case risk. For parameter learning, the
      figure of merit in the Bayesian scenario is defined as the average success
      probability of the learning task over the prior, while the figure of merit
      in the minimax scenario is the worst-case success probability across the
      parameter space (see~\cite{meyerQuantumMetrologyFiniteSample2025,
    huangQueryComplexitiesQuantum2026}).

\subsection{Connecting estimation and learning}
\label{app:comparison_learning_estimation}
We now show that an informative $(\delta,\eta)$-bound follows from a bound on
  the mean squared error of an estimator. For $\delta>0$, applying Markov's
  inequality to the random variable $(\check{\theta}-\theta)^2$ gives
\begin{equation}
    \mathbb{P}\left[ (\check{\theta} - \theta)^2 \ge \delta^2 \right] \le
      \frac{\mathbb{E}[(\check{\theta} - \theta)^2]}{\delta^2} =
      \frac{\text{MSE}}{\delta^2}\,.
\end{equation}
This inequality implies
\begin{equation}
    \mathbb{P}[|\check{\theta} - \theta| > \delta] \le
      \frac{\text{MSE}}{\delta^2} \,,
\end{equation}
which can be rewritten as
\begin{equation}
    \mathbb{P}[|\check{\theta} - \theta| \le \delta] \ge 1 -
      \frac{\text{MSE}}{\delta^2} \,.
\end{equation}
These statements are pointwise in $\theta$ for a fixed measurement. A minimax
  implication requires the MSE bound to hold uniformly in $\theta$ for the same
  measurement, whereas the Bayesian version follows by averaging the above
  result over the prior.

This lower bound reaches the $\frac{3}{4}$ threshold for $\delta = 2
  \sqrt{\text{MSE}}$. Conversely, a $(\delta, \eta)$-statement with a
  non-vanishing error probability is typically insufficient to produce a strong
  bound on the MSE, because it does not control the magnitude of errors outside
  the success window. Explicitly,
\begin{align}
    \text{MSE} &\equiv \int_{\Theta} (\check{\theta} - \theta)^2
      p(\check{\theta})\,\dd\check{\theta} \\
    &= \int_{|\check{\theta} - \theta| > \delta} (\check{\theta} - \theta)^2
      p(\check{\theta})\,\dd\check{\theta} \nonumber\\
    &\quad + \int_{|\check{\theta} - \theta| \le \delta} (\check{\theta} -
      \theta)^2 p(\check{\theta})\,\dd\check{\theta} \,.
\end{align}
In the above equation, $p(\check{\theta})$ abbreviates
  $p(\check{\theta}|\hat\rho_\theta,\mathcal{M})$. For $0<\delta\le |\Theta|$,
  the squared error is at most $\delta^2$ inside the success window and at most
  $|\Theta|^2$ outside it. Since the success probability is at least $\eta$,
  this gives
\begin{align}
    \text{MSE}
    &\le \delta^2\,\mathbb{P}[|\check{\theta}-\theta|\le\delta]
    +|\Theta|^2\,\mathbb{P}[|\check{\theta}-\theta|>\delta] \nonumber\\
    &\le \delta^2\eta+|\Theta|^2(1-\eta) \,.
\end{align}
Thus, an informative MSE bound implies a success-probability guarantee, whereas
  a $(\delta,\eta)$-guarantee with non-negligible failure probability generally
  yields only the loose MSE bound above.

\subsection{Robustness under composition with quantum algorithms}
\label{app:robustness_under_composition}
In general, quantum algorithms produce output states $\hat{\rho}'$ that are
  close in trace distance to the ideal states $\hat{\rho}$ one wishes to
  prepare. In quantum signal
  processing~\cite{lowHamiltonianSimulationQubitization2019}, for example, a
  desired target function is approximated by a polynomial to precision
  $\varepsilon$, producing an output state that is $\varepsilon$-close to the
  ideal state in trace distance,
  $\tfrac12\norm{\hat{\rho}'-\hat{\rho}}_1\le\varepsilon$. In Grover search, the
  target state is reached through discrete rotations, so the algorithm can miss
  the target by a small amount controlled by the rotation step. In Trotterized
  Hamiltonian simulation, the state-preparation error is controlled by the time
  discretization.

The trace-distance guarantee implies that the ideal distribution
  $p(\check{\theta})$ and the perturbed distribution $p'(\check{\theta})$ are
  close in total variation. Trace-distance closeness alone does not control the
  Fisher information or quantum Fisher information, whose comparison generally
  requires derivative information. Median-error figures of merit can provide
  related robustness to
  outliers~\cite{belliardoOptimizingQuantumenhancedBayesian2024,
    ciminiBenchmarkingBayesianQuantum2024}, but this approach is less developed,
      in part because the Cram\'er--Rao bound and Fisher information make the
      MSE more tractable.

We now prove the robustness of parameter learning under perturbations of the
  state $\hat{\rho}_\theta$. More precisely, the success probability in
  Eq.~\eqref{eq:parameter_learning_eta} is $\tfrac{1}{2}$-Lipschitz with respect
  to the trace norm, equivalently $1$-Lipschitz with respect to the trace
  distance. Define $\Delta \hat{\rho}_\theta = \hat{\rho}_\theta -
  \hat{\rho}_\theta'$. Using the definition of the success probability in
  Eq.~\eqref{eq:parameter_learning_eta}, the difference in success probabilities
  can be written as
\begin{multline}
    \mathsf{P}_{\text{PL}} (\theta, \hat{\rho}_\theta, \delta; \mathcal{M}) -
      \mathsf{P}_{\text{PL}} (\theta, \hat{\rho}_\theta', \delta; \mathcal{M}) =
      \\ \int_{\Theta} w_\delta (\check{\theta} - \theta) \trsquare{\Delta
      \hat{\rho}_\theta \hat{M}_{\check{\theta}}}\,\dd\check{\theta} \,.
\end{multline}
Rather than bounding the integrand pointwise, we exploit the fact that the
  entire integral is the expectation value of a single measurement effect. Using
  the linearity of the trace, we commute it with the integral and define
\begin{equation}
    \hat{E}_\theta \equiv \int_{\Theta} w_\delta (\check{\theta} - \theta) \,
      \hat{M}_{\check{\theta}}\,\dd\check{\theta} \,,
    \label{eq:acceptance_effect}
\end{equation}
so that the difference in success probabilities is exactly $\trsquare{\Delta
  \hat\rho_\theta \hat{E}_\theta}$. The operator $\hat{E}_\theta$ is a valid
  effect, i.e., $0 \le \hat{E}_\theta \le \hat{\id}$: it is positive
  semidefinite because $w_\delta(\check{\theta} - \theta) \ge 0$ and
  $\hat{M}_{\check{\theta}} \ge 0$, and it is bounded above by the identity
  because the window function satisfies $w_\delta(\check{\theta} - \theta) \le
  1$ from Eq.~\eqref{eq:window_function} and the POVM obeys the completeness
  relation $\int_{\Theta} \hat{M}_{\check{\theta}}\,\dd\check{\theta} =
  \hat{\id}$.

The crucial observation is that $\Delta \hat\rho_\theta = \hat{\rho}_\theta -
  \hat{\rho}_\theta'$ is traceless, since both states are normalized. Writing
  its decomposition $\Delta \hat\rho_\theta = \hat{P} - \hat{N}$ into positive
  and negative parts $\hat{P}, \hat{N} \ge 0$ with orthogonal support, we obtain
  from tracelessness that $\trsquare{\hat{P}} = \trsquare{\hat{N}} =
  \tfrac{1}{2} \norm{\Delta \hat\rho_\theta}_1$. Because $0 \le \hat{E}_\theta
  \le \hat{\id}$ and $\hat{N} \ge 0$, we have $\trsquare{\hat{N} \hat{E}_\theta}
  \ge 0$ and $\trsquare{\hat{P} \hat{E}_\theta} \le \trsquare{\hat{P}}$, so that
\begin{equation}
    \begin{aligned}
    \trsquare{\Delta \hat\rho_\theta \hat{E}_\theta}
    &= \trsquare{\hat{P} \hat{E}_\theta} - \trsquare{\hat{N} \hat{E}_\theta} \\
    &\le \trsquare{\hat{P}} = \tfrac{1}{2} \norm{\Delta \hat\rho_\theta}_1 \,.
    \end{aligned}
\end{equation}
The same argument applied to $-\Delta \hat\rho_\theta$ yields the matching lower
  bound, so we obtain the Helstrom bound
\begin{align}
    \left| \mathsf{P}_{\text{PL}} (\theta, \hat{\rho}_\theta, \delta;
      \mathcal{M}) - \mathsf{P}_{\text{PL}} (\theta, \hat{\rho}_\theta', \delta;
      \mathcal{M}) \right|
    &\le \frac{1}{2} \norm{\hat{\rho}_\theta' - \hat{\rho}_\theta}_1 \,.
\end{align}
Because this bound is uniform in the POVM, it persists under optimization: for
  the Bayesian success probability, the right-hand side is averaged over the
  prior, whereas for the minimax success probability, it is replaced by the
  supremum over $\theta$. Thus, both optimized quantities are
  $\tfrac{1}{2}$-Lipschitz in the trace-norm metric.

\section{PGM success probability with tolerance}
\label{app:success_pgm}
In this appendix, we derive the expression for the PGM success probability in
  Eq.~\eqref{eq:pgm_success}, which is valid for pure states. We adopt a looser
  definition of success in which hypotheses close to the ground truth are also
  accepted. We call this the ``band'' acceptance probability. The average
  success probability of the PGM, as defined in
  Eq.~\eqref{eq:hypotheses_testing_eta_bayesian_povm}, is
\begin{equation}
    \bar{\mathsf{P}}_{\text{HT}} (\mu; \mathcal{M}_{\text{PG}}) =
      \sum_{k=1}^{K(\delta')} \mu_{k} \trsquare{\hat{\rho}_{k} \hat{M}_{k}} \,,
\end{equation}
with $\hat{M}_k$ given by Eq.~\eqref{eq:effect_PGM}. The success probability
  with tolerance, which has been used in
  Eq.~\eqref{eq:hypotheses_testing_eta_extended}, is
\begin{equation}
    \bar{\mathsf{P}}_{\text{Band}} (\mu; \mathcal{M}_{\text{PG}}) \equiv
      \sum_{k=1}^{K(\delta')} \sum_{\check{k}\,:\,|\check{k}-k| \le a} \mu_k
      \trsquare{\hat{\rho}_{k} \hat{M}_{\check{k}}} \,.
    \label{eq:extended_eta_pgm}
\end{equation}
We define the unnormalized states $\ket{\tilde{\psi}_k} \equiv \sqrt{\mu_k}
  \ket{\psi_k}$. Thinking of the Hilbert space as a column-vector space, we
  build a matrix $V$ where each column is one of the states
  $\ket{\tilde{\psi}_k}$:
\begin{equation}
    V \equiv \big[ |\tilde{\psi}_1\rangle \quad |\tilde{\psi}_2\rangle \quad
      \dots \quad |\tilde{\psi}_{K(\delta')}\rangle \big] \,.
\end{equation}
Because the hypothesis ensemble is finite, the range of $V$ is
  finite-dimensional. All inverse powers in the following calculation are
  therefore restricted to this range, consistently with the support convention
  of Eq.~\eqref{eq:effect_PGM}.

From this matrix we recover $\bar\rho = VV^\dagger = \sum_i
  |\tilde{\psi}_i\rangle \langle \tilde{\psi}_i|$ and the Gram matrix
  $\boldsymbol{G} = V^\dagger V$, whose components are $(\boldsymbol{G})_{ij} =
  \braket{\tilde{\psi}_i | \tilde{\psi}_j} = \sqrt{\mu_i \mu_j} \braket{\psi_i |
  \psi_j}$. The singular-value decomposition of $V$ implies that $\bar\rho$ and
  $\boldsymbol{G}$ have the same nonzero eigenvalues. Since the hypothesis
  states are pure, the success probability can be written as
\begin{equation}
\begin{aligned}
    & \sum_{k=1}^{K(\delta')} \sum_{\check{k}\,:\,|\check{k}-k| \le a} \mu_k
      \trsquare{\hat{\rho}_{k} \hat{M}_{\check{k}}} \\
    &=  \sum_{k=1}^{K(\delta')} \sum_{\check{k}\,:\,|\check{k}-k| \le a}
      \braket{\tilde{\psi}_k | \bar\rho^{-1/2} | \tilde{\psi}_{\check{k}}}
      \braket{\tilde{\psi}_{\check{k}} | \bar\rho^{-1/2} | \tilde{\psi}_k} \\
    &=  \sum_{k=1}^{K(\delta')} \sum_{\check{k}\,:\,|\check{k}-k| \le a} \left|
      \braket{\tilde{\psi}_k | \bar\rho^{-1/2} | \tilde{\psi}_{\check{k}}}
      \right|^2 \,.
\end{aligned}
\end{equation}
It follows straightforwardly from the singular-value decomposition of $V$ that
  $V^\dagger (VV^\dagger)^{-1/2} = (V^\dagger V)^{-1/2} V^\dagger$. Let $\lbrace
  \mathbf{e}_{k}\rbrace_{k=1}^{K(\delta')}$ denote the standard basis vectors.
  This identity allows us to rewrite the inner terms as
\begin{align}
    \braket{\tilde{\psi}_k | \bar\rho^{-1/2} | \tilde{\psi}_{\check{k}}} &=
      \mathbf{e}_k^\dagger V^\dagger \bar\rho^{-1/2} V \mathbf{e}_{\check{k}}
      \notag \\
    &= \mathbf{e}_k^\dagger (\boldsymbol{G}^{-1/2} V^\dagger) V
      \mathbf{e}_{\check{k}} \\
    &= \mathbf{e}_k^\dagger \boldsymbol{G}^{-1/2} (V^\dagger V)
      \mathbf{e}_{\check{k}} \notag \\
    &= \mathbf{e}_k^\dagger \boldsymbol{G}^{-1/2} \boldsymbol{G}
      \mathbf{e}_{\check{k}} \notag \\
    &= \mathbf{e}_k^\dagger \sqrt{\boldsymbol{G}} \mathbf{e}_{\check{k}} \notag
      \\
    &= (\sqrt{\boldsymbol{G}})_{k \check{k}} \,.
    \label{eq:exact_band_success_pure}
\end{align}
The Gram matrix is Hermitian, and so is $\sqrt{\boldsymbol{G}}$. Its diagonal
  elements are real, whereas the off-diagonal elements can be complex and
  therefore require an absolute value. The ``band'' success probability in
  Eq.~\eqref{eq:extended_eta_pgm} can thus be written as
\begin{equation}
    \bar{\mathsf{P}}_{\text{Band}} (\mu; \mathcal{M}_{\text{PG}}) =
      \sum_{k=1}^{K(\delta')} \sum_{\check{k}\,:\,|k-\check{k}|\le a} |
      (\sqrt{\boldsymbol{G}})_{k \check{k}}|^2 \,.
    \label{eq:exact_band_success_pure_result}
\end{equation}
This concludes the proof. If $a=0$, testing is successful only when the exact
  ground truth is identified, i.e., $\check{k}=k$. Under these conditions, the
  double summation reduces to
\begin{equation}
    \bar{\mathsf{P}}_{\text{HT}} (\mu; \mathcal{M}_{\text{PG}}) =
      \sum_{k=1}^{K(\delta')} | (\sqrt{\boldsymbol{G}})_{k k}|^2 \,,
\end{equation}
which yields Eq.~\eqref{eq:pgm_success}.

\section{Bounds on the block success probability}
\label{app:block_bounds}
In this section, we generalize the upper bound on the error probability of
  multi-hypothesis testing presented in~\cite{montanaroPrettySimpleBounds}. We
  use this generalization to compute a lower bound on the ``band'' success
  probability introduced in
  Eq.~\eqref{eq:theorem_statement_testing_to_learning}, which is central to
  evaluating success-probability guarantees for parameter learning. The relevant
  success probability is
\begin{equation}
    \bar{\mathsf{P}}_{\text{Band}}(\mu; \mathcal{M}) \equiv
      \sum_{k=1}^{K(\delta')} \sum_{\check{k}\,:\,|k-\check{k}|\le a} \mu_k
      \trsquare{\hat{\rho}_{k} \hat{M}_{\check{k}}} \,,
    \label{eq:band_success_probability}
\end{equation}
or, equivalently, we can use its corresponding error probability,
  $1-\bar{\mathsf{P}}_{\text{Band}}(\mu; \mathcal{M})$.

If the hypothesis states $\hat{\rho}_k$ are pure, we can lower-bound this
  expression using the performance of the PGM for pure states; see
  Eq.~\eqref{eq:exact_band_success_pure_result}. This quantity is significantly
  more difficult to evaluate for mixed states. To address this, we lower-bound
  the ``block'' success probability achieved by the PGM, defined in
  Eq.~\eqref{eq:block_success_probability}. See also
  Fig.~\ref{fig:matrix_representation}.

\subsection{Lower bound on the block success probability}
\label{subsec:block_error_probability_upper_bound}
We generalize the result of~\cite{montanaroPrettySimpleBounds} to establish an
  upper bound on the block error probability. We first assume that the
  hypothesis states have finite rank and write
\begin{equation}
    \hat{\rho}_i
    =
    \sum_{k=1}^{D}
    \lambda_{ik}\ket{\psi_{ik}}\!\bra{\psi_{ik}} \, ,
    \label{eq:orthonormal_decomposition_hypotheses}
\end{equation}
where $D$ is the maximum rank of the states and zero eigenvalues are appended
  when necessary. The first index labels the hypothesis, whereas the second
  labels its eigenvectors. The weighted pure-state ensemble
  $\{\sqrt{\mu_i\lambda_{ik}}\ket{\psi_{ik}}\}_{i,k}$ has Gram matrix
\begin{equation}
    \boldsymbol{G}_{ik,jl}
    \equiv
    \sqrt{\mu_i\mu_j\lambda_{ik}\lambda_{jl}}\,
    \braket{\psi_{ik}|\psi_{jl}} \, .
    \label{eq:mixed_gram_matrix}
\end{equation}
The pairs $(i,k)$ and $(j,l)$ are the row and column indices, respectively, so
  $\boldsymbol{G}$ has a natural block structure indexed by the hypotheses $i$
  and $j$.

We next express the PGM outcome probabilities in terms of
  $\sqrt{\boldsymbol{G}}$. On the support of $\bar{\rho}$, define
\begin{equation}
    \ket{\nu_{ik}}
    \equiv
    \sqrt{\mu_i\lambda_{ik}}\,
    \bar{\rho}^{-1/2}\ket{\psi_{ik}} \, ,
    \qquad
    \hat{M}_i
    =
    \sum_{k=1}^{D}
    \ket{\nu_{ik}}\!\bra{\nu_{ik}} \, .
    \label{eq:mixed_pgm_vectors}
\end{equation}
These effects satisfy $\sum_i\hat{M}_i=\hat{\Pi}_{\bar{\rho}}$, where
  $\hat{\Pi}_{\bar{\rho}}$ is the support projection of $\bar{\rho}$. Introduce
  the matrix $\boldsymbol{P}$ with entries
\begin{align}
    \boldsymbol{P}_{ik,jl}
    &\equiv
    \sqrt{\mu_j\lambda_{jl}}\,
    \braket{\nu_{ik}|\psi_{jl}} \notag\\
    &=
    \sqrt{\mu_i\mu_j\lambda_{ik}\lambda_{jl}}\,
    \bra{\psi_{ik}}\bar{\rho}^{-1/2}\ket{\psi_{jl}} \, .
    \label{eq:mixed_pgm_amplitude_matrix}
\end{align}
The positivity of $\bar{\rho}^{-1/2}$ implies that $\boldsymbol{P}$ is positive
  semidefinite. Moreover,
\begin{widetext}
    \begin{align}
    (\boldsymbol{P}^2)_{ik,jl}
    &=
    \sqrt{\mu_i\mu_j\lambda_{ik}\lambda_{jl}}\,
    \bra{\psi_{ik}}\bar{\rho}^{-1/2}
    \left(
        \sum_{r,s}
        \mu_r\lambda_{rs}
        \ket{\psi_{rs}}\!\bra{\psi_{rs}}
    \right)
    \bar{\rho}^{-1/2}\ket{\psi_{jl}}
    \notag\\
    &=
    \sqrt{\mu_i\mu_j\lambda_{ik}\lambda_{jl}}\,
    \braket{\psi_{ik}|\psi_{jl}}
    =
    \boldsymbol{G}_{ik,jl} \, .
    \label{eq:mixed_pgm_square_root}
\end{align}
\end{widetext}
Since $\boldsymbol{P}$ is positive semidefinite,
  Eq.~\eqref{eq:mixed_pgm_square_root} shows that
  $\boldsymbol{P}=\sqrt{\boldsymbol{G}}$.

We can now evaluate the probability of obtaining outcome $i$ when the true
  hypothesis is $j$:
\begin{align}
    \mu_j\trsquare{\hat{\rho}_j\hat{M}_i}
    &=
    \sum_{k,l=1}^{D}
    \mu_j\lambda_{jl}
    \left|
        \braket{\nu_{ik}|\psi_{jl}}
    \right|^2 \notag\\
    &=
    \norm{\boldsymbol{P}^{(ij)}}_2^2
    =
    \norm{\sqrt{\boldsymbol{G}}^{(ij)}}_2^2 \, ,
    \label{eq:error_probability_mixed_states_pgm}
\end{align}
where the superscript $(ij)$ denotes the corresponding hypothesis block and
  $\norm{\cdot}_2$ is the Frobenius norm.

For each hypothesis block $I_t$, define the projector
\begin{equation}
    \boldsymbol{R}_t
    \equiv
    \sum_{i\in I_t}\sum_{k=1}^{D}
    \ket{i,k}\!\bra{i,k} \, .
    \label{eq:hypothesis_block_projection}
\end{equation}
Orthogonality of the matrix blocks gives the exact block success probability
\begin{equation}
    \bar{\mathsf{P}}_{\mathrm{Block}}
    (\mu;\mathcal{M}_{\mathrm{PG}})
    =
    \sum_{t=1}^{T}
    \norm{
        \boldsymbol{R}_t
        \sqrt{\boldsymbol{G}}
        \boldsymbol{R}_t
    }_2^2 \, .
    \label{eq:block_success_gram_operator}
\end{equation}
Since $\norm{\sqrt{\boldsymbol{G}}}_2^2 = \operatorname{Tr}(\boldsymbol{G}) =
  \sum_i\mu_i = 1$, the complementary block error probability is
\begin{align}
    1-\bar{\mathsf{P}}_{\mathrm{Block}}
    (\mu;\mathcal{M}_{\mathrm{PG}})
    &=
    \sum_{\substack{t,t'=1\\t\neq t'}}^{T}
    \norm{
        \boldsymbol{R}_t
        \sqrt{\boldsymbol{G}}
        \boldsymbol{R}_{t'}
    }_2^2 \notag\\
    &=
    \sum_{t=1}^{T}
    \norm{
        \boldsymbol{R}_t
        \sqrt{\boldsymbol{G}}
        \boldsymbol{S}_t
    }_2^2 \, ,
    \label{eq:block_error_gram_operator}
\end{align}
where $\boldsymbol{S}_t\equiv\id-\boldsymbol{R}_t$.

Applying the two-block lemma from Appendix~E
  of~\cite{audenaertUpperBoundsError2014} to $\sqrt{\boldsymbol{G}}$ and the
  projection $\boldsymbol{R}_t$ gives
\begin{equation}
    \norm{
        \boldsymbol{R}_t
        \sqrt{\boldsymbol{G}}
        \boldsymbol{S}_t
    }_2^2
    \le
    \frac{1}{2}
    \norm{
        \boldsymbol{R}_t
        \boldsymbol{G}
        \boldsymbol{S}_t
    }_1 \, .
    \label{eq:two_block_pgm_inequality}
\end{equation}

Applying Eq.~\eqref{eq:two_block_pgm_inequality} to
  Eq.~\eqref{eq:block_error_gram_operator} and then using the trace-norm
  triangle inequality first over hypothesis blocks and then over individual
  hypotheses gives
\begin{align}
    1-\bar{\mathsf{P}}_{\mathrm{Block}}
    (\mu;\mathcal{M}_{\mathrm{PG}})
    &\le
    \frac{1}{2}
    \sum_{t=1}^{T}
    \norm{
        \boldsymbol{R}_t
        \boldsymbol{G}
        \boldsymbol{S}_t
    }_1 \notag\\
    &\le
    \frac{1}{2}
    \sum_{\substack{t,t'=1\\t\neq t'}}^{T}
    \norm{
        \boldsymbol{R}_t
        \boldsymbol{G}
        \boldsymbol{R}_{t'}
    }_1 \notag\\
    &\le
    \frac{1}{2}
    \sum_{\substack{t,t'=1\\t\neq t'}}^{T}
    \sum_{\substack{i\in I_t\\j\in I_{t'}}}
    \norm{\boldsymbol{G}^{(ij)}}_1 \, .
    \label{eq:block_error_trace_norm}
\end{align}
Each hypothesis block of the Gram matrix factorizes as
\begin{align}
    \boldsymbol{G}^{(ij)}
    &=
    \left(
        \sum_{k=1}^{D}
        \sqrt{\mu_i\lambda_{ik}}
        \ket{k}\!\bra{\psi_{ik}}
    \right)
    \left(
        \sum_{l=1}^{D}
        \sqrt{\mu_j\lambda_{jl}}
        \ket{\psi_{jl}}\!\bra{l}
    \right) \, .
    \label{eq:mixed_gram_block_factorization}
\end{align}
Let $\boldsymbol{W}_i:\mathbb{C}^{D}\to\mathcal{H}$ denote the isometry defined
  by $\boldsymbol{W}_i\ket{k}=\ket{\psi_{ik}}$.
  Equation~\eqref{eq:mixed_gram_block_factorization} then gives
\begin{equation}
    \boldsymbol{W}_i
    \boldsymbol{G}^{(ij)}
    \boldsymbol{W}_j^\dagger
    =
    \sqrt{\mu_i\mu_j}\,
    \sqrt{\hat{\rho}_i}
    \sqrt{\hat{\rho}_j} \, .
    \label{eq:mixed_gram_block_isometries}
\end{equation}
Multiplication on the left and right by these isometries preserves the nonzero
  singular values. Therefore,
\begin{align}
    \norm{\boldsymbol{G}^{(ij)}}_1
    &=
    \sqrt{\mu_i\mu_j}\,
    \norm{
        \sqrt{\hat{\rho}_i}
        \sqrt{\hat{\rho}_j}
    }_1 \notag\\
    &=
    \sqrt{\mu_i\mu_j}\,
    F(\hat{\rho}_i,\hat{\rho}_j) \, .
    \label{eq:mixed_gram_block_fidelity}
\end{align}
Combining Eqs.~\eqref{eq:block_error_trace_norm}
  and~\eqref{eq:mixed_gram_block_fidelity} yields
\begin{equation}
    \begin{aligned}
    \bar{\mathsf{P}}_{\mathrm{Block}}
    (\mu;\mathcal{M}_{\mathrm{PG}})
    &\ge
    1-\frac{1}{2}
    \sum_{\substack{t,t'=1\\t\neq t'}}^T
    \sum_{\substack{i\in I_t\\j\in I_{t'}}}
    \sqrt{\mu_i\mu_j}\,
    F(\hat{\rho}_i,\hat{\rho}_j) \, .
    \end{aligned}
    \label{eq:lower_bound_success_blocks}
\end{equation}
The finite-rank assumption can be removed by applying the bound to the
  normalized rank-$D$ spectral truncations and taking $D\to\infty$: the
  corresponding Gram operators converge in trace norm and their square roots in
  Hilbert--Schmidt norm, implying convergence of the PGM block success
  probabilities, while continuity of the fidelity under trace-norm convergence
  recovers the right-hand side of Eq.~\eqref{eq:lower_bound_success_blocks}.

\subsection{Worked example}
\label{subsec:application_example}
For this bound to be useful, the fidelity must decay sufficiently fast when the
  hypotheses are far apart. We consider the following exponential decay:
\begin{equation}
    F(\hat{\rho}_i, \hat{\rho}_j) \equiv \mathrm{e}^{-\gamma |i-j|} \,.
    \label{eq:exponential_decay_fidelities}
\end{equation}
For the uniform prior, suppose that $K(\delta')/\ell$ is an integer, so that
  every block has exactly $\ell=a+1$ elements.
  Equation~\eqref{eq:exponential_decay_fidelities} then gives
\begin{equation}
    \begin{aligned}
    &\sum_{\substack{t,t'=1\\t\neq t'}}^T
    \sum_{\substack{i \in I_t\\j \in I_{t'}}}
    \sqrt{\mu_i \mu_j} F(\hat{\rho}_i, \hat{\rho}_j)\\
    &\qquad = \frac{2 \mathrm{e}^{-\gamma}}{\left(1 -
      \mathrm{e}^{-\gamma}\right)^2}
    \left[
    \frac{1 - \mathrm{e}^{-\gamma \ell}}{\ell}
    - \frac{1 - \mathrm{e}^{-\gamma K(\delta')}}{K(\delta')}
    \right] \,.
    \end{aligned}
\end{equation}
Consequently,
\begin{equation}
    \bar{\mathsf{P}}_{\text{Band}} \ge 1
    - \frac{\mathrm{e}^{-\gamma}}{\left(1 - \mathrm{e}^{-\gamma}\right)^2}
    \left[
    \frac{1 - \mathrm{e}^{-\gamma \ell}}{\ell}
    - \frac{1 - \mathrm{e}^{-\gamma K(\delta')}}{K(\delta')}
    \right] \,.
    \label{eq:example_new_bound}
\end{equation}
For $\ell\gg1$ and $K(\delta')\gg \ell$, the leading failure term scales as
  $\mathrm{e}^{-\gamma}/[\ell(1-\mathrm{e}^{-\gamma})^2]$.

Suppose instead that one evaluates the band success probability using only the
  probability of identifying the exact ground truth, without any tolerance. In
  this case, the bound on the success probability is
\begin{equation}
    \bar{\mathsf{P}}_{\text{Band}} \ge 1 - \frac{\mathrm{e}^{-\gamma}}{1 -
      \mathrm{e}^{-\gamma}} \,,
\end{equation}
which does not exhibit the $1/\ell$ suppression of the leading failure term in
  Eq.~\eqref{eq:example_new_bound}.

\subsection{Failure of an out-of-band fidelity penalty}
\label{subsec:out_of_band_fidelity_penalty}
The disjoint-block construction in Eq.~\eqref{eq:lower_bound_success_blocks}
  cannot, in general, be replaced by a fidelity penalty restricted to pairs
  outside the full tolerance band, $|i-j|>a$. This fails even for pure qubit
  states. Consider five hypotheses with uniform prior, $\mu_j=1/5$, and
\begin{equation}
    \ket{\psi_j}
    =
    \cos\!\left[\frac{(j-1)\pi}{8}\right]\ket{0}
    +
    \sin\!\left[\frac{(j-1)\pi}{8}\right]\ket{1}\; ,
    \label{eq:out_of_band_counterexample_states}
\end{equation}
with $j=1,\dots,5$. For $a=3$, the only unordered pair outside the tolerance
  band is $\{1,5\}$. Since $\ket{\psi_1}=\ket{0}$ and $\ket{\psi_5}=\ket{1}$,
  one has $F(\hat{\rho}_1,\hat{\rho}_5)=0$. A PGM bound whose fidelity penalty
  contained only pairs with $|i-j|>a$ would therefore predict unit band success.

This prediction is false. The average state is
\begin{equation}
    \bar{\rho}
    =
    \frac{1}{2}\hat{\id}
    +
    \frac{1+\sqrt{2}}{10}\hat{\sigma}_x ,
\end{equation}
with eigenvalues $\lambda_+=(6+\sqrt{2})/10$ and $\lambda_-=(4-\sqrt{2})/10$.
  Using $(\sqrt{\boldsymbol{G}})_{ij} =
  \sqrt{\mu_i\mu_j}\braket{\psi_i|\bar{\rho}^{-1/2}|\psi_j}$ gives
\begin{equation}
    (\sqrt{\boldsymbol{G}})_{15}
    =
    \frac{1}{10}
    \left(
    \lambda_+^{-1/2}
    -
    \lambda_-^{-1/2}
    \right)
    \neq 0 .
    \label{eq:out_of_band_counterexample_gram}
\end{equation}
The exact pure-state formula in Eq.~\eqref{eq:exact_band_success_pure_result}
  therefore yields
\begin{equation}
    1-\bar{\mathsf{P}}_{\mathrm{Band}}
    (\mu;\mathcal{M}_{\mathrm{PG}})
    =
    2\left|(\sqrt{\boldsymbol{G}})_{15}\right|^2
    >0 .
    \label{eq:out_of_band_counterexample_error}
\end{equation}
Thus, the square root of the Gram matrix couples the two endpoints through the
  intermediate hypotheses, even though the endpoint states are orthogonal. The
  disjoint-block construction avoids this obstruction because its penalty
  retains every pair belonging to different blocks, rather than only pairs
  outside the full tolerance band.

\section{Application of the quantum minimax theorem}
\label{app:application_minimax_theorem}
\subsection{Parameter learning}
\label{app:application_minimax_theorem_pl}
We assume that $\Theta$ is compact and that
  $\lbrace\hat\rho_\theta\rbrace_{\theta\in\Theta}$ is a regular quantum
  statistical model in the sense of~\cite{tanakaQuantumMinimaxTheorem2014}. We
  optimize over all POVMs with outcomes in $\Theta$; this measurement class is
  compact (and hence closed) and convex in the corresponding weak topology. The
  failure loss corresponding to Eq.~\eqref{eq:window_function} is
\begin{equation}
    \ell_\delta(\theta,\check\theta)
    \equiv 1-w_\delta(\check\theta-\theta)
    =\chi_{\{|\check\theta-\theta|>\delta\}} \,.
\end{equation}
This bounded loss is discontinuous at $|\check\theta-\theta|=\delta$ but lower
  semicontinuous because $\{|\check\theta-\theta|>\delta\}$ is open. The quantum
  minimax theorem hypotheses are therefore satisfied.

\subsection{Hypothesis testing}
\label{app:application_minimax_theorem_ht}
For hypothesis testing, the parameter and decision spaces are finite, and the
  full set of POVMs is compact and convex. The failure loss is
\begin{equation}
    \ell(k, \check{k}) \equiv 1-\delta_{k, \check{k}} \,.
\end{equation}
This loss is continuous because its domain is discrete. Thus, the hypotheses of
  the quantum minimax theorem~\cite{tanakaQuantumMinimaxTheorem2014} are
  satisfied.

\section{Equivalence between quantum parameter learning and multi-hypothesis
  testing}
\label{app:theorem_proof}
In this appendix, we prove the equivalence between multi-hypothesis testing and
  parameter learning expressed in
  Eqs.~\eqref{eq:theorem_statement_learning_to_testing}
  and~\eqref{eq:theorem_statement_testing_to_learning}.

\subsection{Parameter learning to hypothesis testing}
\label{app:from_pl_to_ht}
Here, we demonstrate how a hypothesis-testing estimator can be constructed from
  a parameter-learning estimator such that its success probability is at least
  as large as that of the underlying learning task. Let
  $\check{\theta}_{\text{PL}}$ denote the estimator with precision $\delta$
  produced by the learning algorithm. We construct the hypothesis-testing
  estimator $\check{k}$ as follows:
\begin{equation}
    \check{k}
    \equiv
        \operatorname*{arg\,min}_{j\in\lbrace 1,\ldots,K\rbrace}
        |\check{\theta}_{\text{PL}}-\theta_j| \,.
    \label{eq:definition_ht_from_pl}
\end{equation}
Suppose that the learning estimator succeeds when the true hypothesis is $k$, so
  that $|\check{\theta}_{\text{PL}}-\theta_k|\le\delta$. Then, for every
  $\ell\neq k$,
\begin{equation}
    |\check{\theta}_{\text{PL}}-\theta_\ell|
    \ge
    |\theta_k-\theta_\ell|
    -
    |\check{\theta}_{\text{PL}}-\theta_k|
    \ge
    \delta'-\delta
    >
    \delta
    \ge
    |\check{\theta}_{\text{PL}}-\theta_k| \,.
\end{equation}
Thus, $k$ is the unique nearest grid point whenever the learning estimator
  succeeds on hypothesis $k$, and the derived hypothesis-testing estimator
  outputs $k$. Thus, the success probability of correctly identifying the
  hypothesis is at least the probability that the estimator
  $\check{\theta}_{\text{PL}}$ falls within the learning window around the true
  hypothesis $\theta_k$:
\begin{equation}
    \mathsf{P}_{\text{HT}} (k; \mathcal{M}') \ge \mathsf{P}_{\text{PL}}
      ({\theta_{k}}, \delta; \mathcal{M}) \,.
\end{equation}
In this context, $\mathcal{M}$ is the POVM that outputs
  $\check{\theta}_{\text{PL}}$, while $\mathcal{M}'$ outputs the derived
  hypothesis-testing estimator $\check{k}$. Both strategies utilize the same
  underlying physical measurement, differing only in the straightforward
  post-processing stage defined by Eq.~\eqref{eq:definition_ht_from_pl}.

Next, we introduce a discrete prior probability distribution over the
  hypotheses, denoted by $\mu$, and average the hypothesis-testing success
  probability over this prior:
\begin{equation}
    \sum_{k=1}^{K} \mu_k \, \mathsf{P}_{\text{HT}} (k; \mathcal{M}') \ge
      \sum_{k=1}^{K} \mu_k \, \mathsf{P}_{\text{PL}} ({\theta_{k}}, \delta;
      \mathcal{M}) \,.
\end{equation}
The left-hand side of this equation represents the Bayesian success probability
  for hypothesis testing under a specific measurement. Because the optimal
  hypothesis-testing device maximizes this success probability over all valid
  measurements, we can replace the left-hand side with the optimal Bayesian
  success probability defined in Eq.~\eqref{eq:hypotheses_testing_eta_bayesian},
  yielding the inequality:
\begin{equation}
    \bar{\eta}_{\text{HT}} (\mu) \ge \sum_{k=1}^{K} \mu_k \,
      \mathsf{P}_{\text{PL}} ({\theta_{k}}, \delta; \mathcal{M}) \,.
\end{equation}
Since this construction holds for any parameter-learning estimator and
  measurement scheme, we can also maximize the right-hand side over
  $\mathcal{M}$ to find
\begin{equation}
    \bar{\eta}_{\text{HT}} (\mu) \ge \sup_{\mathcal{M}} \sum_{k=1}^{K} \mu_k \,
      \mathsf{P}_{\text{PL}} (\theta_k, \delta; \mathcal{M}) \,.
    \label{eq:bayesian_ht_upper_bound_discrete_learning}
\end{equation}
Minimizing both sides over the discrete prior $\mu$ yields
\begin{equation}
    \inf_{\mu} \bar{\eta}_{\text{HT}} (\mu) \ge \inf_{\mu} \sup_{\mathcal{M}}
      \sum_{k=1}^{K} \mu_k \, \mathsf{P}_{\text{PL}} (\theta_k, \delta;
      \mathcal{M}) \,.
\end{equation}
On the right-hand side, the discrete prior can be embedded into a probability
  measure on $\Theta$ defined as
\begin{equation}
    \tilde{\mu} \equiv \sum_{k=1}^{K} \mu_k \delta_{\theta_k} \,,
    \label{eq:discrete_form_mu}
\end{equation}
where $\delta_{\theta_k}$ is the Dirac probability measure concentrated at
  $\theta_k$. This allows us to rewrite the inequality as
\begin{multline}
    \inf_{\mu} \bar{\eta}_{\text{HT}} (\mu) \ge \inf_{\tilde{\mu}}
      \sup_{\mathcal{M}} \int_{\theta_{\min}}^{\theta_{\max}}
      \mathsf{P}_{\text{PL}} (\theta, \delta; \mathcal{M}) \,\dd \tilde{\mu}
      (\theta) \,.
\end{multline}
Crucially, minimizing over the restricted class of probability measures
  supported on $\Omega$ in Eq.~\eqref{eq:discrete_form_mu} yields an infimum
  that is at least as large as the infimum obtained by minimizing over all
  probability measures on $\Theta$. Therefore, we can write
\begin{equation}
    \inf_{\mu} \bar{\eta}_{\text{HT}} (\mu) \ge \inf_{\mu} \sup_{\mathcal{M}}
      \int_{\theta_{\min}}^{\theta_{\max}} \mathsf{P}_{\text{PL}} (\theta,
      \delta; \mathcal{M}) \,\dd {\mu} (\theta) \,,
\end{equation}
where the minimization on the right-hand side is now performed over all
  probability measures on $\Theta$. This right-hand expression is precisely the
  worst-prior Bayesian success probability of quantum parameter learning:
\begin{equation}
    \inf_{\mu} \bar{\eta}_{\text{HT}} (\mu) \ge \inf_{\mu}
      \bar{\eta}_{\text{PL}} (\mu, \delta) \,.
\end{equation}
Finally, applying the quantum minimax theorem as formulated
  in~\cite{tanakaQuantumMinimaxTheorem2014} and summarized in
  Eqs.~\eqref{eq:pl_minimax_theorem} and~\eqref{eq:ht_quantum_minimax}, we
  convert both worst-prior success probabilities into their equivalent minimax
  success probabilities and obtain
\begin{equation}
    \eta_{\text{PL}} (\delta) \le \eta_{\text{HT}} \,.
\end{equation}
This result survives the limit $\delta'\to2\delta^+$, completing the proof of
  Eq.~\eqref{eq:theorem_statement_learning_to_testing}.

We now return to Eq.~\eqref{eq:bayesian_ht_upper_bound_discrete_learning} and
  derive the Bayesian upper bound for a uniform prior on $\Theta$. Fix
  $\delta'>2\delta$ and consider the family of equally spaced grids
\begin{equation}
    \theta_k^r \equiv \theta_{\min}+r+(k-1)\delta' \,,
    \qquad k=1,\ldots,K(\delta') \,,
    \label{eq:shifted_grids_bayesian_upper_bound}
\end{equation}
with $0\le r<\delta'$. The grid in Eq.~\eqref{eq:set_hypothesis_definition}
  corresponds to $r=\delta'/2$. Apart from the endpoint $\theta_{\max}$, which
  has zero measure under the uniform prior, every $\theta\in\Theta$ belongs to
  exactly one of the shifted grids in
  Eq.~\eqref{eq:shifted_grids_bayesian_upper_bound}. We denote the optimal
  Bayesian hypothesis-testing success probability for the uniform prior on the
  grid with offset $r$ by $\bar{\eta}_{\mathrm{HT}}^r(\mu_U)$.

Rewriting the uniform average over $\Theta$ as an average first over the offset
  $r$ and then over the points of the corresponding grid gives
\begin{align}
    \bar{\eta}_{\mathrm{PL}}(\mu_U,\delta)
    &= \sup_{\mathcal{M}}
    \frac{1}{\delta'}\int_0^{\delta'}\dd r\,
    \frac{1}{K(\delta')}\sum_{k=1}^{K(\delta')}
    \mathsf{P}_{\mathrm{PL}}(\theta_k^r,\delta;\mathcal{M})
    \notag\\
    &\le \frac{1}{\delta'}\int_0^{\delta'}\dd r\,
    \sup_{\mathcal{M}}
    \frac{1}{K(\delta')}\sum_{k=1}^{K(\delta')}
    \mathsf{P}_{\mathrm{PL}}(\theta_k^r,\delta;\mathcal{M})
    \notag\\
    &\le \frac{1}{\delta'}\int_0^{\delta'}\dd r\,
    \bar{\eta}_{\mathrm{HT}}^r(\mu_U) \,.
    \label{eq:bayesian_learning_to_testing_shift_average}
\end{align}
The first inequality allows the learning measurement to depend on $r$, which is
  equivalent to revealing the grid offset before the measurement and hence
  defines an easier learning problem.
  Equation~\eqref{eq:bayesian_ht_upper_bound_discrete_learning} therefore gives
  the second inequality. The final integral averages the resulting
  hypothesis-testing upper bound over the offset introduced to decompose the
  original continuous uniform prior.

If the optimal hypothesis-testing success probability is invariant under shifts
  of the grid, then $\bar{\eta}_{\mathrm{HT}}^r(\mu_U)$ is independent of $r$.
  Equation~\eqref{eq:bayesian_learning_to_testing_shift_average} then reduces to
\begin{equation}
    \bar{\eta}_{\mathrm{PL}}(\mu_U,\delta)
    \le
    \bar{\eta}_{\mathrm{HT}}(\mu_U) \,,
    \label{eq:bayesian_learning_to_testing_shift_invariant}
\end{equation}
which proves
  Eqs.~\eqref{eq:theorem_statement_learning_to_testing_bayesian_upper}
  and~\eqref{eq:theorem_statement_learning_to_testing_%
bayesian_shift_invariant}.

\subsection{Hypothesis testing to parameter learning}
\label{app:from_ht_to_pl}
In this appendix, we show how to construct a parameter-learning estimator from a
  hypothesis-testing estimator and relate the success probabilities of the two
  tasks. We define the learning estimator as
\begin{equation}
    \check{\theta}_{\text{PL}} \equiv \theta_{\check{k}} \,.
    \label{eq:definition_pl_from_ht}
\end{equation}
We first evaluate the success probability of this learning estimator under the
  assumption that the ground truth $\theta$ coincides with one of the hypothesis
  values $\theta_k$ in Eq.~\eqref{eq:set_hypothesis_definition}, so that
  $\hat\rho_\theta \in \lbrace \hat{\rho}_{\theta_k} \rbrace_{\theta_k \in
  \Omega}$. In parameter learning, however, the encoded value $\theta \in
  \Theta$ need not coincide exactly with any hypothesis. We account for this
  discrepancy below using continuity of the statistical model.

\begin{figure}[htbp]
  \centering
  \includegraphics[width=0.15\textwidth]{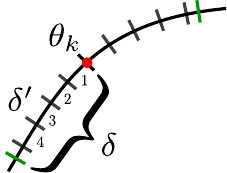}
  \caption{\justifying The red point is the ground-truth hypothesis $\theta_k$,
    while the dark gray ticks are the hypotheses in $\Omega$. An outcome
    $\check{k}$ is accepted whenever $|\check{k}-k|\le a$, with $a$ defined in
    Eq.~\eqref{eq:definition_a}.}
  \label{fig:explanation_ht_to_pl}
\end{figure}

When performing hypothesis testing with $k$ as the ground truth, we will accept
  as successful an estimator $\check{k}$ such that
\begin{equation}
    |\check{k}-k|\le \left\lfloor \frac{\delta}{\delta'}-\frac{1}{2}
      \right\rfloor \equiv a \,,
    \label{eq:definition_a}
\end{equation}
because this condition implies
\begin{equation}
    |\check{\theta}_{\text{PL}} - \theta_{{k}}|
    \le a\delta'
    \le \delta - \frac{\delta'}{2} \,.
    \label{eq:ht_requirement}
\end{equation}
When $a\ge1$, this condition accepts neighboring labels in addition to the true
  label; when $a=0$, it reduces to exact label recovery. This relation is
  illustrated in Fig.~\ref{fig:explanation_ht_to_pl}.
We conclude that the success probability of the learning estimator
  $\check{\theta}_\text{PL}$ is at least as large as the hypothesis-testing
  probability associated with this band-success condition:
\begin{equation}
    \mathsf{P}_{\text{PL}} (\theta_{k}, \delta; \mathcal{M}') \ge
      \mathsf{P}_{\text{Band}} (k; \mathcal{M}) \equiv
      \sum_{\check{k}\,:\,|\check{k}-k| \le a} \trsquare{\hat{\rho}_{\theta_k}
      \hat{M}_{\check{k}}} \,.
    \label{eq:hypotheses_testing_eta_extended}
\end{equation}
Here, $\mathsf{P}_{\text{Band}}(k;\mathcal{M})$ is the band success probability
  for a \emph{fixed} ground-truth hypothesis $k$ and measurement $\mathcal{M}$,
  i.e., the probability that the returned estimator $\check{k}$ falls within the
  tolerance band $|\check{k}-k|\le a$. It is not averaged over a prior; the
  prior-averaged version, denoted
  $\bar{\mathsf{P}}_{\text{Band}}(\mu;\mathcal{M})$, is introduced in
  Eq.~\eqref{eq:band_success_probability}, with the overbar marking the prior
  average. In this expression, $\hat{M}_{\check{k}}$ denotes the POVM elements
  of the measurement $\mathcal{M}$ implementing hypothesis testing, while
  $\mathcal{M}'$ denotes the learning POVM obtained from $\mathcal{M}$ by the
  trivial classical post-processing in Eq.~\eqref{eq:definition_pl_from_ht}.
  Nevertheless, we retain the distinction for clarity.

We now generalize this expression to an arbitrary $\theta \in \Theta$, rather
  than only $\theta_k \in \Omega$. By the $\delta'$-net property of $\Omega$,
  for every value $\theta \in \Theta$ there is a hypothesis within distance
  $\delta'/2$. We denote this unique selected hypothesis by
  $\theta_{k(\theta)}$, where
\begin{equation}
    k(\theta)
    \equiv
    \min\left\lbrace
    K(\delta'),
    1+\left\lfloor
    \frac{\theta-\theta_{\min}}{\delta'}
    \right\rfloor
    \right\rbrace.
    \label{eq:nearest_grid_tie_rule}
\end{equation}
The triangle inequality gives
\begin{equation}
\begin{aligned}
    |\check{\theta}_{\text{PL}} - \theta | &\le |\theta_{\check{k}} -
      \theta_{k(\theta)}| + | \theta_{k(\theta)} - \theta | \\
    &\le \left(\delta-\frac{\delta'}{2}\right) + \frac{\delta'}{2} \\
    &= \delta \,.
\end{aligned}
\end{equation}
The first inequality is the triangle inequality; we then used
  Eq.~\eqref{eq:ht_requirement} and the definition of $\theta_{k(\theta)}$.
  Written in terms of $k(\theta)$, parameter learning succeeds when the
  estimator $\check{k}$ returned by the hypothesis-testing procedure is close to
  $k(\theta)$, i.e., $|\theta_{\check{k}} - \theta_{k(\theta)}| \le \delta -
  \delta'/2$. However, the physical hypothesis-testing procedure acts on the
  actual state $\hat{\rho}_\theta$, not on the nearby grid state
  $\hat{\rho}_{\theta_{k(\theta)}}$. Hence, the hypothesis-testing procedure
  must produce a good estimator $\check{k}$ for $k(\theta)$ when acting on
  $\hat{\rho}_\theta$, which gives the lower bound
\begin{equation}
\begin{aligned}
    \mathsf{P}_{\text{PL}} (\theta, \delta; \mathcal{M}') &\ge
      \sum_{\check{k}\,:\,|\check{k}-k(\theta)| \le a}
      \trsquare{\hat{\rho}_{\theta} \hat{M}_{\check{k}}} \,.
\end{aligned}
\label{eq:hypotheses_testing_eta_extended_k_theta}
\end{equation}
We now rewrite this expression in terms of $\hat{\rho}_{\theta_{k(\theta)}}$.
  Define the success probability of the hypothesis-testing task executed on the
  state encoded with $\theta_{k(\theta)}$ as
\begin{equation}
     \mathsf{P}_{\text{Band}} (k(\theta); \mathcal{M}) \equiv
       \sum_{\check{k}\,:\,|\check{k}-k(\theta)| \le a}
       \trsquare{\hat{\rho}_{\theta_{k(\theta)}} \hat{M}_{\check{k}}} \,.
\end{equation}
By computing the difference between these two probabilities, we get
\begin{equation}
\begin{aligned}
    \bigg| \mathsf{P}_{\text{Band}} & (k(\theta); \mathcal{M}) -
      \sum_{\check{k}\,:\,|\check{k}-k(\theta)| \le a}
      \trsquare{\hat{\rho}_{\theta} \hat{M}_{\check{k}}} \bigg| \\
    &= \left| \trsquare{\left( \hat{\rho}_{\theta_{k(\theta)}} -
      \hat{\rho}_{\theta} \right) \hat{P}} \right| \\
    &\le \frac{1}{2} \| \hat{\rho}_{\theta} - \hat{\rho}_{\theta_{k(\theta)}}
      \|_1
    = D \left( \hat\rho_{\theta}, \hat\rho_{\theta_{k(\theta)}} \right) \,.
\end{aligned}
\end{equation}
From this continuity result we deduce
\begin{equation}
    \mathsf{P}_{\text{PL}} (\theta, \delta; \mathcal{M}')
    \ge \mathsf{P}_{\text{Band}} (k(\theta); \mathcal{M})
    - D \left( \hat\rho_{\theta}, \hat\rho_{\theta_{k(\theta)}} \right) \,.
    \label{eq:trace_distance_link}
\end{equation}
For the uniform prior, we average this pointwise inequality. The Fuchs--van de
  Graaf inequality $D(\hat\rho,\hat\sigma) \le
  \sqrt{1-F^2(\hat\rho,\hat\sigma)}$ then gives
\begin{align}
    \bar{\mathsf{P}}_{\text{PL}} &(\mu_U,\delta;\mathcal{M}') \\
    &\ge \frac{1}{|\Theta|}
    \sum_{k=1}^{K(\delta')}
    \int_{\theta_k-\delta'/2}^{\theta_k+\delta'/2}
    \dd\theta\,
    \left[
    \mathsf{P}_{\text{Band}}(k;\mathcal{M})
    -D\!\left(\hat\rho_\theta,\hat\rho_{\theta_k}\right)
    \right]
    \notag\\
    &\ge \frac{1}{K(\delta')}
    \sum_{k=1}^{K(\delta')}
    \mathsf{P}_{\text{Band}}(k;\mathcal{M})
    \notag\\
    &\quad -\frac{1}{|\Theta|}
    \sum_{k=1}^{K(\delta')}
    \int_{\theta_k-\delta'/2}^{\theta_k+\delta'/2}
    \dd\theta\,
    \sqrt{1-F^2\!\left(\hat\rho_\theta,\hat\rho_{\theta_k}\right)} \,.
    \label{eq:fidelity_link_averaged_uniform}
\end{align}
Here we used $|\Theta|=K(\delta')\delta'$ and the fact that the index
  $k(\theta)$ is stepwise constant on the partition of $\Theta$ with step
  $\delta'$ identified by the hypotheses. Optimizing the left-hand side over all
  learning POVMs and the discrete term over the hypothesis-testing POVM yields
\begin{equation}
    \begin{aligned}
    \bar\eta_{\text{PL}}(\mu_U,\delta)
    &\ge \frac{1}{K(\delta')}
    \sup_{\mathcal M}
    \sum_{k=1}^{K(\delta')}
    \mathsf P_{\text{Band}}(k;\mathcal M)\\
    &\quad-\frac{1}{|\Theta|}
    \sum_{k=1}^{K(\delta')}
    \int_{\theta_k-\delta'/2}^{\theta_k+\delta'/2}
    \dd\theta\,
    \sqrt{1-F^2\!\left(\hat\rho_\theta,\hat\rho_{\theta_k}\right)} \,.
    \end{aligned}
    \label{eq:lower_bound_eta_bayesian_uniform_fidelity}
\end{equation}
Substituting the definition of the band success probability proves
  Eq.~\eqref{eq:theorem_statement_testing_to_learning_bayesian}; optimizing over
  the admissible grid spacings proves
  Eq.~\eqref{eq:theorem_statement_testing_to_learning_bayesian_optimized}. The
  reference spacing in Eq.~\eqref{eq:delta_prime_bayesian_convenient} gives one
  admissible choice within this optimization.

For the minimax result and for a simpler arbitrary-prior Bayesian corollary, we
  use the $L$-Lipschitz condition
\begin{equation}
    D \left( \hat\rho_{\theta}, \hat\rho_{\theta_{k(\theta)}} \right)
    \le L |\theta - \theta_{k(\theta)}|
    \le L \delta'/2 \,.
\end{equation}
Combining this bound with Eq.~\eqref{eq:trace_distance_link} gives
\begin{equation}
    \mathsf{P}_{\text{PL}} (\theta, \delta; \mathcal{M}')
    \ge \mathsf{P}_{\text{Band}} (k(\theta); \mathcal{M}) - L\delta'/2 \,.
    \label{eq:lipschitz_link}
\end{equation}
Let $\mu$ be any probability measure on $\Theta$. Averaging the success
  probabilities on the left and right in Eq.~\eqref{eq:lipschitz_link} over
  $\mu$ gives
\begin{multline}
    \int_{\theta_{\min}}^{\theta_{\max}} \dd \mu (\theta) \,
      \mathsf{P}_{\text{PL}} (\theta, \delta; \mathcal{M}') \\ \ge
      \int_{\theta_{\min}}^{\theta_{\max}} \dd \mu (\theta) \,
      \mathsf{P}_{\text{Band}} (k(\theta); \mathcal{M})  - L\delta'/2 \,.
    \label{eq:lipschitz_link_averaged}
\end{multline}
The index $k(\theta)$ is stepwise constant on the partition of $\Theta$
  identified by the hypotheses. We define the induced discrete prior by
  $\mu_k\equiv\mu(\lbrace \theta\in\Theta : k(\theta)=k \rbrace)$.

Consequently, $\mu_k\ge0$ and $\sum_k\mu_k=1$ for every probability measure
  $\mu$, which gives us
\begin{equation}
    \int_{\theta_{\min}}^{\theta_{\max}}
    \dd\mu(\theta)\,
    \mathsf{P}_{\text{Band}}(k(\theta);\mathcal{M})
    =
    \sum_{k=1}^{K(\delta')}
    \mu_k\,
    \mathsf{P}_{\text{Band}}(k;\mathcal{M}).
    \label{eq:band_average_discrete_prior}
\end{equation}
Applying Eq.~\eqref{eq:band_average_discrete_prior} to
  Eq.~\eqref{eq:lipschitz_link_averaged} gives
\begin{equation}
    \begin{aligned}
    \int_{\theta_{\min}}^{\theta_{\max}} \dd \mu (\theta) \,
      \mathsf{P}_{\text{PL}} (\theta, \delta; \mathcal{M}')
    &\ge \sum_{k=1}^{K(\delta')} \mu_k \, \mathsf{P}_{\text{Band}} (k;
      \mathcal{M})
    \\
    &\quad - L \delta'/2 \,.
    \end{aligned}
\end{equation}
On the right-hand side, the summation is now over the index $k$, which does not
  depend on $\theta$. The left-hand side is the Bayesian learning success
  probability $\bar{\mathsf{P}}_{\text{PL}} (\mu, \delta; \mathcal{M}')$:
\begin{equation}
    \bar{\mathsf{P}}_{\text{PL}} (\mu, \delta; \mathcal{M}') \ge
      \sum_{k=1}^{K(\delta')} \mu_k \, \mathsf{P}_{\text{Band}} (k; \mathcal{M})
      - L \delta'/2 \,.
\end{equation}
We now optimize the Bayesian learning success probability over the full set of
  POVMs $\lbrace \mathcal{M}'' \rbrace$, which gives
\begin{equation}
\begin{aligned}
    \bar{\eta}_{\text{PL}} (\mu, \delta) &= \sup_{\mathcal{M}''}
      \bar{\mathsf{P}}_{\text{PL}} (\mu, \delta; \mathcal{M}'') \ge
      \bar{\mathsf{P}}_{\text{PL}} (\mu, \delta; \mathcal{M}') \\
    &\ge \sum_{k=1}^{K(\delta')} \mu_k \, \mathsf{P}_{\text{Band}} (k;
      \mathcal{M}) - L \delta'/2 \,.
\end{aligned}
\end{equation}
At this point, we are also free to optimize over the measurement $\mathcal{M}$
  in the hypothesis-testing task:
\begin{equation}
    \bar{\eta}_{\text{PL}} (\mu, \delta) \ge \sup_{\mathcal{M}}
      \sum_{k=1}^{K(\delta')} \mu_k \, \mathsf{P}_{\text{Band}} (k; \mathcal{M})
      - L \delta'/2 \,.
    \label{eq:lower_bound_eta_bayesian}
\end{equation}
We take the worst-case prior and apply the quantum minimax theorem to the
  learning part to obtain
\begin{equation}
    {\eta}_{\text{PL}} (\delta) \ge \inf_\mu \sup_{\mathcal{M}}
      \sum_{k=1}^{K(\delta')} \mu_k \, \mathsf{P}_{\text{Band}} (k; \mathcal{M})
      - L \delta'/2 \,.
    \label{eq:lower_bound_eta_minimax}
\end{equation}
Combining this result with Eq.~\eqref{eq:hypotheses_testing_eta_extended}
  completes the proof and yields
\begin{equation}
    \begin{split}
        \eta_{\text{PL}} (\delta) &\ge \inf_{\mu} \sup_{\mathcal{M}}
          \sum_{k=1}^{K(\delta')} \mu_k \sum_{\check{k}\,:\,|\check{k}-k| \le a}
          \trsquare{\hat{\rho}_{k} \hat{M}_{\check{k}}} - L \delta'/2 \\
        &= \inf_{\mu} \sup_{\mathcal{M}} \sum_{k=1}^{K(\delta')}
          \sum_{\check{k}\,:\,|\check{k}-k|\le a} \mu_k \trsquare{\hat{\rho}_{k}
          \hat{M}_{\check{k}}} - L\delta'/2 \,,
    \end{split}
    \label{eq:lower_bound_eta_learning}
\end{equation}
where the summation is now over both indices $k$ and $\check{k}$. This is
  Eq.~\eqref{eq:theorem_statement_testing_to_learning}. It remains only to
  ensure that the Lipschitz correction does not make the bound trivial. If
  $L\delta'/2$ were too large, the right-hand side of the above equation could
  become negative and hence uninformative. The condition $\delta' \le 0.1/L$
  ensures that $L\delta'/2 \le 0.05$. Therefore, a multi-hypothesis-testing
  procedure with guaranteed success probability at least $0.80$ gives a
  corresponding learning procedure with success probability at least $0.75$.
  Thus, for any choice of $\delta'\le2\delta$, with the additional condition
  $\delta'\le0.1/L$, a sufficiently accurate hypothesis-testing procedure
  simulates a learning procedure with precision $\delta$ and a similar success
  probability.

\section{Alternative upper bounds}
\label{app:alternative_upper_bounds}
This appendix collects alternative upper bounds on the success probability of
  parameter learning. Some of these bounds are global in nature; others are
  local and therefore not expected to be very tight. We first combine Fano's
  inequality with Holevo's theorem at the level of hypothesis testing and obtain
  parameter-learning bounds in terms of the average-state entropy, the entropy
  of the Gram matrix for pure states, the quantum Fisher information, and
  quantum relative entropies between hypotheses. We then derive a fidelity-based
  bound from a result by Montanaro, present the Holevo--Curlander upper bound,
  and compare selected bounds in a simplified scenario in which all pairs of
  hypothesis states have the same overlap or fidelity.

\subsection{Fano's inequality and Holevo's theorem}
\label{subsubsec:fano_holevo_theorem}
Fano's inequality bounds the probability of correctly decoding a classical
  random variable from the mutual information between the true value and the
  estimate~\cite{fanoTransmissionInformationStatistical1961}. For a fixed POVM
  $\mathcal{M}$ and prior $\mu$, the hypothesis-testing experiment induces the
  joint probability distribution
\begin{equation}
    p_\mu(k,\check{k}|\mathcal{M}) \equiv \mu_k p(\check{k}|k,\mathcal{M}) \,,
    \label{eq:joint_distribution_ht}
\end{equation}
with marginal $p_\mu(\check{k}|\mathcal{M})$ defined in
  Eq.~\eqref{eq:prob_estimator_ht}. We denote the corresponding
  measurement-dependent mutual information by
\begin{equation}
    I_\mu(k;\check{k}\,|\,\mathcal{M}) \equiv \sum_{k,\check{k}=1}^K
      p_\mu(k,\check{k}|\mathcal{M}) \log_2
      \frac{p(\check{k}|k,\mathcal{M})}{p_\mu(\check{k}|\mathcal{M})} \,.
    \label{eq:mutual_information_ht}
\end{equation}
For the uniform prior $\mu_U$, with $\mu_k^U=1/K$, Fano's inequality implies
\begin{equation}
    \bar{\mathsf{P}}_{\text{HT}}(\mu_U;\mathcal{M}) \le \frac{1 +
      I_{\mu_U}(k;\check{k}\,|\,\mathcal{M})}{\log_2 K} \,.
    \label{eq:fano_ht_povm}
\end{equation}
Since $\eta_{\text{HT}} \le \bar{\eta}_{\text{HT}}(\mu_U)$, optimizing
  Eq.~\eqref{eq:fano_ht_povm} over all POVMs gives
\begin{equation}
    \eta_{\text{HT}} \le \frac{1 + \sup_{\mathcal{M}}
      I_{\mu_U}(k;\check{k}\,|\,\mathcal{M})}{\log_2 K} \,.
    \label{eq:fano_ht_minimax}
\end{equation}
This is the standard communication interpretation of hypothesis testing: the
  index $k$ is a classical message encoded into the quantum state
  $\hat{\rho}_k$, and the POVM followed by the estimator is the decoding channel
  from $k$ to $\check{k}$. If the encoding and measurement convey substantially
  less than $\log_2 K$ bits about the message, then the probability of correctly
  decoding $k$ must be small.

Holevo's theorem bounds the accessible information in this communication
  problem. For the ensemble $\lbrace(\mu_k,\hat{\rho}_k)\rbrace_{k=1}^K$ with
  average state $\bar{\rho}=\sum_{k=1}^K \mu_k\hat{\rho}_k$, one
  has~\cite{Hol73}
\begin{equation}
    \begin{aligned}
    \sup_{\mathcal{M}} I_\mu(k;\check{k}\,|\,\mathcal{M})
    &\le \chi\left(\lbrace(\mu_k,\hat{\rho}_k)\rbrace_{k=1}^K\right) \\
    &\equiv S(\bar{\rho}) - \sum_{k=1}^K \mu_k S(\hat{\rho}_k) \,,
    \end{aligned}
    \label{eq:holevo_information_bound}
\end{equation}
where $S(\cdot)$ is the von Neumann entropy measured in bits. Combining
  Eqs.~\eqref{eq:fano_ht_minimax} and~\eqref{eq:holevo_information_bound} yields
  the general mixed-state hypothesis-testing bound
\begin{equation}
    \eta_{\text{HT}} \le \frac{1 +
      \chi\left(\lbrace(1/K,\hat{\rho}_k)\rbrace_{k=1}^K\right)}{\log_2 K} \,.
    \label{eq:fano_holevo_ht}
\end{equation}
If all states in the ensemble coincide, then $\chi=0$, reflecting the fact that
  the measurement outcome carries no information about $k$. More generally, the
  Holevo information upper-bounds the accessible information about the
  hypothesis label.

\subsection{Average-state and pure-state Gram-matrix bounds}
\label{subsubsec:fano_entropy_gram_matrix}
We now combine the Fano hypothesis-testing bound with the reduction from
  parameter learning to hypothesis testing. For a learning precision $\delta$,
  we discretize the parameter interval using the limiting spacing
  $\delta'=2\delta$. The induced hypothesis-testing task contains $K(\delta')$
  states $\hat{\rho}_k\equiv\hat{\rho}_{\theta_k}$, with $\theta_k\in\Omega$ as
  defined in Eq.~\eqref{eq:set_hypothesis_definition}. Since
  Eq.~\eqref{eq:theorem_statement_learning_to_testing} gives
  $\eta_{\text{PL}}(\delta)\le\eta_{\text{HT}}$, any upper bound on the success
  probability of this multi-hypothesis-testing task with a uniform prior is also
  an upper bound on the minimax learning success probability.

For general mixed states, Eqs.~\eqref{eq:fano_holevo_ht}
  and~\eqref{eq:holevo_information_bound}, together with $\chi\le
  S(\bar{\rho})$, give
\begin{equation}
    \eta_{\text{PL}} (\delta) \le \frac{1 + S(\bar{\rho})}{\log_2 K(\delta')}
      \,.
    \label{eq:upper_bound_entropy_pure_states_1}
\end{equation}
If the states are pure, their individual entropies vanish. Moreover, the nonzero
  eigenvalues of the average state $\bar{\rho}$ coincide with those of the Gram
  matrix defined in Eq.~\eqref{eq:gram_matrix}, as shown in
  Appendix~\ref{app:success_pgm}. Denoting these eigenvalues by $\lbrace
  \lambda_k \rbrace_{k=1}^{K(\delta')}$, we obtain
\begin{align}
    \eta_{\text{PL}} (\delta) &\le \frac{1 + S(\boldsymbol{G})}{\log_2
      K(\delta')} \label{eq:upper_bound_entropy_pure_states_2} \\
    &=  \frac{1 - \sum_{k=1}^{K(\delta')} \lambda_k \log_2 \lambda_k }{\log_2
      K(\delta')} \label{eq:upper_bound_entropy_pure_states_3} \,.
\end{align}
The entropy bound becomes tighter as the average ensemble state approaches a
  pure state and can outperform Eq.~\eqref{eq:upper_bound_eta_pl_pgm} when
  $K(\delta')$ is large. A simplified comparison with the PGM and Montanaro
  bounds is given in Sec.~\ref{subsec:comparison_upper_bounds}.

\subsection{Fisher-information bound on the Holevo information}
\label{subsubsec:fano_quantum_fisher_info}
In~\cite{goreckiMutualInformationBounded2025a}, the authors established a
  connection between the mutual information and the Fisher information of a
  measurement, which can be extended to a relation between the Holevo
  information and the quantum Fisher information. Consider the hypothesis
  ensemble $\lbrace \hat{\rho}_{\theta_k} \rbrace_{k=1}^{K(\delta')}$. This
  ensemble can be embedded within a continuous model $\lbrace \hat{\rho}_\theta
  \rbrace_{\theta \in \Theta}$ by defining a discrete prior consisting of Dirac
  delta functions centered at the positions $\lbrace \theta_k
  \rbrace_{k=1}^{K(\delta')}$. The support of this prior is entirely contained
  within the interval $[\theta_{\min}, \theta_{\max}]$, allowing us to apply
  Eq.~(15) from~\cite{goreckiMutualInformationBounded2025a}, which yields
\begin{equation}
    \chi\left( \lbrace \hat{\rho}_k \rbrace_{k=1}^{K(\delta')} \right) \le
      \log_2 \left( 1 + \frac{1}{2} \int_{\theta_{\min}}^{\theta_{\max}}
      \sqrt{\mathcal{F}(\theta)} \, \dd\theta \right) \,,
\end{equation}
where $\mathcal{F}(\theta)$ is the quantum Fisher information of the state
  $\hat{\rho}_\theta$ with respect to $\theta$. This establishes a link between
  the quantum Fisher information and the Holevo information. The right-hand side
  of this expression is, moreover, not bounded by the dimensionality of the
  Hilbert space. Combining this result with the relationship between the
  hypothesis-testing error and the parameter-learning error, as expressed in
  Eq.~\eqref{eq:theorem_statement_learning_to_testing}, we obtain
\begin{equation}
    \eta_{\text{PL}} (\delta) \le \frac{1 + \log_2 \left( 1 + \frac{1}{2}
      \int_{\theta_{\min}}^{\theta_{\max}} \sqrt{\mathcal{F}(\theta)} \,
      \dd\theta \right)}{\log_2 K(\delta')} \,,
    \label{eq:upper_bound_eta_pl_fano_qfi}
\end{equation}
where $K(\delta')$ is the number of hypotheses. While this bound is quite loose,
  even compared with Eq.~\eqref{eq:pairwise_success_probability}, and will not
  yield an optimal scaling, it could be useful because it requires evaluating
  only the quantum Fisher information of the model, and it is therefore a local
  bound. Furthermore, note that for a given set of discrete hypotheses $\lbrace
  \hat{\rho}_k \rbrace_{k=1}^{K(\delta')}$, we can compute $\mathcal{F}(\theta)$
  using any continuous parametrization $\lbrace \hat{\rho}_\theta'
  \rbrace_{\theta \in \Theta}$ that passes through these discrete states, and
  subsequently minimize the bound over the choice of parametrization, without
  necessarily having to use the original parametrization of the state of the
  probe $\hat\rho_\theta$.

\subsection{Relative-entropy bound}
\label{subsubsec:fano_pairwise_entropy}
For a given measurement $\mathcal{M}$ and the uniform prior $\mu_U$, with
  $\mu_k^U=1/K(\delta')$, the mutual information $I_{\mu_U}(k, \check{k})$ is
  bounded from above by the average Kullback--Leibler (KL) divergence over
  ordered pairs of outcome distributions conditioned on the ground truth:
\begin{equation}
    \begin{aligned}
    I_{\mu_U}(k, \check{k})
    &\le \frac{1}{K(\delta')^2} \sum_{k, k'=1}^{K(\delta')}
    \\
    &\qquad{}\times D_{\text{KL}} \left[ p(\check{k} \mid k, \mathcal{M})
      \parallel p(\check{k} \mid k', \mathcal{M}) \right] \,,
    \end{aligned}
    \label{eq:classical_kl_upper_bound_mutual_information}
\end{equation}
where the probability $p(\check{k} \mid k, \mathcal{M})$ is defined in
  Eq.~\eqref{eq:born_rule_ht}. We now connect these KL divergences to the
  quantum ensemble. Introduce the measurement channel
  $\mathcal{E}_{\mathcal{M}}$, which performs the POVM measurement on the state
  and records the outcome in an auxiliary system:
\begin{equation}
    \mathcal{E}_{\mathcal{M}} (\hat{\rho}) = \sum_{\check{k}=1}^{K(\delta')}
      \hat{\tilde{\rho}}_{\check{k}} \otimes \ket{\check{k}} \! \bra{\check{k}}
      \,,
    \label{eq:measurement_channel_definition}
\end{equation}
where $\hat{\tilde{\rho}}_{\check{k}}$ is the unnormalized post-measurement
  state, and the states $\lbrace \ket{\check{k}}
  \rbrace_{\check{k}=1}^{K(\delta')}$ form an orthonormal basis. Let $S$ denote
  the probe system and $A$ denote the ancilla that records the measurement
  outcome. Tracing out the probe yields a state that exactly encodes the
  measurement-outcome distribution:
\begin{equation}
    \trsys{\mathcal{E}_{\mathcal{M}} (\hat{\rho})} =
      \sum_{\check{k}=1}^{K(\delta')} \trsquare{\hat{\rho} \hat{M}_{\check{k}}}
      \ket{\check{k}} \! \bra{\check{k}} \,.
    \label{eq:partial_trace_outcome_distribution}
\end{equation}
The quantum relative entropy between these reduced ancilla states is exactly
  equal to the classical KL divergence between the two outcome distributions:
\begin{align}
    D_{\text{KL}}&\left( p(\check{k} \mid k, \mathcal{M}) \parallel p(\check{k}
      \mid k', \mathcal{M}) \right) \nonumber \\
    &= S\left( \trsys{\mathcal{E}_{\mathcal{M}} (\hat{\rho}_k)} \parallel
      \trsys{\mathcal{E}_{\mathcal{M}} (\hat{\rho}_{k'})} \right)
      \label{eq:kl_to_quantum_relative_entropy_step1} \\
    &\le S\left( \mathcal{E}_{\mathcal{M}} (\hat{\rho}_k) \parallel
      \mathcal{E}_{\mathcal{M}} (\hat{\rho}_{k'}) \right)
      \label{eq:kl_to_quantum_relative_entropy_step2} \\
    &\le S\left( \hat{\rho}_k \parallel \hat{\rho}_{k'} \right) \,.
      \label{eq:kl_to_quantum_relative_entropy_step3}
\end{align}
Both inequalities follow from the data-processing inequality: the first arises
  from extending to a larger system (i.e., omitting the partial trace over $S$),
  while the second arises from applying the measurement map
  $\mathcal{E}_{\mathcal{M}}$. Substituting this inequality into
  Eq.~\eqref{eq:classical_kl_upper_bound_mutual_information} yields
\begin{equation}
    I_{\mu_U}(k, \check{k}) \le \frac{1}{K(\delta')^2} \sum_{k,
      k'=1}^{K(\delta')} S\left( \hat{\rho}_k \parallel \hat{\rho}_{k'} \right)
      \,,
    \label{eq:mutual_information_quantum_relative_entropy_bound}
\end{equation}
which leads to the following upper bound on the success probability of parameter
  learning:
\begin{equation}
    \eta_{\text{PL}} (\delta) \le \frac{1 + \frac{1}{K(\delta')^2} \sum_{k,
      k'=1}^{K(\delta')} S(\hat{\rho}_k \parallel \hat{\rho}_{k'})}{\log_2
      K(\delta')} \,.
    \label{eq:fano_pairwise_divergence_bound}
\end{equation}
Note that this upper bound is uninformative for ensembles of pure states, since
  the relative entropy between two distinct pure states diverges to infinity,
  thereby rendering the bound trivial.

\begin{table*}[htbp]
\centering
\begin{ruledtabular}
\begin{tabular}{cccc}
\textbf{Bound origin and reference} & \textbf{Equation} & \textbf{Section} &
  \textbf{Requires} \\
\hline
Montanaro's upper bound on hypothesis
  testing~\cite{montanaroLowerBoundProbability2008} &
  Eq.~\eqref{eq:lower_bound_error_montanaro} &
  Sec.~\ref{subsubsec:montanaros_bound} & $F(\hat{\rho}_\theta,
  \hat{\rho}_{\theta'})$ \\
Fano's inequality~\cite{fanoTransmissionInformationStatistical1961} and quantum
  Fisher information & Eq.~\eqref{eq:upper_bound_eta_pl_fano_qfi} &
  Sec.~\ref{subsubsec:fano_quantum_fisher_info} & $\mathcal{F}(\theta)$ \\
Fano's inequality~\cite{fanoTransmissionInformationStatistical1961} and quantum
  relative entropy & Eq.~\eqref{eq:fano_pairwise_divergence_bound} &
  Sec.~\ref{subsubsec:fano_pairwise_entropy} & $S(\hat{\rho}_\theta \parallel
  \hat{\rho}_{\theta'})$ \\
Fano's inequality~\cite{fanoTransmissionInformationStatistical1961} and
  average-state entropy & Eqs.~\eqref{eq:upper_bound_entropy_pure_states_1}
  and~\eqref{eq:upper_bound_entropy_pure_states_2} &
  Sec.~\ref{subsubsec:fano_entropy_gram_matrix} & Diagonalizing $\bar{\rho}$;
  $\boldsymbol{G}$ for pure states \\
Holevo--Curlander bound~\cite{tysonTwosidedEstimatesMinimumerror2009} &
  Eqs.~\eqref{eq:holevo_curlander_bound}
  and~\eqref{eq:holevo_curlander_bound_pure_gram} &
  Sec.~\ref{subsubsec:holevo_curlander_bound} & Operating on
  $\hat{\rho}_\theta$; $\boldsymbol{G}$ for pure states \\
\end{tabular}
\end{ruledtabular}
\caption{\justifying Summary of alternative upper bounds on the success
  probability for learning. The columns give the origin of each bound, the
  equation and section where it is discussed, and the quantity required to
  evaluate it.}
\label{table:alternative_upper_bounds}
\end{table*}

\subsection{Fidelity bound for mixed states (Montanaro)}
\label{subsubsec:montanaros_bound}
We now discuss a fidelity-based bound valid for mixed states.
  In~\cite{montanaroLowerBoundProbability2008}, the following upper bound on the
  Bayesian success probability of hypothesis testing was derived:
\begin{equation}
    \bar{\eta}_{\text{HT}}(\mu) \le 1 - \sum_{k>k'} \mu_{k} \mu_{k'} F^2
      (\hat{\rho}_k, \hat{\rho}_{k'}) \,.
    \label{eq:lower_bound_error_bayesian}
\end{equation}
Define the squared-fidelity matrix $(\boldsymbol{F}_2)_{k k'}\equiv
  F^2(\hat{\rho}_k, \hat{\rho}_{k'})$ for the set of states corresponding to the
  discrete hypotheses. To find the worst-case prior for hypothesis testing, we
  begin by upper-bounding the minimax hypothesis-testing success probability:
\begin{align}
    \inf_{\mu} \, & \bar{\eta}_{\text{HT}}(\mu) \le 1 - \sup_{\mu} \sum_{k > k'}
      \mu_k \mu_{k'} F^2(\hat{\rho}_k, \hat{\rho}_{k'}) \\
    &= 1 - \sup_{\mu} \frac{1}{2} \left( \sum_{k,k'=1}^{K(\delta')} \mu_k
      \mu_{k'} F^2(\hat{\rho}_k, \hat{\rho}_{k'}) - \sum_{k=1}^{K(\delta')}
      \mu_k^2 \right) \\
    &= 1 - \frac{1}{2} \sup_{\ket{\mu}} \braket{\mu | \boldsymbol{F}_2 - \idmat
      | \mu} \label{eq:lower_bound_error_montanaro} \\
    & \quad \; \text{subject to} \; \sum_{k=1}^{K(\delta')} \mu_k = 1, \; \mu_k
      \geq 0 \,.
\end{align}
In the above equations, we introduced the probability vector $\ket{\mu} \equiv
  (\mu_1, \mu_2, \dots, \mu_{K(\delta')}) \in \mathbb{R}^{K(\delta')}$. This
  problem represents the maximization of the quadratic form $\boldsymbol{F}_2 -
  \idmat$ over the probability simplex $\lbrace \mu \mid \sum_{k=1}^{K(\delta')}
  \mu_k = 1, \mu_k \ge 0 \rbrace$. This is a non-convex quadratic optimization
  on a convex domain, since $\boldsymbol{F}_2 - \idmat$ is indefinite. In
  general, it must be solved numerically, but a valid upper bound can be
  obtained by inserting either the uniform distribution or the probability
  distribution induced by the eigenvector with the largest eigenvalue of
  $\boldsymbol{F}_2$.

\subsection{Upper bound from the Holevo--Curlander inequality}
\label{subsubsec:holevo_curlander_bound}
Combining the Holevo--Curlander upper
  bound~\cite{tysonTwosidedEstimatesMinimumerror2009,
    hadiasharOptimalLowerBounds2024} for the success probability of hypothesis
      testing with Eq.~\eqref{eq:theorem_statement_learning_to_testing} yields
\begin{equation}
    \eta_{\text{PL}} (\delta) \le \inf_\mu
      \trsquare{\sqrt{\sum_{k=1}^{K(\delta')} \mu_k^2 \hat{\rho}_k^2}} \,.
    \label{eq:holevo_curlander_bound}
\end{equation}
For a pure-state ensemble $\hat{\rho}_k=\ket{\psi_k}\!\bra{\psi_k}$, this bound
  can be evaluated directly from the Gram matrix $\boldsymbol{G}$ of
  Eq.~\eqref{eq:gram_matrix}. Let
  $\boldsymbol{M}_\mu\equiv\operatorname{diag}(\mu_1,\ldots,\mu_{K(\delta')})$.
  Since $\hat{\rho}_k^2=\hat{\rho}_k$, the nonzero eigenvalues of
  $\sum_k\mu_k^2\hat{\rho}_k$ coincide with those of
  $\boldsymbol{M}_\mu^{1/2}\boldsymbol{G}\boldsymbol{M}_\mu^{1/2}$. Therefore,
\begin{equation}
    \eta_{\text{PL}} (\delta)
    \le
    \inf_\mu
    \trsquare{\sqrt{\boldsymbol{M}_\mu^{1/2} \boldsymbol{G}
      \boldsymbol{M}_\mu^{1/2}}} \,.
    \label{eq:holevo_curlander_bound_pure_gram}
\end{equation}
For the uniform prior $\mu_U$, this quantity reduces to
  $K(\delta')^{-1/2}\trsquare{\sqrt{\boldsymbol{G}}}$. Thus, although
  Eq.~\eqref{eq:holevo_curlander_bound} generally requires direct manipulation
  of the density matrices, its pure-state specialization depends only on the
  full complex Gram matrix.

\subsection{Comparison between the upper bounds}
\label{subsec:comparison_upper_bounds}
We now compare the upper bounds in Eqs.~\eqref{eq:upper_bound_eta_pl_pgm},
  \eqref{eq:upper_bound_eta_pl_entropy_fano},
  \eqref{eq:lower_bound_error_montanaro},
  and~\eqref{eq:holevo_curlander_bound_pure_gram} for a problem where the
  pairwise overlaps and fidelities of all the hypotheses are the same. We write
  $K\equiv K(\delta')\ge2$, assume a uniform prior, and take every off-diagonal
  overlap or fidelity to have the same real nonnegative value $f$.

\subsubsection{PGM-based upper bound}
First consider Eq.~\eqref{eq:upper_bound_eta_pl_pgm}. For a uniform distribution
  over hypotheses, the Gram matrix is
  $\boldsymbol{G}=K^{-1}[(1-f)\idmat+f\boldsymbol{J}]$, where $\boldsymbol{J}$
  is the all-ones matrix. Under these conditions,
  Eq.~\eqref{eq:upper_bound_eta_pl_pgm} can be evaluated analytically, giving
\begin{equation}
    \begin{aligned}
    \eta_{\text{PL}} (\delta)
    &\le \frac{1}{K} \sqrt{1 + (K-1)f} \\
    &\quad + \left(1 - \frac{1}{K}\right) \sqrt{1-f} \,.
    \end{aligned}
    \label{eq:f_real_overlap_pgm}
\end{equation}
The right-hand side decreases monotonically with $f$. Solving for the value at
  which it equals $3/4$ gives the exact finite-$K$ contour
\begin{equation}
    \eta_{\mathrm{PL}}(\delta)\ge\frac{3}{4}
    \quad\Longrightarrow\quad
    f
    \le
    1-
    \left(
    \frac{3}{4}
    -
    \sqrt{\frac{7}{16(K-1)}}
    \right)^2.
    \label{eq:finite_k_pgm_threshold}
\end{equation}
The right-hand side of Eq.~\eqref{eq:finite_k_pgm_threshold} approaches $7/16$
  only in the limit $K\to\infty$.

\subsubsection{Montanaro's bound on multi-hypothesis testing}
\label{sec:montanaros_bound}
We now consider Eq.~\eqref{eq:lower_bound_error_montanaro}, which we write in
  terms of $\boldsymbol{F}_2 - \idmat = f^2 ( \boldsymbol{J} - \idmat)$:
\begin{equation}
    \eta_{\text{PL}} (\delta) \le 1 - \frac{1}{2} \sup_{\ket{\mu}} \braket{\mu |
      \boldsymbol{F}_2 - \idmat | \mu} \,.
\end{equation}
The constraint $\sum_k \mu_k = 1$ lets us rewrite the quadratic form as $f^2
  \left[ 1 - \sum_k \mu_k^2 \right]$, so that its maximization is equivalent to
  the minimization of $\sum_k \mu_k^2$. By the Cauchy--Schwarz inequality,
  $\sum_k \mu_k^2 \ge 1/K$ for any probability distribution, with equality if
  and only if $\mu$ is uniform. We therefore obtain the exact result
\begin{equation}
    \eta_{\text{PL}} (\delta) \le 1 - \frac{f^2}{2} \left( 1 - \frac{1}{K}
      \right) \,.
    \label{eq:upper_bound_eta_pl_f_exact}
\end{equation}
Solving for the value at which this bound equals $3/4$ gives
\begin{equation}
    \eta_{\mathrm{PL}}(\delta)\ge\frac{3}{4}
    \quad\Longrightarrow\quad
    f
    \le
    \sqrt{\frac{K}{2(K-1)}}.
    \label{eq:finite_k_montanaro_threshold}
\end{equation}
The right-hand side of Eq.~\eqref{eq:finite_k_montanaro_threshold} approaches
  $1/\sqrt{2}$ only in the limit $K\to\infty$.

\subsubsection{Fano single-fidelity formula}
\label{subsubsec:fano_single_fidelity_formula}
For a uniform pure-state ensemble, consider the symmetric Gram-matrix ansatz
  $\boldsymbol{G}=K^{-1}[(1-f)\idmat+f\boldsymbol{J}]$, where $K\equiv
  K(\delta')\ge2$, $f$ is the common real nonnegative off-diagonal overlap, and
  $\boldsymbol{J}$ is the all-ones matrix. Its eigenvalues are
\begin{equation}
    \lambda_+ = \frac{1+(K-1)f}{K} \, ,
    \qquad
    \lambda_- = \frac{1-f}{K} \, ,
\end{equation}
where $\lambda_-$ has multiplicity $K-1$. The Gram-matrix entropy is exactly
\begin{equation}
    \begin{aligned}
    S(\boldsymbol{G})
    &=
    -\lambda_+\log_2\lambda_+
    -(K-1)\lambda_-\log_2\lambda_- \\
    &=
    h_2(\lambda_+)
    +(1-\lambda_+)\log_2(K-1),
    \end{aligned}
    \label{eq:exact_single_fidelity_gram_entropy}
\end{equation}
where $h_2(p)\equiv-p\log_2p-(1-p)\log_2(1-p)$ is the binary entropy.

Substituting Eq.~\eqref{eq:exact_single_fidelity_gram_entropy} into
  Eq.~\eqref{eq:upper_bound_entropy_pure_states_2} gives the exact finite-$K$
  formula
\begin{equation}
    \eta_{\text{PL}}(\delta)
    \le
    \frac{
    1+h_2\!\left(\frac{1+(K-1)f}{K}\right)
    +\frac{(K-1)(1-f)}{K}\log_2(K-1)
    }{
    \log_2 K} \; .
    \label{eq:upper_bound_eta_pl_entropy_fano}
\end{equation}

\subsubsection{Holevo--Curlander single-fidelity formula}
\label{subsubsec:holevo_curlander_single_fidelity_formula}
We finally consider the pure-state Holevo--Curlander bound in
  Eq.~\eqref{eq:holevo_curlander_bound_pure_gram}.

For the uniform prior with all hypotheses having the same real overlap, we use
  the same expression for the Gram matrix used for the PGM, i.e.,
  $\boldsymbol{G}=K^{-1}[(1-f)\idmat+f\boldsymbol{J}]$. The matrix inside the
  square root in Eq.~\eqref{eq:holevo_curlander_bound_pure_gram} becomes
\begin{equation}
    \boldsymbol{M}_{\mu_U}^{1/2}
    \boldsymbol{G}
    \boldsymbol{M}_{\mu_U}^{1/2}
    =
    \frac{1}{K^2}
    \left[
    (1-f)\idmat+f\boldsymbol{J}
    \right] \,.
\end{equation}
Its eigenvalues are $[1+(K-1)f]/K^2$ and $(1-f)/K^2$, where the latter has
  multiplicity $K-1$. Equation~\eqref{eq:holevo_curlander_bound_pure_gram}
  therefore gives the exact single-fidelity formula
\begin{equation}
    \begin{aligned}
    \eta_{\mathrm{PL}}(\delta)
    &\le
    \frac{1}{K}\sqrt{1+(K-1)f} \\
    &\quad
    +\left(1-\frac{1}{K}\right)\sqrt{1-f} \,.
    \end{aligned}
    \label{eq:holevo_curlander_single_fidelity}
\end{equation}
This expression coincides with the PGM-based upper bound in
  Eq.~\eqref{eq:f_real_overlap_pgm}. Hence, within the symmetric single-fidelity
  ansatz, the Holevo--Curlander and PGM bounds have the same $3/4$ contour,
  although the two bounds need not coincide for general ensembles.

\subsubsection{Comparison between the bounds}
\begin{figure}[htbp]
  \centering
  \includegraphics[width=0.45\textwidth]{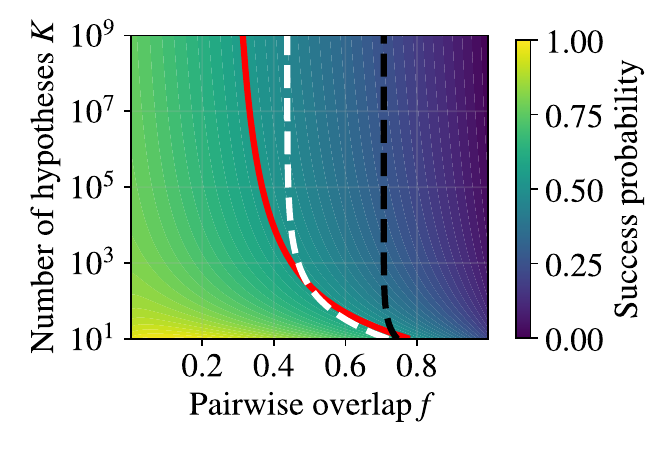}
  \caption{\justifying Exact finite-$K$ Fano upper bound on the success
    probability of the learning task in
    Eq.~\eqref{eq:upper_bound_eta_pl_entropy_fano} as a function of the number
    of hypotheses $K(\delta')$ and the common off-diagonal fidelity $f$. The red
    line is the Fano threshold. The dashed white and black curves are the exact
    PGM and Montanaro thresholds in Eqs.~\eqref{eq:finite_k_pgm_threshold}
    and~\eqref{eq:finite_k_montanaro_threshold}, respectively.}
  \label{fig:weak_dependence}
\end{figure}

We conclude the appendix with a graphical comparison between the three distinct
  formulas. Figure~\ref{fig:weak_dependence} shows the Fano upper bound of
  Eq.~\eqref{eq:upper_bound_eta_pl_entropy_fano}, together with the PGM,
  Montanaro, and Fano threshold contours. The PGM contour also represents the
  Holevo--Curlander bound in Eq.~\eqref{eq:holevo_curlander_single_fidelity}. No
  single formula is uniformly the tightest. For small numbers of hypotheses,
  Eq.~\eqref{eq:f_real_overlap_pgm} can be smaller than the entropy-based bound,
  whereas Fano's bound can become smaller as the number of hypotheses increases.

\section{Exact coarse-grained reformulation of the success probability with
  tolerance}
\label{app:coarse_grained_reformulation}
In this appendix, we show that the discrete success probability with tolerance
  equals a known positive multiple of the success probability of an ordinary
  multi-hypothesis-testing problem for a coarse-grained ensemble. This
  construction uses overlapping moving windows of hypotheses and is distinct
  from the disjoint-block relaxation employed in
  Sec.~\ref{subsubsec:achievable_success_mixed}. We rewrite the success
  probability from Eq.~\eqref{eq:theorem_statement_testing_to_learning} as
\begin{align}
    \bar{\mathsf{P}}_{\text{Band}}(\mu; \mathcal{M}) &= \sum_{k=1}^{K(\delta')}
      \mu_k \sum_{j : |k - j| \le a} \trsquare{\hat{\rho}_{k} \hat{M}_j}
      \nonumber \\
    &= \sum_{j=1}^{K(\delta')} \sum_{k : |k - j| \le a} \mu_k
      \trsquare{\hat{\rho}_{k} \hat{M}_j} \nonumber \\
    &= \sum_{j=1}^{K(\delta')} \trsquare{\left( \sum_{k : |k - j| \le a} \mu_k
      \hat{\rho}_k \right) \hat{M}_j} \,.
\end{align}
We now define nonnegative weights $\mu_j'$ and, whenever $\mu_j'>0$, normalized
  states $\hat{\sigma}_j$ such that
\begin{equation}
    \mu_j' \hat{\sigma}_j \equiv \sum_{k : |k - j| \le a} \mu_k \hat{\rho}_k \,.
    \label{eq:modified_ensemble_definition}
\end{equation}

We define the normalizing constant $Z \equiv \sum_{j=1}^{K(\delta')} \mu_j'$,
  which allows us to rewrite the band success probability as
\begin{equation}
\begin{split}
    \bar{\mathsf{P}}_{\text{Band}}(\mu; \mathcal{M}) &= Z
      \sum_{j=1}^{K(\delta')} \trsquare{\frac{1}{Z} \left( \sum_{k : |k - j| \le
      a} \mu_k \hat{\rho}_k \right) \hat{M}_j} \\
    &= Z \bar{\mathsf{P}}_{\text{HT}}(\nu; \mathcal{M}) \,.
\end{split}
\end{equation}
This connects the success probability with tolerance to the ordinary
  hypothesis-testing success probability for the ensemble $\lbrace (\nu_j,
  \hat{\sigma}_j) \rbrace_{j=1}^{K(\delta')}$, with $\nu_j \equiv \mu_j'/Z$. The
  identity is exact and introduces no pairwise or measurement relaxation.
  Evaluating the resulting optimal success probability nevertheless requires
  solving the ordinary multi-hypothesis measurement optimization for the
  coarse-grained states.

\section{Auxiliary calculations for single-shot learning of a fully connected
  Ising Hamiltonian}
\label{app:ising_qfi_lipschitz}
This appendix collects the auxiliary results used for the single-shot
  Ising-learning strategy of Sec.~\ref{subsec:hamiltonian_learning_ising}: the
  effective dimension of the output-state family, the exact overlaps entering
  the Gram matrix, the output-state quantum Fisher information (QFI) for the
  prescribed input, and the Lipschitz constant required for the achievability
  analysis.

\subsection{Effective dimension and exact overlaps}
\label{app:ising_effective_dimension_overlap}
Let us define the $J$-independent generator
\begin{equation}
    \hat{H}_0
    = 2\hat{S}_z^2
    = \frac{N}{2}\hat{\id}
    + \sum_{1\le i<j\le N}\hat{Z}_i\hat{Z}_j \,.
    \label{eq:ising_generator}
\end{equation}
The operator $\hat{H}_0$ differs from the $J$-independent operator multiplying
  $J$ in Eq.~\eqref{eq:ising_hamiltonian_main} only by the identity term
  $N\hat{\id}/2$. This term contributes a $J$-dependent global phase and
  therefore does not affect the encoded state of the probe. For the prescribed
  probe state $\ket{+}^{\otimes N}$, the encoded state is
\begin{equation}
    \ket{\psi(J)} = \mathrm{e}^{-\mathrm{i}J t \hat{H}_0} \ket{+}^{\otimes N}
      \,.
    \label{eq:ising_initial_state}
\end{equation}

We now compute the exact overlap between two states with parameters $J_1$ and
  $J_2$. Defining $\Delta J \equiv J_1-J_2$, one has
\begin{equation}
\begin{aligned}
    \braket{\psi(J_1) | \psi(J_2)} &= \left( \bra{+}^{\otimes N}
      \mathrm{e}^{\mathrm{i} J_1 t \hat{H}_0} \right) \left(
      \mathrm{e}^{-\mathrm{i} J_2 t \hat{H}_0} \ket{+}^{\otimes N} \right) \\
    &= \bra{+}^{\otimes N} \mathrm{e}^{\mathrm{i} \Delta J t \hat{H}_0}
      \ket{+}^{\otimes N} \,.
\end{aligned}
\end{equation}
The initial state is an equal superposition of computational basis states,
\begin{equation}
    \ket{+}^{\otimes N} = \frac{1}{2^{N/2}} \sum_{x \in \{0,1\}^N} \ket{x} \,.
\end{equation}
Because $\hat{H}_0$ is a function of $\hat{S}_z$, it is diagonal in the
  computational basis. Acting on a basis state $\ket{x}$ with Hamming weight
  $m$, it gives
\begin{equation}
\begin{split}
    \hat{H}_0 \ket{x} & = 2\hat{S}_z^2\ket{x} \\
    & = \frac{1}{2}\left(\sum_{i=1}^N (-1)^{x_i}\right)^2\ket{x}
    = \frac{(N-2m)^2}{2}\ket{x} \,.
\end{split}
\end{equation}
The overlap can therefore be written as a binomial average over Hamming weights:
\begin{equation}
\begin{aligned}
    \braket{\psi(J_1) | \psi(J_2)}
    = \frac{1}{2^N}
    \sum_{m=0}^N \binom{N}{m}
    \exp\left( \mathrm{i} \Delta J t \frac{(N-2m)^2}{2} \right) \,.
    \label{eq:exact_overlap_ising}
\end{aligned}
\end{equation}
This expression determines all entries of the Gram matrix used for the pretty
  good measurement.

\subsection{Quantum Fisher information}
\label{app:ising_qfi}
We derive here the QFI quoted in Eq.~\eqref{eq:ising_qfi}. The encoded state
  $\ket{\psi(J)} = \mathrm{e}^{-\mathrm{i}Jt\hat{H}_0}\ket{+}^{\otimes N}$ is
  generated unitarily from a $J$-independent input. The generator with respect
  to $J$ is $\hat{G} = t\hat{H}_0$, and for a pure state the QFI is four times
  the variance of the generator,
\begin{equation}
    \mathcal{F}(J) = 4\,\mathrm{Var}_{\psi_0}(\hat{G}) = 4 t^2 \,
      \mathrm{Var}_{\psi_0}(\hat{H}_0) \,.
    \label{eq:ising_qfi_variance}
\end{equation}
This is independent of $J$ because the generator does not depend on it. The
  variance is most easily evaluated by rewriting the generator in terms of the
  total magnetization $\hat{S} \equiv \sum_{i=1}^N \hat{Z}_i = 2\hat{S}_z$,
\begin{equation}
\begin{aligned}
    \hat{H}_0 = 2\hat{S}_z^2 = \frac{1}{2}\hat{S}^2 \,.
\end{aligned}
\label{eq:ising_H0_as_magnetization}
\end{equation}
On the initial state $\ket{+}^{\otimes N}$, each $\hat{Z}_i$ takes the values
  $\pm 1$ with equal probability and the spins are independent, i.e., each
  $\hat{Z}_i$ is a Rademacher random variable. Consequently, the single-spin
  moments are $\braket{\hat{Z}_i} = \braket{\hat{Z}_i^3} = 0$ and
  $\braket{\hat{Z}_i^2} = \braket{\hat{Z}_i^4} = 1$ (the latter because
  $\hat{Z}_i^2 = \hat{\id}$), and expectations of distinct spins factorize. The
  required moments of $\hat{S}$ then follow by collecting the surviving index
  combinations. For $\braket{\hat{S}^2} = \sum_{i,j} \braket{\hat{Z}_i
  \hat{Z}_j}$, only the diagonal terms $i = j$ survive, giving
\begin{equation}
    \braket{\hat{S}^2} = \sum_{i=1}^N \braket{\hat{Z}_i^2} = N \,.
\end{equation}
For $\braket{\hat{S}^4} = \sum_{i,j,k,l} \braket{\hat{Z}_i \hat{Z}_j \hat{Z}_k
  \hat{Z}_l}$, a term is nonzero only when every index appears an even number of
  times, so either all four indices coincide ($N$ terms, each equal to
  $\braket{\hat{Z}_i^4} = 1$) or they form two distinct pairs ($3$ ways to pair
  the four slots, times $N(N-1)$ ordered choices of the two values, each term
  equal to $\braket{Z^2}^2 = 1$), yielding
\begin{equation}
\begin{split}
    \braket{\hat{S}^4} & = N + 3 N(N-1) \\
    & = 3N^2 - 2N \,.
\end{split}
\end{equation}
Therefore,
\begin{equation}
\begin{split}
    \mathrm{Var}_{\psi_0}(\hat{H}_0) &=
      \frac{1}{4}\,\mathrm{Var}_{\psi_0}(\hat{S}^2) = \frac{1}{4}\left( 3N^2 -
      2N - N^2 \right) \\
    &= \frac{N(N-1)}{2} \,.
\end{split}
    \label{eq:ising_H0_variance}
\end{equation}
Substituting into Eq.~\eqref{eq:ising_qfi_variance} yields the output-state QFI
  quoted in the main text, $\mathcal{F}(J) = 2 \, N(N-1) \, t^2$, independent of
  $J$ and scaling as $\mathcal{F} \sim 2 N^2 t^2$ for large $N$.

\subsection{Lipschitz constant}
\label{app:lipschitz_constant_ising}
The achievability result requires the Lipschitz constant $L$ of the prescribed
  family of states with respect to the trace distance, i.e., the smallest
  constant such that $D(\hat{\rho}_{J_1}, \hat{\rho}_{J_2}) \le L\,|J_1 - J_2|$.
  We show that for this strategy $L$ is fixed by the same variance that
  determines the output-state QFI, and equals $\tfrac{1}{2}\sqrt{\mathcal{F}}$.

Because the states are pure and generated by the $J$-independent generator
  $\hat{G} = t\hat{H}_0$, the rate of change of $\hat{\rho}_J =
  \ket{\psi(J)}\!\bra{\psi(J)}$ is $\partial_J \hat{\rho}_J = -i[\hat{G},
  \hat{\rho}_J]$. For a pure state, the Hermitian derivative
  $-i[\hat{G},\hat{\rho}_J]$ has rank at most two and nonzero eigenvalues $\pm
  \Delta G$, where $\Delta G \equiv \sqrt{\mathrm{Var}_{\psi_0}(\hat{G})}$, so
  the speed at which the state moves in trace distance is
\begin{equation}
    \frac{1}{2}\| \partial_J \hat{\rho}_J \|_1 = \Delta G =
      \frac{1}{2}\sqrt{\mathcal{F}} \,,
    \label{eq:ising_trace_velocity}
\end{equation}
where we used $\| \partial_J \hat{\rho}_J \|_1 = 2\Delta G$ and the pure-state
  identity $\mathcal{F} = 4\Delta G^2$. This is independent of $J$ since
  $\hat{G}$ is. Integrating along the path from $J_1$ to $J_2$ and using the
  triangle inequality for the trace norm,
\begin{equation}
\begin{split}
    D(\hat{\rho}_{J_1}, \hat{\rho}_{J_2}) & = \frac{1}{2}\left\|
      \int_{J_1}^{J_2} \partial_J \hat{\rho}_J \, \dd J \right\|_1 \\
    & \le \frac{1}{2}\int_{J_1}^{J_2} \| \partial_J \hat{\rho}_J \|_1 \, \dd J =
      \frac{1}{2}\sqrt{\mathcal{F}}\,|J_1 - J_2| \,,
\end{split}
\end{equation}
so that, using $\mathcal{F} = 2N(N-1)t^2$ from Eq.~\eqref{eq:ising_qfi}, we
  obtain a rigorous, global Lipschitz constant
\begin{equation}
    L = \frac{1}{2}\sqrt{\mathcal{F}} = t\sqrt{\frac{N(N-1)}{2}} \,.
    \label{eq:ising_lipschitz}
\end{equation}

\section{Joint-state overlap, phase-averaged fidelity, and Lipschitz constant
  for dissipative frequency learning}
\label{app:dissipative_fidelity}
This appendix derives the resonant rotating-frame Lindbladian used in
  Sec.~\ref{subsec:dissipative_frequency_learning}. Within this rotating-wave
  model, we then obtain the pure joint spin--field overlap in
  Eq.~\eqref{eq:joint_overlap_dissipative_main} directly from the two-sided
  master equation for arbitrary candidate Rabi frequencies. A
  photon-number-resolved deformation of this equation is subsequently used to
  show that, after a uniform average over the drive phase, the same expression
  gives the fidelity in Eq.~\eqref{eq:fidelity_dissipative_main} on the
  restricted domain $0\le\Omega_i,\Omega_j\le\gamma/2$. The resulting direct-sum
  decomposition also permits an exact evaluation of the zero-tolerance and band
  PGM success probabilities for the phase-averaged mixed states. Finally, the
  common overlap-and-fidelity formula yields a fidelity-derived Lipschitz
  constant for the two state families.

\subsection{Spin Lindbladian in the resonant rotating frame}
\label{app:dissipative_spin_lindbladian}
Let $\omega_{\mathrm{d}}$ denote the known angular frequency of both the spin
  transition and the resonant classical drive. With $\hbar=1$, the
  laboratory-frame spin Hamiltonian is
\begin{equation}
    \hat{H}_{\mathrm{lab}}(t)
    =
    \frac{\omega_{\mathrm{d}}}{2}\hat{\sigma}_z
    +
    \Omega
    \cos\left(\omega_{\mathrm{d}}t+\phi\right)
    \hat{\sigma}_x.
    \label{eq:dissipative_lab_frame_hamiltonian}
\end{equation}
This convention makes $\Omega$ the resonant Rabi angular frequency. Within the
  vacuum Markov approximation, spontaneous emission at rate $\gamma$ gives
\begin{equation}
    \begin{aligned}
        \frac{\dd\hat{\rho}_{\mathrm{lab}}}{\dd t}
        &=
        -\mathrm{i}
        \left[
        \hat{H}_{\mathrm{lab}}(t),
        \hat{\rho}_{\mathrm{lab}}
        \right]
        \\
        &\quad+
        \gamma
        \left(
        \hat{\sigma}_-\hat{\rho}_{\mathrm{lab}}\hat{\sigma}_+
        -
        \frac{1}{2}
        \left\{
        \hat{\sigma}_+\hat{\sigma}_-,
        \hat{\rho}_{\mathrm{lab}}
        \right\}
        \right) \; .
    \end{aligned}
    \label{eq:dissipative_lab_frame_lindbladian}
\end{equation}
Define the resonant rotating frame by
\begin{equation}
    \hat{R}_{\mathrm{d}}(t)
    \equiv
    \myexp{-\mathrm{i}\omega_{\mathrm{d}}t\hat{\sigma}_z/2},
    \qquad
    \hat{\rho}(t)
    \equiv
    \hat{R}_{\mathrm{d}}(t)^\dagger
    \hat{\rho}_{\mathrm{lab}}(t)
    \hat{R}_{\mathrm{d}}(t).
    \label{eq:dissipative_rotating_frame_definition}
\end{equation}
The Hamiltonian generating $\hat{\rho}(t)$ is
\begin{equation}
    \hat{H}_{\mathrm{rot}}(t)
    =
    \hat{R}_{\mathrm{d}}(t)^\dagger
    \hat{H}_{\mathrm{lab}}(t)
    \hat{R}_{\mathrm{d}}(t)
    -
    \mathrm{i}
    \hat{R}_{\mathrm{d}}(t)^\dagger
    \dot{\hat{R}}_{\mathrm{d}}(t).
    \label{eq:dissipative_rotating_frame_generator}
\end{equation}
Using $\hat{\sigma}_x=\hat{\sigma}_++\hat{\sigma}_-$ and
$\hat{R}_{\mathrm{d}}^\dagger\hat{\sigma}_\pm\hat{R}_{\mathrm{d}}
=\myexp{\pm\mathrm{i}\omega_{\mathrm{d}}t}\hat{\sigma}_\pm$, the two terms
  proportional to $\hat{\sigma}_z$ in
  Eq.~\eqref{eq:dissipative_rotating_frame_generator} cancel. The exact resonant
  rotating-frame Hamiltonian is therefore
\begin{equation}
    \begin{aligned}
    \hat{H}_{\mathrm{rot}}(t)
    ={}&
    \frac{\Omega}{2}
    \left(
    \myexp{-\mathrm{i}\phi}\hat{\sigma}_+
    +
    \myexp{\mathrm{i}\phi}\hat{\sigma}_-
    \right)
    \\
    &+
    \frac{\Omega}{2}
    \left(
    \myexp{\mathrm{i}(2\omega_{\mathrm{d}}t+\phi)}
    \hat{\sigma}_+
    +
    \myexp{-\mathrm{i}(2\omega_{\mathrm{d}}t+\phi)}
    \hat{\sigma}_-
    \right).
    \end{aligned}
    \label{eq:dissipative_exact_rotating_frame_hamiltonian}
\end{equation}
For $|\Omega|,\gamma\ll\omega_{\mathrm{d}}$, the second line of
  Eq.~\eqref{eq:dissipative_exact_rotating_frame_hamiltonian} averages to zero
  on the timescale of the spin evolution. The rotating-wave Hamiltonian is thus
\begin{equation}
    \hat{H}_{\Omega,\phi}
    =
    \frac{\Omega}{2}
    \left(
    \myexp{-\mathrm{i}\phi}\hat{\sigma}_+
    +
    \myexp{\mathrm{i}\phi}\hat{\sigma}_-
    \right).
    \label{eq:dissipative_rotating_wave_hamiltonian}
\end{equation}
The jump operator transforms as
$\hat{R}_{\mathrm{d}}^\dagger\hat{\sigma}_-\hat{R}_{\mathrm{d}}
=\myexp{-\mathrm{i}\omega_{\mathrm{d}}t}\hat{\sigma}_-$.
This scalar phase cancels between the two sides of the emission term and leaves
  $\hat{\sigma}_+\hat{\sigma}_-$ invariant. Consequently, the amplitude-damping
  dissipator is unchanged, and the rotating-frame spin Lindbladian is
\begin{equation}
    \frac{\dd\hat{\rho}}{\dd t}
    =
    -\mathrm{i}
    \left[
    \hat{H}_{\Omega,\phi},
    \hat{\rho}
    \right]
    +
    \gamma
    \left(
    \hat{\sigma}_-\hat{\rho}\hat{\sigma}_+
    -
    \frac{1}{2}
    \left\{
    \hat{\sigma}_+\hat{\sigma}_-,
    \hat{\rho}
    \right\}
    \right).
    \label{eq:dissipative_spin_lindbladian_app}
\end{equation}
Equation~\eqref{eq:dissipative_spin_lindbladian_app} is
  Eq.~\eqref{eq:master_equation_qubit}. At the level of the joint spin--field
  pure state, the phase multiplying the jump operator can equivalently be
  removed by a known rotating-frame transformation of the output field. Both
  rotating-frame transformations are independent of $\Omega$ and therefore
  preserve all pairwise overlaps and fidelities considered below.

\subsection{Two-sided master equation and pure joint-state overlap}
\label{app:known_phase_joint_overlap}
When the drive phase is known, we set $\phi=0$. For two candidate Rabi
  frequencies $\Omega_i$ and $\Omega_j$, define
\begin{equation}
    \hat{H}_k
    \equiv
    \hat{H}_{\Omega_k,0}
    =
    \frac{\Omega_k}{2}\hat{\sigma}_x,
    \qquad
    k\in\{i,j\} \; ,
    \label{eq:dissipative_phase_zero_hamiltonians}
\end{equation}
with $\Delta\Omega_{ij} \equiv \Omega_i-\Omega_j$. The spin is initialized in
  $\ket{g}$, the field is initialized in the vacuum, and the only decay channel
  is $\sqrt{\gamma}\hat{\sigma}_-$. Let
\begin{equation}
    \hat{X}^{ij}(t)
    \equiv
    \trenv{
    \ket{\Psi_{\Omega_i,0}(t)}
    \!\bra{\Psi_{\Omega_j,0}(t)}
    }
    \label{eq:dissipative_two_sided_cross_operator}
\end{equation}
be the spin-only operator obtained by tracing out the environment after evolving
  the ket and bra sides at frequencies $\Omega_i$ and $\Omega_j$, respectively.
  The ket side evolves under $\hat{H}_i$, whereas the bra side evolves under
  $\hat{H}_j$. To derive its equation of motion explicitly, consider an
  infinitesimal time interval $\dd t$. To first order in $\dd t$, for candidate
  Rabi frequency $\Omega_k$, the Kraus operators corresponding to no emission
  and one-photon emission are
\begin{equation}
    \begin{aligned}
        \hat{K}^{(k)}_0
        &=
        \hat{\id}
        -
        \left(
        \mathrm{i}\hat{H}_k
        +
        \frac{\gamma}{2}\hat{\sigma}_+\hat{\sigma}_-
        \right)
        \dd t,
        \\
        \hat{K}_1
        &=
        \sqrt{\gamma\,\dd t}\,
        \hat{\sigma}_-.
    \end{aligned}
    \label{eq:dissipative_infinitesimal_operators}
\end{equation}
The vacuum and one-photon states of the output increment are orthogonal. Tracing
  over the environment~\cite{gammelmarkFisherInformationQuantum2014} gives
\begin{equation}
    \hat{X}^{ij}(t+\dd t)
    =
    \hat{K}^{(i)}_0
    \hat{X}^{ij}(t)
    \hat{K}^{(j)\dagger}_0
    +
    \hat{K}_1
    \hat{X}^{ij}(t)
    \hat{K}^{\dagger}_1
    +
    \mathcal{O}(\dd t^2).
    \label{eq:dissipative_two_sided_increment}
\end{equation}
Expanding Eq.~\eqref{eq:dissipative_two_sided_increment} and taking the limit
  $\dd t\to0$ yields the two-sided master
  equation~\cite{molmerHypothesisTestingOpen2015}
\begin{equation}
    \begin{aligned}
    \frac{\dd\hat{X}^{ij}}{\dd t}
    ={}&
    -\mathrm{i}
    \left(
    \hat{H}_i\hat{X}^{ij}
    -
    \hat{X}^{ij}\hat{H}_j
    \right)
    \\
    &+
    \gamma
    \left(
    \hat{\sigma}_-\hat{X}^{ij}\hat{\sigma}_+
    -
    \frac{1}{2}\hat{\sigma}_+\hat{\sigma}_-\hat{X}^{ij}
    -
    \frac{1}{2}\hat{X}^{ij}\hat{\sigma}_+\hat{\sigma}_-
    \right) \; ,
    \end{aligned}
    \label{eq:dissipative_two_sided_master_equation}
\end{equation}
and $\hat{X}^{ij}(0) = \ket{g}\!\bra{g}$. When $\Omega_i\ne\Omega_j$, the
  coherent term is not a commutator, and
  Eq.~\eqref{eq:dissipative_two_sided_master_equation} is neither trace
  preserving nor a master equation for a density operator. Instead, the trace of
  $\hat{X}^{ij}(t)$ is precisely the overlap of the two pure joint
  states~\cite{molmerHypothesisTestingOpen2015}:
\begin{equation}
    \trsquare{\hat{X}^{ij}(t)}
    =
    \braket{\Psi_{\Omega_j,0}(t)|\Psi_{\Omega_i,0}(t)} \,.
    \label{eq:dissipative_cross_operator_trace}
\end{equation}
We now solve Eq.~\eqref{eq:dissipative_two_sided_master_equation} explicitly. In
  the ordered basis $\{\ket{e},\ket{g}\}$, write
\begin{equation}
    \hat{X}^{ij}
    =
    \begin{pmatrix}
        x_{ee} & x_{eg} \\
        x_{ge} & x_{gg}
    \end{pmatrix}.
\end{equation}
Substitution of Eq.~\eqref{eq:dissipative_phase_zero_hamiltonians} into
  Eq.~\eqref{eq:dissipative_two_sided_master_equation} gives all four component
  equations:
\begin{equation}
    \begin{aligned}
    \dot{x}_{ee}
    &=
    -\frac{\mathrm{i}}{2}
    \left(
    \Omega_i x_{ge}
    -
    \Omega_j x_{eg}
    \right)
    -
    \gamma x_{ee},
    \\
    \dot{x}_{gg}
    &=
    -\frac{\mathrm{i}}{2}
    \left(
    \Omega_i x_{eg}
    -
    \Omega_j x_{ge}
    \right)
    +
    \gamma x_{ee},
    \\
    \dot{x}_{eg}
    &=
    -\frac{\mathrm{i}}{2}
    \left(
    \Omega_i x_{gg}
    -
    \Omega_j x_{ee}
    \right)
    -
    \frac{\gamma}{2}x_{eg},
    \\
    \dot{x}_{ge}
    &=
    -\frac{\mathrm{i}}{2}
    \left(
    \Omega_i x_{ee}
    -
    \Omega_j x_{gg}
    \right)
    -
    \frac{\gamma}{2}x_{ge}.
    \end{aligned}
    \label{eq:dissipative_two_sided_components}
\end{equation}
The initial conditions are $x_{gg}(0)=1$ and
$x_{ee}(0)=x_{eg}(0)=x_{ge}(0)=0$.
  Equation~\eqref{eq:dissipative_two_sided_components} preserves the real linear
  subspace in which $x_{ee}$ and $x_{gg}$ are real while $x_{eg}$ and $x_{ge}$
  are purely imaginary. Define the real variables

\begin{equation}
    S(t)
    \equiv
    \trsquare{\hat{X}^{ij}(t)} \; ,
\end{equation}
and
\begin{equation}
    C(t)
    \equiv
    -\mathrm{i}
    \left(
    \bra{e}\hat{X}^{ij}(t)\ket{g}
    +
    \bra{g}\hat{X}^{ij}(t)\ket{e}
    \right) \; .
\end{equation}
Adding the first two equations in
  Eq.~\eqref{eq:dissipative_two_sided_components} gives the equation for $S$.
  Adding the last two equations and multiplying by $-\mathrm{i}$ gives the
  equation for $C$. The resulting closed system is
\begin{equation}
    \dot S
    =
    \frac{\Delta\Omega_{ij}}{2}C,
    \qquad
    \dot C
    =
    -\frac{\Delta\Omega_{ij}}{2}S
    -
    \frac{\gamma}{2}C,
    \label{eq:known_phase_joint_closed_system}
\end{equation}
with $S(0)=1$ and $C(0)=0$. Eliminating $C$ gives
\begin{equation}
    \ddot S
    +
    \frac{\gamma}{2}\dot S
    +
    \frac{\Delta\Omega_{ij}^2}{4}S
    =
    0,
    \qquad
    S(0)=1,
    \quad
    \dot S(0)=0.
    \label{eq:known_phase_joint_damped_oscillator}
\end{equation}
The characteristic roots of Eq.~\eqref{eq:known_phase_joint_damped_oscillator}
  are
\begin{equation}
    r_\pm
    =
    -\frac{\gamma}{4}
    \pm
    \mathrm{i}\nu_{ij},
    \qquad
    \nu_{ij}^2
    =
    \frac{\Delta\Omega_{ij}^2}{4}
    -
    \frac{\gamma^2}{16},
    \label{eq:known_phase_joint_characteristic_roots}
\end{equation}
where $\nu_{ij}$ agrees with Eq.~\eqref{eq:dissipative_overlap_rate}. Imposing
  the initial conditions determines both variables:
\begin{equation}
    \begin{aligned}
    S(t)
    &=
    \myexp{-\gamma t/4}
    \left[
    \cos\left(\nu_{ij}t\right)
    +
    \frac{\gamma}{4\nu_{ij}}
    \sin\left(\nu_{ij}t\right)
    \right],
    \\
    C(t)
    &=
    -\frac{\Delta\Omega_{ij}}{2\nu_{ij}}
    \myexp{-\gamma t/4}
    \sin\left(\nu_{ij}t\right).
    \end{aligned}
    \label{eq:known_phase_joint_solution}
\end{equation}
Combining the first line of Eq.~\eqref{eq:known_phase_joint_solution} with
  Eq.~\eqref{eq:dissipative_cross_operator_trace} gives
\begin{equation}
    \begin{aligned}
    \braket{\Psi_{\Omega_j,0}(t)|\Psi_{\Omega_i,0}(t)}
    &=
    \myexp{-\gamma t/4}
    \Biggl[
    \cos\left(\nu_{ij}t\right)
    \\
    &\qquad+
    \frac{\gamma}{4\nu_{ij}}
    \sin\left(\nu_{ij}t\right)
    \Biggr].
    \end{aligned}
    \label{eq:joint_overlap_dissipative_app}
\end{equation}
Within the rotating-wave model of
  Eq.~\eqref{eq:dissipative_spin_lindbladian_app},
  Eq.~\eqref{eq:joint_overlap_dissipative_app} is valid for arbitrary real
  $\Omega_i$ and $\Omega_j$. When $|\Delta\Omega_{ij}|<\gamma/2$, one has
  $\nu_{ij}=\mathrm{i}\lambda_{ij}$, with $\lambda_{ij}$ defined in
  Eq.~\eqref{eq:overdamped_fidelity_rate}, and the trigonometric functions are
  replaced by their hyperbolic analytic continuations. At
  $|\Delta\Omega_{ij}|=\gamma/2$, the limit of the overlap is $\myexp{-\gamma
  t/4}(1+\gamma t/4)$. Thus Eq.~\eqref{eq:joint_overlap_dissipative_app} proves
  the all-frequency overlap in Eq.~\eqref{eq:joint_overlap_dissipative_main}.

To analyze the phase-averaged family, we next resolve the two-sided evolution by
  the emitted photon number.

\subsection{Photon-number-resolved two-sided master equation}
\label{app:photon_resolved_two_sided}
We introduce a counting field $z$ as an auxiliary generating variable that will
  help us separate the sectors of the output containing different numbers of
  emitted photons. Let $\hat{X}^{ij}_m(t)$ denote the spin operator obtained
  from trajectories containing exactly $m$ emissions, with the ket and bra sides
  evolved at frequencies $\Omega_i$ and $\Omega_j$, respectively. Assigning a
  factor $z$ to every emission event weights such a trajectory by $z^m$, and the
  corresponding tilted spin operator $\hat{X}^{ij}_z(t)$ can be written as
\begin{equation}
    \hat{X}^{ij}_z(t)
    =
    \sum_{m=0}^{\infty}
    z^m\hat{X}^{ij}_{m}(t) \; .
    \label{eq:photon_resolved_expansion}
\end{equation}
With each emission marked by $z$, the infinitesimal evolution of the tilted
  operator is
\begin{equation}
    \hat{X}^{ij}_z(t+\dd t)
    =
    \hat{K}^{(i)}_0
    \hat{X}^{ij}_z(t)
    \hat{K}^{(j)\dagger}_0
    +
    z \hat{K}_1
    \hat{X}^{ij}_z(t)
    \hat{K}^{\dagger}_1
    +
    \mathcal{O}(\dd t^2).
    \label{eq:dissipative_two_sided_increment_tilted}
\end{equation}
See~\cite{garrahanThermodynamicsQuantum2010} for an application of this
  technique to regular master equations. Notice that here we apply it to the
  two-sided master equation of Ref.~\cite{molmerHypothesisTestingOpen2015}. The
  tilted operator obeys the differential equation
\begin{equation}
    \begin{aligned}
    \frac{\dd \hat{X}^{ij}_z}{\dd t}
    ={}&
    -\mathrm{i}
    \left(
    \hat{H}_i\hat{X}^{ij}_z
    -
    \hat{X}^{ij}_z\hat{H}_j
    \right)
    \\
    &+
    \gamma
    \left(
    z\hat{\sigma}_-\hat{X}^{ij}_z\hat{\sigma}_+
    -
    \frac{1}{2}\hat{\sigma}_+\hat{\sigma}_-\hat{X}^{ij}_z
    -
    \frac{1}{2}\hat{X}^{ij}_z\hat{\sigma}_+\hat{\sigma}_-
    \right),
    \end{aligned}
    \label{eq:tilted_two_sided_master_equation}
\end{equation}
with $\hat{X}^{ij}_z(0)=\ket{g}\!\bra{g}$. Setting $z=1$ assigns unit weight to
  every photon number and recovers
  Eq.~\eqref{eq:dissipative_two_sided_master_equation}, whereas $z=0$ retains
  only the no-emission sector.

Substituting the photon-number-resolved expansion of
  Eq.~\eqref{eq:photon_resolved_expansion} into
  Eq.~\eqref{eq:tilted_two_sided_master_equation} and equating the coefficients
  of $z^m$ on both sides gives the hierarchy
\begin{equation}
    \begin{aligned}
    \frac{\dd \hat{X}^{ij}_{m}}{\dd t}
    &=
    -\mathrm{i}
    \left(
    \hat{H}_i\hat{X}^{ij}_{m}
    -
    \hat{X}^{ij}_{m}\hat{H}_j
    \right) \\
    &-
    \frac{\gamma}{2}
    \left(
    \hat{\sigma}_+\hat{\sigma}_-\hat{X}^{ij}_{m}
    +
    \hat{X}^{ij}_{m}\hat{\sigma}_+\hat{\sigma}_-
    \right) +
    \gamma
    \hat{\sigma}_-
    \hat{X}^{ij}_{m-1}
    \hat{\sigma}_+,
    \end{aligned}
    \label{eq:photon_resolved_two_sided_hierarchy}
\end{equation}
where $\hat{X}^{ij}_{0}(0)=\ket{g}\!\bra{g}$, $\hat{X}^{ij}_{m}(0)=0$ for
  $m\ge1$, and $\hat{X}^{ij}_{-1}\equiv0$.

We can write an alternative expression for the photon-number-resolved operators
  $\hat{X}^{ij}_{m}$, which starts from the decomposition into sectors of the
  pure joint state of the probe and the bosonic environment:
\begin{equation}
    \ket{\Psi_{\Omega_k,0}(t)}
    =
    \sum_{m=0}^{\infty}
    \left(
    \ket{g}\ket{\psi^g_{\Omega_k,m}(t)}_E
    +
    \ket{e}\ket{\psi^e_{\Omega_k,m}(t)}_E
    \right),
    \label{eq:joint_state_photon_decomposition}
\end{equation}
where both field vectors in the $m$th term belong to the $m$-photon sector.
  Uniqueness of the hierarchy in
  Eq.~\eqref{eq:photon_resolved_two_sided_hierarchy} and the definition in
  Eq.~\eqref{eq:dissipative_two_sided_cross_operator} give
\begin{equation}
    \hat{X}^{ij}_{m}(t)
    =
    \trenv{
    \ket{\Psi_{\Omega_i,m}(t)}
    \!\bra{\Psi_{\Omega_j,m}(t)}
    },
    \label{eq:photon_resolved_cross_operator}
\end{equation}
where
\begin{equation}
    \ket{\Psi_{\Omega_k,m}(t)}
    \equiv
    \ket{g}\ket{\psi^g_{\Omega_k,m}(t)}_E
    +
    \ket{e}\ket{\psi^e_{\Omega_k,m}(t)}_E.
\end{equation}

\subsection{Uniform phase average and total-excitation-sector fidelity}
\label{app:phase_averaged_joint_fidelity}
For an initial drive phase $\phi$, the joint state of the probe and the
  environment can be written as
\begin{equation}
    \ket{\Psi_{\Omega,\phi}(t)}
    =
    \myexp{-\mathrm{i}\phi\hat{N}_{\mathrm{tot}}}
    \ket{\Psi_{\Omega,0}(t)},
    \qquad
    \hat{N}_{\mathrm{tot}}
    \equiv
    \hat{\sigma}_+\hat{\sigma}_-+\hat{N}_E,
    \label{eq:dissipative_phase_covariance}
\end{equation}
where $\hat{N}_E$ is the photon-number operator of the complete output field on
  $[0,t]$. To verify Eq.~\eqref{eq:dissipative_phase_covariance}, let $\hat
  b(s)$ denote the field annihilation operator. Conjugation by
  $\myexp{-\mathrm{i}\phi\hat N_{\mathrm{tot}}}$ gives
  $\hat\sigma_+\mapsto\myexp{-\mathrm{i}\phi}\hat\sigma_+$,
  $\hat\sigma_-\mapsto\myexp{\mathrm{i}\phi}\hat\sigma_-$, $\hat
  b^\dagger(s)\mapsto\myexp{-\mathrm{i}\phi}\hat b^\dagger(s)$, and $\hat
  b(s)\mapsto\myexp{\mathrm{i}\phi}\hat b(s)$. Hence the rotating-wave coupling
  is invariant, the phase-zero drive Hamiltonian is mapped to $\hat
  H_{\Omega,\phi}$, and the initial state $\ket{g}\ket{\mathrm{vac}}_E$ is
  invariant. A uniform average over the unobserved initial drive phase therefore
  removes coherences between distinct eigenspaces of $\hat{N}_{\mathrm{tot}}$:
\begin{equation}
    \begin{aligned}
    \hat{\rho}^{SE}_{\Omega}(t)
    &=
    \int_0^{2\pi}
    \frac{\dd\phi}{2\pi}\,
    \ket{\Psi_{\Omega,\phi}(t)}
    \!\bra{\Psi_{\Omega,\phi}(t)}
    \\
    &=
    \bigoplus_{n=0}^{\infty}
    \ket{\Xi_{\Omega,n}(t)}
    \!\bra{\Xi_{\Omega,n}(t)}.
    \end{aligned}
    \label{eq:phase_averaged_joint_direct_sum}
\end{equation}
From the photon-number decomposition in
  Eq.~\eqref{eq:joint_state_photon_decomposition}, the vector in the
  total-excitation-$n$ sector is
\begin{equation}
    \ket{\Xi_{\Omega,n}(t)}
    \equiv
    \ket{g}\ket{\psi^g_{\Omega,n}(t)}_E
    +
    \ket{e}\ket{\psi^e_{\Omega,n-1}(t)}_E \; ,
    \label{eq:total_excitation_sector_vector}
\end{equation}
with $\ket{\psi^e_{\Omega,-1}(t)}_E \equiv 0$. The two terms have the same total
  excitation number: the first contains a ground-state spin and $n$ photons,
  whereas the second contains an excited spin and $n-1$ photons. For two
  candidate frequencies, define the sector overlap
\begin{equation}
    c_n^{ij}(t)
    \equiv
    \braket{\Xi_{\Omega_j,n}(t)|\Xi_{\Omega_i,n}(t)}.
    \label{eq:total_excitation_sector_overlap}
\end{equation}
Equations~\eqref{eq:photon_resolved_cross_operator}
  and~\eqref{eq:total_excitation_sector_vector} give
\begin{equation}
    c_n^{ij}(t)
    =
    \bra{g}\hat{X}^{ij}_{n}(t)\ket{g}
    +
    \bra{e}\hat{X}^{ij}_{n-1}(t)\ket{e} \, ,
    \label{eq:total_excitation_overlap_from_photon_coefficients}
\end{equation}
with $\hat{X}^{ij}_{-1}(t)\equiv0$.

The fidelity is additive over common orthogonal direct sums. Since every sector
  block in Eq.~\eqref{eq:phase_averaged_joint_direct_sum} is rank one, its
  contribution is the absolute overlap of the corresponding unnormalized
  vectors. Therefore,
\begin{equation}
    \begin{aligned}
    F &\left(
    \hat{\rho}^{SE}_{\Omega_i}(t),
    \hat{\rho}^{SE}_{\Omega_j}(t)
    \right)
    = \sum_{n=0}^{\infty}
    \left|c_n^{ij}(t)\right| \; .
    \end{aligned}
\label{eq:phase_averaged_joint_fidelity_coefficients}
\end{equation}

\subsection{Sector positivity on the restricted frequency interval}
\label{app:phase_averaged_joint_positivity}
The signs of the sector overlaps can be determined directly from the
  photon-emission trajectories, without evaluating the photon-resolved
  hierarchy. Define the non-Hermitian no-emission Hamiltonian and propagator for
  each candidate frequency on the spin by
\begin{equation}
    \hat{H}_{\mathrm{eff},k}
    \equiv
    \hat{H}_k
    -
    \frac{\mathrm{i}\gamma}{2}\hat{\sigma}_+\hat{\sigma}_-,
    \qquad
    \hat{V}_k(t)
    \equiv
    \myexp{-\mathrm{i}\hat{H}_{\mathrm{eff},k}t},
\end{equation}
and write
\begin{equation}
    \hat{V}_k(t)\ket{g}
    =
    G_k(t)\ket{g}
    +
    E_k(t)\ket{e}.
\end{equation}
Defining
\begin{equation}
    \chi_k
    \equiv
    \sqrt{
    \Omega_k^2
    -
    \frac{\gamma^2}{4}
    },
\end{equation}
direct exponentiation gives
\begin{equation}
    G_k(t)
    =
    \myexp{-\gamma t/4}
    \left[
    \cos\left(\frac{\chi_k t}{2}\right)
    +
    \frac{\gamma}{2\chi_k}
    \sin\left(\frac{\chi_k t}{2}\right)
    \right],
    \label{eq:no_emission_ground_amplitude}
\end{equation}
and
\begin{equation}
    E_k(t)
    =
    -\mathrm{i}
    \frac{\Omega_k}{\chi_k}
    \myexp{-\gamma t/4}
    \sin\left(\frac{\chi_k t}{2}\right).
    \label{eq:no_emission_excited_amplitude}
\end{equation}
These expressions are understood by analytic continuation when $\chi_k$ is
  imaginary and by continuity at $\chi_k=0$. On the domain
  $0\le\Omega_k\le\gamma/2$, define
\begin{equation}
    \kappa_k
    \equiv
    \sqrt{
    \frac{\gamma^2}{4}
    -
    \Omega_k^2
    }
    \ge 0,
\end{equation}
so that $\chi_k=\mathrm{i}\kappa_k$.
  Equations~\eqref{eq:no_emission_ground_amplitude}
  and~\eqref{eq:no_emission_excited_amplitude} then become
\begin{equation}
    \begin{aligned}
    G_k(t)
    &=
    \myexp{-\gamma t/4}
    \left[
    \cosh\left(\frac{\kappa_k t}{2}\right)
    +
    \frac{\gamma}{2\kappa_k}
    \sinh\left(\frac{\kappa_k t}{2}\right)
    \right]
    \ge 0, \\
    E_k(t)
    &=
    -\mathrm{i}
    \frac{\Omega_k}{\kappa_k}
    \myexp{-\gamma t/4}
    \sinh\left(\frac{\kappa_k t}{2}\right).
    \end{aligned}
    \label{eq:overdamped_no_emission_amplitudes}
\end{equation}
The expressions at $\kappa_k=0$ are defined by continuity. For $m\ge1$, let
  $\boldsymbol{t}_m\equiv(t_1,\ldots,t_m)$ belong to the ordered simplex
  $\Delta_m(t)\equiv\{0<t_1<\cdots<t_m<t\}$, set $t_0\equiv0$, and define the
  corresponding $m$-photon time ket by
\begin{equation}
    \ket{\boldsymbol{t}_m}_E
    \equiv
    \hat{b}^{\dagger}(t_m)\cdots
    \hat{b}^{\dagger}(t_1)
    \ket{\mathrm{vac}}_E \, .
\end{equation}
We take $\Delta_0(t)$ to be a singleton with unit measure and
  $\ket{\boldsymbol{t}_0}_E\equiv\ket{\mathrm{vac}}_E$. The photon-resolved
  field vectors in Eq.~\eqref{eq:joint_state_photon_decomposition} are defined
  by
\begin{equation}
    \ket{\psi^s_{\Omega,m}(t)}_E
    =
    \int_{\Delta_m(t)}
    A^s_{\Omega,m}(\boldsymbol{t}_m)
    \ket{\boldsymbol{t}_m}_E
    \,\dd\boldsymbol{t}_m,
    \;
    s\in\{g,e\}.
    \label{eq:ordered_time_field_amplitude_expansion}
\end{equation}
Each emission applies $\sqrt{\gamma}\hat{\sigma}_-$ and resets the spin to
  $\ket{g}$. Hence, for $m\ge1$, the trajectory amplitude ending in the excited
  state is
\begin{equation}
    A^e_{\Omega,m}(t_1,\ldots,t_m)
    =
    \gamma^{m/2}
    E_\Omega(t-t_m)
    \prod_{r=1}^{m}
    E_\Omega(t_r-t_{r-1}).
    \label{eq:ordered_m_photon_excited_amplitude}
\end{equation}
The corresponding ground-state amplitude is given by
\begin{equation}
    A^g_{\Omega,m}(t_1,\ldots,t_m)
    =
    \gamma^{m/2}
    G_\Omega(t-t_m)
    \prod_{r=1}^{m}
    E_\Omega(t_r-t_{r-1}) \; .
    \label{eq:ordered_n_photon_ground_amplitude}
\end{equation}
For $m=0$, the amplitudes are $A^g_{\Omega,0}=G_\Omega(t)$ and
  $A^e_{\Omega,0}=E_\Omega(t)$.
  Equations~\eqref{eq:total_excitation_sector_vector}
  and~\eqref{eq:ordered_time_field_amplitude_expansion} therefore give, for
  $n\ge1$,
\begin{equation}
    \begin{aligned}
    c_n^{ij}(t)
    &={}
    \int_{\Delta_n(t)}
    \left[A^g_{\Omega_j,n}(\boldsymbol{t}_n)\right]^*
    A^g_{\Omega_i,n}(\boldsymbol{t}_n)
    \,\dd\boldsymbol{t}_n
    \\
    &{}+
    \int_{\Delta_{n-1}(t)}
    \left[A^e_{\Omega_j,n-1}(\boldsymbol{t}_{n-1})\right]^*
    A^e_{\Omega_i,n-1}(\boldsymbol{t}_{n-1})
    \,\dd\boldsymbol{t}_{n-1}.
    \end{aligned}
    \label{eq:sector_overlap_from_trajectory_amplitudes}
\end{equation}
For $n=0$, one instead has $c_0^{ij}(t) = G_{\Omega_j}(t)^*G_{\Omega_i}(t) =
  G_{\Omega_j}(t)G_{\Omega_i}(t) \ge 0$ on the stated domain.

In the total-excitation-$n$ sector, the ground-state component contains $n$
  photons and therefore $n$ factors of $E_\Omega$. The excited-state component
  contains $n-1$ photons but also the final factor $E_\Omega(t-t_{n-1})$, and
  therefore contains the same number of such factors. It follows that every
  trajectory amplitude in this sector has the common phase $(-\mathrm{i})^n$:
\begin{equation}
    A^g_{\Omega,n}
    =
    (-\mathrm{i})^n a^g_{\Omega,n},
    \qquad
    A^e_{\Omega,n-1}
    =
    (-\mathrm{i})^n a^e_{\Omega,n-1},
    \label{eq:total_excitation_trajectory_common_phase}
\end{equation}
where the functions $a^g_{\Omega,n}$ and $a^e_{\Omega,n-1}$ are pointwise
  nonnegative. Consequently, the sector overlap is a sum of integrals of
  products of nonnegative functions:
\begin{equation}
    \begin{aligned}
    c_n^{ij}(t)
    &=
    \int_{\Delta_n(t)}
    a^g_{\Omega_j,n}(\boldsymbol{t})\,
    a^g_{\Omega_i,n}(\boldsymbol{t})\,
    \dd\boldsymbol{t}
    \\
    &\quad+
    \int_{\Delta_{n-1}(t)}
    a^e_{\Omega_j,n-1}(\boldsymbol{t})\,
    a^e_{\Omega_i,n-1}(\boldsymbol{t})\,
    \dd\boldsymbol{t}
    \\
    &\ge0,
    \qquad n\ge1.
    \end{aligned}
    \label{eq:total_excitation_overlap_nonnegative}
\end{equation}
This argument requires the individual conditions
  $0\le\Omega_i,\Omega_j\le\gamma/2$. The absolute values in
  Eq.~\eqref{eq:phase_averaged_joint_fidelity_coefficients} may therefore be
  removed on the stated domain, and
\begin{equation}
    \begin{aligned}
    F\left(
    \hat{\rho}^{SE}_{\Omega_i}(t),
    \hat{\rho}^{SE}_{\Omega_j}(t)
    \right)
    &=
    \sum_{n=0}^{\infty}c_n^{ij}(t)
    \\
    &= \trsquare{\hat{X}^{ij}_{z=1}(t)}
    \\
    &=
    \braket{\Psi_{\Omega_j,0}(t)|\Psi_{\Omega_i,0}(t)}.
    \end{aligned}
    \label{eq:total_excitation_overlap_sum_app}
\end{equation}
Using $\nu_{ij}=\mathrm{i}\lambda_{ij}$ in
  Eq.~\eqref{eq:joint_overlap_dissipative_app}, we therefore obtain
\begin{equation}
    \begin{aligned}
    F\left(
    \hat{\rho}^{SE}_{\Omega_i}(t),
    \hat{\rho}^{SE}_{\Omega_j}(t)
    \right) &= \myexp{-\gamma t/4}
    \Biggl[
    \cosh\left(\lambda_{ij}t\right)
    \\
    &\qquad+
    \frac{\gamma}{4\lambda_{ij}}
    \sinh\left(\lambda_{ij}t\right)
    \Biggr].
    \end{aligned}
    \label{eq:phase_averaged_joint_fidelity_overdamped}
\end{equation}
At $\lambda_{ij}=0$, Eq.~\eqref{eq:phase_averaged_joint_fidelity_overdamped} is
  understood by continuity and equals $\myexp{-\gamma t/4}(1+\gamma t/4)$. Thus,
  for $0\le\Omega_i,\Omega_j\le\gamma/2$, the analytic continuation of the
  phase-zero joint-state overlap equals the fidelity of the phase-averaged mixed
  states. Outside this domain, Eq.~\eqref{eq:joint_overlap_dissipative_app}
  remains valid for the phase-zero pure-state overlap, but it need not equal the
  phase-averaged fidelity.

\subsection{Exact PGM performance for the phase-averaged states}
\label{app:phase_averaged_joint_pgm}
The direct-sum decomposition in Eq.~\eqref{eq:phase_averaged_joint_direct_sum}
  permits an exact evaluation of the PGM performance for the phase-averaged
  mixed states. Consider the discrete ensemble
  $\{(\mu_k,\hat{\rho}^{SE}_{\Omega_k}(t))\}_{k=1}^{K(\delta')}$. For each $k$,
  define $p_{kn}(t)\equiv\braket{\Xi_{\Omega_k,n}(t)|\Xi_{\Omega_k,n}(t)}$.
  Because distinct total-excitation sectors are orthogonal, the normalized
  vector $p_{kn}(t)^{-1/2}\ket{\Xi_{\Omega_k,n}(t)}$, whenever $p_{kn}(t)>0$, is
  an eigenvector of $\hat{\rho}^{SE}_{\Omega_k}(t)$ with eigenvalue $p_{kn}(t)$.
  Thus, Eq.~\eqref{eq:phase_averaged_joint_direct_sum} is a spectral
  decomposition, with vanishing eigenvalues omitted.

For each total-excitation sector, define the $K(\delta')\times K(\delta')$ Gram
  matrix
\begin{equation}
    \begin{aligned}
    \big(\boldsymbol{G}^{(n)}\big)_{kk'}
    &\equiv
    \sqrt{\mu_k\mu_{k'}}\,
    \braket{\Xi_{\Omega_k,n}(t)|\Xi_{\Omega_{k'},n}(t)}
    \\
    &=
    \sqrt{\mu_k\mu_{k'}}\,c_n^{k'k}(t) \, .
    \end{aligned}
    \label{eq:phase_averaged_sector_gram}
\end{equation}
The reversal of the indices on $c_n^{k'k}$ follows from the convention in
  Eq.~\eqref{eq:total_excitation_sector_overlap}. Applying the mixed-state
  Gram-matrix construction in Eq.~\eqref{eq:mixed_gram_matrix} to the above
  spectral decomposition and ordering the composite eigenvector labels by
  total-excitation sector gives
\begin{equation}
    \begin{aligned}
    \big(\boldsymbol{G}^{\mathrm{mix}}\big)_{(k,n),(k',m)}
    &=
    \delta_{nm}\big(\boldsymbol{G}^{(n)}\big)_{kk'},
    \\
    \boldsymbol{G}^{\mathrm{mix}}
    &=
    \bigoplus_{n=0}^{\infty}\boldsymbol{G}^{(n)},
    \\
    \sqrt{\boldsymbol{G}^{\mathrm{mix}}}
    &=
    \bigoplus_{n=0}^{\infty}\sqrt{\boldsymbol{G}^{(n)}} \, .
    \end{aligned}
    \label{eq:phase_averaged_mixed_gram_direct_sum}
\end{equation}
The resulting prior-weighted PGM outcome probability is
\begin{equation}
    \mu_k\trsquare{
        \hat{\rho}^{SE}_{\Omega_k}(t)\hat{M}_{\check{k}}
    }
    =
    \sum_{n=0}^{\infty}
    \left|
        \big(\sqrt{\boldsymbol{G}^{(n)}}\big)_{\check{k}k}
    \right|^2 \, .
    \label{eq:phase_averaged_pgm_transition_probability}
\end{equation}
Consequently, the exact Bayesian PGM success probability is
\begin{equation}
    \bar{\mathsf{P}}_{\mathrm{HT}}
    (\mu;\mathcal{M}_{\mathrm{PG}})
    =
    \sum_{n=0}^{\infty}
    \sum_{k=1}^{K(\delta')}
    \left|
        \big(\sqrt{\boldsymbol{G}^{(n)}}\big)_{kk}
    \right|^2 \, ,
    \label{eq:phase_averaged_pgm_success}
\end{equation}
whereas the exact band success probability is
\begin{equation}
    \bar{\mathsf{P}}_{\mathrm{Band}}
    (\mu;\mathcal{M}_{\mathrm{PG}})
    =
    \sum_{n=0}^{\infty}
    \sum_{k=1}^{K(\delta')}
    \sum_{\check{k}\,:\,|\check{k}-k|\le a}
    \left|
        \big(\sqrt{\boldsymbol{G}^{(n)}}\big)_{\check{k}k}
    \right|^2 \, .
    \label{eq:phase_averaged_pgm_band_success}
\end{equation}
The positive square root must be taken separately in each total-excitation
  sector before the squared moduli are summed. Unlike the reduction of
  Eq.~\eqref{eq:phase_averaged_joint_fidelity_overdamped},
  Eqs.~\eqref{eq:phase_averaged_pgm_success}
  and~\eqref{eq:phase_averaged_pgm_band_success} do not require sector
  positivity or the restriction $0\le\Omega_k\le\gamma/2$; outside that
  interval, one retains the signed sector overlaps without replacing them by
  their absolute values.

The sector Gram matrices can be computed directly from the
  photon-number-resolved two-sided hierarchy. For each pair of hypotheses,
  Eq.~\eqref{eq:total_excitation_overlap_from_photon_coefficients} gives
\begin{equation}
    \begin{aligned}
    \big(\boldsymbol{G}^{(n)}\big)_{kk'}
    ={}&
    \sqrt{\mu_k\mu_{k'}}
    \left[
        \bra{g}\hat{X}^{k'k}_{n}(t)\ket{g}
    \right.
    \\
    &\left.
        {}+
        \bra{e}\hat{X}^{k'k}_{n-1}(t)\ket{e}
    \right],
    \end{aligned}
    \label{eq:phase_averaged_sector_gram_from_hierarchy}
\end{equation}
with $\hat{X}^{k'k}_{-1}(t)\equiv0$. One solves
  Eq.~\eqref{eq:photon_resolved_two_sided_hierarchy} for every hypothesis pair
  up to a chosen excitation sector $N$, constructs each positive-semidefinite
  matrix $\boldsymbol{G}^{(n)}$, computes its square root, and accumulates the
  entries selected by Eq.~\eqref{eq:phase_averaged_pgm_band_success}. This
  procedure requires only the two-dimensional spin operators in the hierarchy
  and finite $K(\delta')\times K(\delta')$ matrix square roots.

\subsection{Common fidelity-derived Lipschitz constant}
\label{app:dissipative_common_lipschitz}
We finally derive a Lipschitz constant with respect to the trace distance
  $D(\hat{\rho},\hat{\sigma}) \equiv
  \tfrac{1}{2}\norm{\hat{\rho}-\hat{\sigma}}_1$. Write
  $\Delta\equiv\Delta\Omega_{ij}$ and make the dependence on the difference of
  the frequencies in Eq.~\eqref{eq:known_phase_joint_closed_system} explicit as
  $S(t,\Delta)$ and $C(t,\Delta)$. Those equations imply
\begin{equation}
    \frac{\dd}{\dd t}
    \left(
    S(t,\Delta)^2+C(t,\Delta)^2
    \right)
    =
    -\gamma C(t,\Delta)^2.
    \label{eq:dissipative_overlap_norm_derivative}
\end{equation}
Since $S(0,\Delta)=1$ and $C(0,\Delta)=0$, integration gives
\begin{equation}
    1-S(t,\Delta)^2
    =
    C(t,\Delta)^2
    +
    \gamma
    \int_0^t
    C(s,\Delta)^2\,\dd s.
    \label{eq:dissipative_overlap_deficit_identity}
\end{equation}
The solution for the second variable is
\begin{equation}
    C(s,\Delta)
    =
    -\frac{\Delta}{2\nu}
    \myexp{-\gamma s/4}
    \sin(\nu s),
    \qquad
    \nu^2
    =
    \frac{\Delta^2}{4}
    -
    \frac{\gamma^2}{16},
    \label{eq:dissipative_auxiliary_solution}
\end{equation}
with analytic continuation when $\nu$ is imaginary. For $\Delta\ne0$, define
  $Y_\Delta(s)\equiv C(s,\Delta)/\Delta$. We claim that, for every nonzero real
  gap and every $s\ge0$,
\begin{equation}
    \left|Y_\Delta(s)\right|
    \le
    \frac{1-\myexp{-\gamma s/2}}{\gamma}.
    \label{eq:dissipative_auxiliary_uniform_bound}
\end{equation}
If $|\Delta|>\gamma/2$, then $\nu$ is real and $|\sin(\nu s)|\le|\nu|s$.
  Equation~\eqref{eq:dissipative_auxiliary_solution} therefore gives
  $|Y_\Delta(s)|\le\myexp{-\gamma s/4}s/2$. With $u=\gamma s/4$, the comparison
  with the right-hand side of Eq.~\eqref{eq:dissipative_auxiliary_uniform_bound}
  reduces to $u\le\sinh u$, which is true for $s\ge0$. If $0<|\Delta|<\gamma/2$,
  write $\nu=\mathrm{i}\mu$, where $0<\mu\le\gamma/4$. The monotonicity of
  $\sinh x/x$ on $x>0$ gives
\begin{equation}
    \begin{aligned}
    \left|Y_\Delta(s)\right|
    &=
    \myexp{-\gamma s/4}
    \frac{\sinh(\mu s)}{2\mu} \\
    &\le
    \myexp{-\gamma s/4}
    \frac{\sinh(\gamma s/4)}{\gamma/2}
    =
    \frac{1-\myexp{-\gamma s/2}}{\gamma}.
    \end{aligned}
\end{equation}
The threshold case follows by continuity. Substituting
  Eq.~\eqref{eq:dissipative_auxiliary_uniform_bound} into
  Eq.~\eqref{eq:dissipative_overlap_deficit_identity} yields
\begin{align}
    \frac{1-S(t,\Delta)^2}{\Delta^2}
    &\le
    \frac{\left(1-\myexp{-\gamma t/2}\right)^2}{\gamma^2}
    +
    \gamma
    \int_0^t
    \frac{\left(1-\myexp{-\gamma s/2}\right)^2}{\gamma^2}
    \,\dd s
    \nonumber\\
    &=
    \frac{
    \gamma t
    -
    2
    +
    2\,\myexp{-\gamma t/2}
    }{\gamma^2}.
    \label{eq:dissipative_overlap_global_quadratic_bound}
\end{align}
The overlap $S(t,\Delta)$ is real for every $\Delta$. For the known-phase pure
  states, the trace distance is
\begin{equation}
    D\left(
    \ket{\Psi_{\Omega_i,0}(t)},
    \ket{\Psi_{\Omega_j,0}(t)}
    \right)
    =
    \sqrt{1-S(t,\Delta)^2}.
\end{equation}
Equation~\eqref{eq:dissipative_overlap_global_quadratic_bound} therefore proves
  the global Lipschitz bound
\begin{equation}
    D\left(
    \ket{\Psi_{\Omega_i,0}(t)},
    \ket{\Psi_{\Omega_j,0}(t)}
    \right)
    \le
    L(t)\left|\Omega_i-\Omega_j\right| \; ,
    \label{eq:lipschitz_dissipative_common}
\end{equation}
with
\begin{equation}
    L(t) \equiv \frac{1}{\gamma} \sqrt{\gamma t-2+2\,\myexp{-\gamma t/2}} \; .
\end{equation}

For the phase-averaged mixed states on $0\le\Omega_i,\Omega_j\le\gamma/2$,
  Eqs.~\eqref{eq:total_excitation_overlap_sum_app}
  and~\eqref{eq:phase_averaged_joint_fidelity_overdamped} give $F=S$. The
  Fuchs--van de Graaf inequality then yields
\begin{equation}
    \begin{aligned}
    D\left(
    \hat{\rho}^{SE}_{\Omega_i}(t),
    \hat{\rho}^{SE}_{\Omega_j}(t)
    \right)
    &\le
    \sqrt{
    1-
    F^2\left(
    \hat{\rho}^{SE}_{\Omega_i}(t),
    \hat{\rho}^{SE}_{\Omega_j}(t)
    \right)
    }
    \\
    &=
    \sqrt{1-S(t,\Delta)^2}
    \le
    L(t)\left|\Omega_i-\Omega_j\right|.
    \end{aligned}
    \label{eq:phase_averaged_dissipative_lipschitz}
\end{equation}
Thus, $L(t)$ in Eq.~\eqref{eq:lipschitz_dissipative_common} is the exact global
  trace-distance Lipschitz constant for the pure-state family with known drive
  phase and a valid fidelity-derived global Lipschitz constant for the
  phase-averaged mixed family on the restricted interval. The latter statement
  does not assert that this is the smallest trace-distance Lipschitz constant
  for the mixed family, because the Fuchs--van de Graaf inequality need not be
  saturated.

\end{document}